# Bandedge-state-limited single-photon emission from volumetric quantum design of 2D colloidal quantum wells

Xiao Liang[1,5], Bo Wang[2,5], Yue Yu[2], Pedro Ludwig Hernandez-Martinez[1], Zengshan Xing[2], Lu Ding[3], Vijay Kumar Sharma[1], Tze Chien Sum[2]*, & Hilmi Volkan Demir[1,2,4]*

[1]LUMINOUS! Center of Excellence for Semiconductor Lighting and Displays, The Photonics Institute, School of Electrical and Electronic Engineering, Nanyang Technological University, Singapore, 639798, Singapore

[2]Division of Physics and Applied Physics, School of Physical and Mathematical Sciences, Nanyang Technological University, Singapore 637371, Singapore

[3]Institute of Materials Research and Engineering, A*STAR (Agency for Science, Technology and Research), 2 Fusionopolis Way, #08-03 Innovis, 138634, Singapore

[4]UNAM—Institute of Materials Science and Nanotechnology, The National Nanotechnology Research Center, Department of Electrical and Electronics Engineering, Department of Physics, Bilkent University, Bilkent, Ankara, 06800, Turkey

[5]These authors contribute equally: Xiao Liang, Bo Wang

*Corresponding author email: hvdemir@ntu.edu.sg; Tzechien@ntu.edu.sg

## Abstract

Present-day solution-processable single-photon sources are dominated by three-dimensionally confined colloidal quantum emitters, yet their particle-to-particle variation in single-exciton properties limits reproducibility and scalability. Here, to avoid such heterogeneity, we demonstrate reliable room-temperature single-photon emission from atomically flat two-dimensional (2D) colloidal quantum wells (CQWs) with inherently uniform one-dimensional quantum confinement, despite their long-standing limitations of efficient multiexciton emission and pronounced exciton-surface susceptibility. We resolve these challenges through volumetric quantum design (VQD) of CQWs, yielding a highly localized, single bandedge state. This design laterally confines the bandedge excitonic domain within the exciton coherent area and vertically decouples it from surface states via a thick, strain-relieved quantum-barrier shell that preserves strong confinement, overcoming the daunting thickness-confinement trade-off in 2D CQWs. Statistical single-particle spectroscopy reveal that VQD-CQWs deliver near-blinking-free (on-time: >99.5%) and fluence-insensitive antibunching ($g^{(2)}(0)$: ~0.041), protected by a bandedge-state-filling bottleneck, together with linear polarization of up to 73% under cavity-free conditions, originating from synergistic transition-dipole and electric-field anisotropies. These advances establish 2D CQWs as a viable, homogenous and scalable platform for quantum technologies.

**Keywords:** 2D colloidal quantum wells; single-photon purity; near-blinking-free emission; linear polarization; bandedge excitonic state; quantum barrier.

## Introduction

Single-photon sources, capable of deterministically delivering one photon into a defined optical mode, are indispensable for scalable quantum technologies[1], ranging from quantum communication[2] to fault-tolerant computation[3]. In the solid-state benchmark, epitaxial III-V quantum dots embedded in optical microcavities can approach near-ideal single-photon performance[4], yet their reliance on sophisticated molecular-beam epitaxy and cryogenic operation imposes inherent constraints on large-scale manufacturability and deployment[5]. Colloidal quantum emitters have consequently emerged as attractive candidates owing to their solution processability, low cost, and scalable wet-chemical synthesis[6]. Three-dimensionally (3D) confined colloidal quantum dots (QDs), including perovskite-[7], CdSe-[8], and InP-based systems[9], currently represent the dominant solution-processable single-photon platforms, with notable advances in purity, brightness, and coherence at the single-dot level[10, 11, 12]. Nevertheless, intrinsic size dispersion and structural heterogeneity give rise to substantial dot-to-dot variability in emission energy and exciton dynamics[13], complicating hybrid integration of nanophotonic devices that require precise mode overlap[14] and constraining the scalability of quantum network architectures relying on multi-emitter interference[15].

Quasi-two-dimensional (quasi-2D) II–VI colloidal quantum wells (CQWs) represent a unique excitonic platform that bridges 2D-materials photophysics with colloidal semiconductor chemistry (Fig. 1a)[16]. Typically, CQWs feature monolayer-quantized one-dimensional (1D) quantum confinement defined by vertical thickness and reduced dielectric screening, favoring tightly bound excitons with large oscillator strengths and rapid recombination rate[17]. Critically, monolayer-precise thickness greatly suppresses ensemble inhomogeneity, with single-particle-like emission linewidths already evident at the ensemble level[18], a fundamental distinction from conventional 3D confined colloidal QDs. These features are broadly recognized as key advantages of 2D hosts for quantum emission. Yet, the same 2D nature also introduces fundamental challenges. Since the lateral dimensions of typical CQWs largely exceed the phonon-limited room-temperature exciton coherent area, a single platelet can host multiple spatially decoupled bandedge states[19, 20, 21], increasing the likelihood of multiple bandedge exciton occupation (Fig. 1b). In addition, in-plane momentum conservation restricts Auger recombination pathways and prolongs multiexciton lifetime, which has significantly advanced lasing and optical gain applications[22, 23] but proves unfavorable for antibunched photon emission. Furthermore, the atomically thin geometry and large surface area render excitons highly susceptible to surface trapping, leading to strong blinking[24]. Conventional "giant-shell" strategies that isolate excitonic wavefunctions from surface states in colloidal QDs would substantially weaken thickness-governed quantum confinement in CQWs, imposing a challenging fundamental trade-off between strong antibunching and blinking-free emission[25]. To date, 2D CQWs simultaneously delivering highly pure, near-blinking-free, and polarized room-temperature single-photon emission remain unachieved.

In this work, we realize, for the first time, room-temperature single-photon emission from CdSe-based 2D CQWs that combines fluence-insensitive single-photon purity ($g^{(2)}(0)$: ~0.041), near-blinking-free emission (on-time: >99.5%), and strong linear polarization (degree of linear polarization (DOLP): 59~73%) via a wavefunction-guided volumetric quantum design (VQD) framework (Fig. 1). Through comprehensive structural and spectroscopic analysis, we reveal that VQD-CQWs host a single bandedge excitonic state per platelet, achieved by laterally confining the bandedge domain within the exciton coherent area, which is an essential condition for high-purity antibunching. Along the out-of-plane direction of CQWs, we design a multilayered shell architecture featuring an embedded two-monolayer wide-bandgap ZnS that serves as a quantum barrier. This compact quantum-barrier shell (QBS) design integrates multiple functions that include strain regulation, confinement preservation, and wavefunction isolation, effectively overcoming the daunting thickness-confinement trade-off in 2D systems and proving critical for the simultaneous realization of narrow emission linewidth, high single-photon purity and near-blinking-free performance.

Furthermore, we show that the high single-photon purity of VQD-CQWs remains insensitive to the excitation power, thanks to a bandedge-state-filling bottleneck protection mechanism inherent to the single-state design, thereby enabling reliable operation at saturation power to boost single-photon brightness. This attribute is essential for scalable quantum network applications, where the success probability of multi-photon protocols decreases exponentially with reduced single-photon emission probability[26]. In contrast, excitation-power-induced purity degradation persistently limits conventional spontaneous parametric down-conversion (SPDC) sources as well as certain colloidal emitter systems[26, 27, 28, 29]. Finally, theoretical calculations and optical simulations uncover that the observed linear polarization originates from the combined effects of lateral-confinement-induced transition-dipole anisotropy and thickness-enhanced electric-field anisotropy induced by the dielectric contrast. Collectively, these results establish VQD-CQWs as a viable and scalable platform for quantum technologies.

## Results

### Conceptual framework of volumetric quantum design

As mentioned earlier, phonon-induced decoherence limits the room-temperature exciton coherent area in CdSe CQWs to ~tens of $nm^2$, whereas chemically synthesized CQWs typically exceed 100 $nm^2$ in lateral dimensions. Consequently, a single CdSe CQW can host multiple spatially separated, decoherent bandedge excitonic states, which compromises antibunching. High-purity single-photon emission from CQWs therefore requires confining the excitonic bandedge domain within the coherent area. This condition can be met by introducing a lateral potential gradient that restricts the bandedge domain, ensuring that only a single bandedge state is accessible (Fig. 1c).

The single bandedge state, being spin-degenerate, accommodates at most two excitons. Consequently, efficient Auger-mediated quenching of biexciton emission becomes essential, which is decisively governed by out-of-plane confinement in 2D systems. Moreover, the shell architecture critically determines the degree of exciton coupling to surface states and thus the extent of photoluminescence (PL) intermittency. A structurally uniform shell that preserves strong quantum confinement while providing sufficient thickness to spatially isolate excitons from surface is therefore critical, although some of these requirements are inherently competing in 2D systems.

Under these considerations, we established a wavefunction-guided VQD framework for CdSe-based CQWs along both the in-plane and out-of-plane directions. The theoretical calculation is based on a six-band $k \cdot p$ model[30], with detailed methodologies provided in the Methods and Supplementary Information. The overall designed heterostructure is illustrated in Fig. 1d-g, with the yz cross-section and the corresponding conduction- and valence-band potential profiles shown in Fig. 1f-g, respectively.

Specifically, to achieve a single bandedge state, we start with an in-plane architecture using a 4.5-monolayer CdSe as the core. The lateral potential gradient is introduced by a CdSe to CdS compositional transition (Fig. 1e). The bandedge domain is thereby defined by the size of CdSe region within the gradient core. It should be noted that the relatively small conduction band offset in CdSe/$CdSe_xS_{1-x}$/CdS gradient core with a small CdSe size can induce a quasi-type-II character[31], manifested as a pronounced reduction in electron-hole wavefunction overlap (Fig. 1i). Indeed, this spatial overlap is a critical parameter, as high wavefunction overlap accelerates radiative recombination while simultaneously promoting Auger-mediated biexciton quenching[32, 33]. To mitigate this limitation, we tune the gradient-core with a high aspect ratio to create an effective geometrical boundary that limits electron delocalization (Fig. 1i-j). This increases the calculated overlap from 49% to 77%, consistent with the experimentally observed shortening in PL radiative lifetime (Supplementary Fig. 1). Moreover, enhanced lateral confinement and high-aspect-ratio geometry are also crucial for the linearly polarized emission, as will be discussed later.

For the out-of-plane design, a wide-bandgap ZnS shell (Supplementary Fig. 2) is, in principle, the most favorable choice over conventional CdS[34] (Supplementary Fig. 3) or $Cd_yZn_{1-y}S$ gradient (from CdS to ZnS, Supplementary Fig. 4) shells[35], as wavefunction calculations in Supplementary Fig. 5 reveal enhanced exciton localization and confinement. Experimentally, however, the direct growth of a thick ZnS shell on the CdSe-based gradient core (~12% lattice mismatch) results in pronounced spectral broadening (from ~22 to ~32 nm) as the ZnS thickness increases from 2 to 5 monolayers (Supplementary Fig. 6a-b), in direct correlation with the increasing shell non-uniformity observed under spherical aberration-corrected scanning transmission electron microscopy (STEM, Supplementary Fig. 6d-f).

The atomic-resolution imaging further enables region-resolved lattice-distortion analysis through local Fourier transforms (Supplementary Fig. 6g-j), revealing pronounced in-plane lattice compression within the CdSe-rich core (Supplementary Fig. 6c), together with a measurable deviation from the intrinsic cubic symmetry of the zinc blende lattice. The observed structural non-uniformity and lattice distortions are characteristics of lattice-mismatched planar heteroepitaxy, where elastic strain energy accumulates in the coherently strained layer and, beyond a critical thickness, the system minimizes its total free energy by activating strain-relief pathways that introduce morphological and structural non-uniformity, following the Stranski-Krastanov (S-K) growth paradigm[36, 37]. Accordingly, the emission broadening observed here can be attributed to strain-induced heterogeneity and the resulting local band-structure fluctuations.

In general, uniform and coherent thick-layer epitaxy requires lattice mismatch substantially below 10%, often restricted to only a few percent[38]. Although direct epitaxy of a thick, uniform ZnS shell is hindered by the large lattice mismatch, we observe that ZnS layers remain structurally coherent at modest thicknesses of 2~3 monolayers, as confirmed by detailed aberration-corrected STEM analysis (Supplementary Fig. 7). This behavior is consistent with the wetting-layer regime in S-K type growth, in which coherent thin layer can form prior to strain relaxation through non-uniform growth[39].

These observations motivate the strain-relieved quantum-barrier shell (QBS) design (Supplementary Fig. 8), in which an atomically flat, thin ZnS layer with a wide bandgap is embedded within the shell stack to function as a quantum barrier (Fig. 1f-g). Subsequent overgrowth with materials of larger lattice constants establishes a core@barrier@outer-shell architecture with alternating lattice parameters (Supplementary Fig. 8b), enabling strain redistribution and relaxation. In principle, this alternating process can be iteratively applied to construct ultra-thick shells while preserving structural quality and can be regarded as a universal strategy for 2D CQWs system (Supplementary Fig. 8d).

Under the QBS scheme, quantum confinement and wavefunction localization are governed by the ZnS barrier thickness and the band structure of the outer shell (Supplementary Fig. 9-12). Wavefunction calculations show that when CdS is applied as the outer shell, significant electron tunneling occurs across a 2~3 monolayer ZnS barrier, owing to the large ZnS/CdS conduction-band offset (Supplementary Fig. 10). In contrast, adopting a gradient $Cd_yZn_{1-y}S$ outer shell introduces a gradually increasing conduction-band potential while preserving CdSe@ZnS@CdS heterointerfaces for strain regulation (Supplementary Fig. 11). Within this architecture, electron delocalization becomes primarily controlled by the ZnS barrier thickness, and calculations indicate that 2~3 monolayers are sufficient to effectively localize the electron wavefunction away from the surface (Fig. 1h and Supplementary Fig. 12). Notably, based on the calculated bandgap, the designed QBS configuration with a 2~3 monolayer ZnS barrier preserves quantum confinement at a level comparable to that of a thick ZnS shell (Fig. 1k and Supplementary Fig. 13), resolving the competing demands of strain regulation, confinement preservation, and wavefunction isolation.

## Strain-relieved quantum barrier architecture

The structural evolution under the strain-relieved shell growth strategy is schematically illustrated in Fig. 2a, with representative atomic-resolution aberration-corrected STEM images at each stage shown in Fig. 2b-g. Following the VQD framework, we first synthesize high-aspect ratio CdSe/$CdSe_xS_{1-x}$/CdS gradient-core (GC) CQWs (Supplementary Fig. 14). Anisotropic growth is achieved using anhydrous cadmium acetate under a high anion-to-cation feeding ratio, promoting preferential <100> elongation while suppressing <110> lateral expansion[40]. The CdSe core size is precisely tuned by controlling the CdSe growth duration and the timing of sulfur precursor injection, consistent with our previous report[41, 42]. Statistical analysis over more than 50 platelets yield an average aspect ratio of ~6. Energy-dispersive spectroscopy (EDS) indicates a CdSe core size of ~6 nm x 2.5 nm with a spatial area of ~15 $nm^2$, in agreement with the reduced normalized intensity of the first excitonic absorption relative to 4.5-monolayer CdSe CQWs without a potential gradient (Fig. 2h). Meanwhile, the excitonic FWHM of the GC remains comparable to those of pure CdSe, further indicating that the strong excitonic absorption peak in conventional CdSe CQWs arises from the sum of multiple incoherent bandedge transitions at room temperature. Otherwise, spectral broadening would be expected in GC due to the loss of collective oscillator strength by reducing the CdSe size.

Along the out-of-plane direction, a uniform, atomically thin two-monolayer ZnS shell deposited on the GC, denoted GC@QB, is clearly resolved in Fig. 2c. Subsequently, two additional gradient-shell growth cycles are implemented to complete surface passivation, yielding GC@QB/1S and GC@QB/2S, following the strain-relieved shell growth strategy. Detailed structural, compositional, and ensemble optical characterizations are provided in Supplementary Fig. 15-18. Owing to the Z-contrast in dark-field STEM, the ZnS quantum barrier can be well resolved (Fig. 2f-g and Supplementary Fig. 16d-f). The final GC@QB/2S, with a total thickness of ~5.2 nm (Fig. 2g), preserves a sharp bandedge excitonic absorption and a narrow emission linewidth of 69 meV (Fig. 2i), together with near-unity quantum yield and near-monoexponential ensemble PL decay (Fig. 2j-k). Compared with conventional core@gradient-shell CQWs (~4.0 nm thick) widely used in lasing and light-emitting diodes applications[35, 43], GC@QB/2S exhibits a faster radiative lifetime (13.3 *vs.* 15.8 ns) despite its much greater thickness (Fig. 2k), thanks to the effective confinement preservation enabled by the QBS architecture.

Compared to the pronounced PL linewidth broadening observed during direct thick ZnS growth, the strain-relieved QBS architecture instead leads to linewidth narrowing (Fig. 2l), especially during the early stage of each $Cd_yZn_{1-y}S$ gradient-shell growth cycle, consistent with strain relaxation at the CdSe@ZnS@CdS interfaces featuring alternating lattice constants. Local Fourier transform analysis of atomic-resolution STEM images from individual CQWs reveals a well-preserved symmetric zinc blende lattice within the core region (Supplementary Fig. 16k-l and Supplementary Fig. 17g-i), whereas samples subjected to direct thick ZnS growth exhibit evident in-plane lattice compression. At the ensemble level,

XRD analysis (Fig. 2m) provides further evidence of strain relaxation and improved shell quality, as shown by narrowing of the diffraction peak FWHM and a reduced slope extracted from Williamson-Hall fitting (Fig. 2n). It is also noteworthy that employing pure CdS as the outer shell yields an ultranarrow room-temperature emission linewidth of 17.8 nm (Supplementary Fig. 19), representing, to the best of our knowledge, the narrowest value reported for 2D core@shell CQWs synthesized via hot-injection growth. Although this configuration is not applicable for the single-photon objective pursued here, it underscores the effectiveness of strain management in governing emission properties within the QBS architecture.

## Nearly blinking-free antibunched emission

To further elucidate the important roles of the in-plane bandedge-state limitation and the out-of-plane QBS in governing single-photon performance, we prepared two control samples: one without in-plane potential gradient (CdSe@QB) and the other one without the ZnS quantum barrier (GC/1S). Detailed structural and compositional characterizations of these controls are provided in Supplementary Fig. 20-21, and the key parameters of the VQD-CQW series (GC@QB, GC@QB/1S, GC@QB/2S) and control samples are summarized in Supplementary Table 1. Single-photon purity and PL on-time fraction were then statistically evaluated for these samples by measuring more than 20 individual platelets per species, and the results are summarized in Fig. 3a, with representative PL mapping of isolated CQWs, additional antibunching and PL on-time data shown in Supplementary Fig. 22-28. These comparisons explicitly demonstrate that both the in-plane bandedge states (CdSe@QB vs GC@QB) and the wide-bandgap ZnS quantum-barrier design (GC/1S vs GC@QB/1S) are decisive for achieving ultrahigh-purity single-photon emission. In particular, GC@QB reaches $g^{(2)}(0)$ of 0.03, which, to the best of our knowledge, is the lowest value reported across the 2D CQW family. Notably, GC@QB/2S simultaneously exhibits high single-photon purity (~96%) while also providing near-blinking-free behavior, which will be discussed next.

We further investigated the blinking mechanism through fluorescence lifetime and intensity distribution (FLID) analysis. Distinct brightness-lifetime correlations were observed among different structures (Fig. 3b-f and Supplementary Fig. 29), indicating fundamentally different blinking mechanisms[44]. Specifically, B type (hot carrier) blinking is identified for both CdSe@QB (Fig. 3b) and GC@QB (Fig. 3c), differing primarily in blinking severity. This behavior suggests that hot carriers are captured by high-energy traps prior to relaxation to the bandedge (schematized in Fig. 3k). Moreover, the suppressed blinking in GC@QB is attributed to enhanced electron and hole localization reshaped by the in-plane potential gradient, which reduces the probability of exciton trapping by defects at the surface or platelet periphery, as supported by the calculated charge probability density distributions in Fig. 3g-3h.

GC/1S and GC@QB/1S, which differ in the presence of the ZnS quantum barrier, display A-type (Fig. 3d) and C-type (Fig. 3e) blinking behaviors associated with charging induced Auger recombination and surface-state-mediated dynamics, respectively. This assignment is further supported by the fitting results from power-dependent transient absorption (TA) measurements, as shown in Supplementary Fig. 30-33. GC/1S without the quantum barrier exhibits a pronounced trion decay component (Fig. 3i), whereas the excitonic dynamics of GC@QB/1S is mainly contributed by the single-exciton and biexciton processes (Fig. 3j). These observations demonstrate that the confinement introduced by the ZnS quantum barrier effectively isolates the bandedge exciton from the surface states and mitigates exciton-surface electrostatic interactions, as depicted in Fig. 3l. In the absence of the barrier, increased electron delocalization in GC/1S facilitates long-lived trapping, generating a positively charged recombination center consistent with the elevated trion process observed in TA, and activates efficient nonradiative Auger recombination channel that quenches emission (Fig. 3m-3n).

Based on the C-type blinking mechanism identified in GC@QB/1S, the reason for the near nonblinking behavior observed in GC@QB/2S (Fig. 3f) is thereby more effective surface-state passivation achieved through additional gradient-shell overgrowth. Moreover, individual GC@QB/2S also shows exceptional photostability. Our long-term PL trace measurements demonstrate near blinking free behavior for over 40 minutes (in air) without any encapsulation (Supplementary Fig. 34), showcasing their potential for practical applications. Additionally, we highlight the high particle-to-particle consistency in our statistical single-photon measurements, enabled by the intrinsic uniformity of these atomically flat 2D CQWs.

## State-filling-protected antibunching

As discussed earlier, single-photon brightness, quantified as the probability of generating one photon per excitation cycle, directly governs the scalability of quantum protocols[26]. As shown in Fig. 4a, exciton generation under per excitation event follows Poissonian statistics, therefore increasing excitation power not only enhances the probability of single-exciton generation that boosts brightness, but also drives the system into the multiexciton regime. In VQD-CQWs, the greatly enlarged absorption cross-section (Supplementary Fig. 35) further amplifies this effect and increases brightness (Fig. 4b), which is advantageous for practical implementation. However, the concomitant increase in multiexciton population raises concerns that single-photon purity may be compromised, as widely observed in conventional SPDC technique and certain colloidal systems where power-dependent multiexciton excitation degrades antibunching[26, 27, 28, 29].

Our power-dependent $g^{(2)}(0)$ measurements (Fig. 4c) reveal that the single-photon purity of VQD-CQWs (GC@QB/2S) remain essentially insensitive to excitation power, even at saturation. To elucidate the underlying mechanism, we analyzed TA dynamics at two representative fluences (Fig. 4d and Supplementary Fig. 36). In the low-excitation regime (<N>: 0.57), the 530 nm bleach associated with shell excitation decays synchronously with the rise of the bandedge exciton bleach, evidencing efficient exciton transfer from the shell to the CdSe core (Fig. 4e-f). By contrast, at high excitation density (<N>: 164), this correlation disappears: the bandedge bleach remains pinned at a saturation plateau despite the rapid decay of the shell bleach, suggesting the shell-to-core exciton transfer is kinetically hindered under multiexciton population conditions. These observations, together with the structural features of VQD-CQWs, point to a bandedge-state-filling bottleneck protection mechanism (Fig. 4g-h), where the single bandedge-state accommodates at most two excitons and, once saturated, cannot be rapidly cleared due to its relatively slow decay compared to the ~100 ps exciton transfer timescale, thereby blocks further population transfer and diverts excess excitons into ultrafast exciton-exciton annihilation or nonradiative Auger recombination pathways. This mechanism is further supported by power-dependent, time-resolved PL measurements on a thick GC@QB/2S film. Under femtosecond excitation, a broadband high-energy emission (530~580 nm) emerges immediately after the onset of amplified spontaneous emission (Fig. 4i), a signature of an average population of more than one bandedge excitons per platelet. Streak camera analysis uncovers its cascaded temporal correlation with bandedge biexciton decay (Fig. 4j-k), consistent with the TA dynamics.

To further validate our single-bandedge-state design, we quantitatively calculated the number of bandedge states from TA measurements (Fig. 4l-m). The state number was obtained by dividing the saturation bleach amplitude measured immediately after hot-carrier relaxation by the single-exciton bleach amplitude at long delay time, with a correction for the fraction of platelets occupied by one exciton, for which more calculation details are provided in

Supplementary Fig. 37-38. Our designed GC@QB/2S yields a value of 0.74 (≈1), indicating a single bandedge-state per platelet (Fig. 4l), in stark contrast to the control sample, conventional CdSe/QB without an in-plane gradient, which hosts 2.89 (≈3) bandedge states (Fig. 4m). Collectively, the single bandedge-state design is essential for achieving strong and fluence-insensitive antibunching via the state-filling protected mechanism.

## Emergence of linear polarization emission

II–VI based CQWs with an isotropic zinc blende structure typically exhibit nearly isotropic in-plane transition dipoles, rendering strong linearly polarized emission challenging[45]. Even in the presence of shape anisotropy, dielectric-contrast-induced electric-field anisotropy is substantially weakened by their ultrathin geometry[30]. Consequently, the highest reported DOLP remains limited to 15~25%, leading to longstanding doubt regarding their suitability for single-photon applications requiring well-defined polarization modes.

We demonstrate that our VQD-CQWs (GC@QB/2S) exhibit a high DOLP up to 73%. To directly probe their emission anisotropy, we employed defocused wide-field microscopy[46] (Fig. 5a), which projects the angular radiation distribution of single emitters into characteristic far-field intensity patterns and enables the visualization of a large number of spatially isolated CQWs simultaneously, providing statistical insights into polarization property at the batch level. As shown in Fig. 5b, GC@QB/2S exhibits pronounced bilobed defocused emission patterns, indicating a highly anisotropic far-field radiation distribution. In contrast, conventional squarish CQWs display nearly isotropic donut-like patterns (Fig. 5c), corresponding to a nearly isotropic in-plane emission. Optical simulations of defocused emission (Supplementary Fig. 39-40) capture the formation of defocused imaging patterns during the defocusing process and reproduce the gradual evolution toward bilobed features as the in-plane radiation anisotropy increases, in qualitative agreement with the experimental observations. Consistently, polarization-resolved single-photon measurements on individual GC@QB/2S identify DOLP values of 59~73% (Fig. 5d-e), in agreement with the anisotropic radiation characteristics observed from defocused imaging.

Further theoretical calculations and simulations indicate that the pronounced linear polarization in GC@QB/2S originates from two synergistic mechanisms. First, in zinc blende CdSe, the conduction-bandedge is derived from isotropic Cd 5s orbitals, whereas the valence-bandedge is contributed by the Se 4p orbitals, composed of three orthogonal components denoted as $p_x$, $p_y$, and $p_z$. In 2D CQWs, the emission polarization is therefore governed by the relative weights of the $p_x$ and $p_y$ components near the valence bandedge. We analyzed the valence-band fine structure under lateral confinement within a six-band $k \cdot p$ framework based on the Luttinger-Kohn Hamiltonian, allowing orbital-resolved calculations of the bandedge hole states. Comparative calculations between structures with strong and weak hole confinement (Supplementary Fig. 41) show that strong hole confinement along the short axis lifts the in-plane $p_x/p_y$ degeneracy near the bandedge, as depicted in Fig. 5f. The corresponding orbital-projected eigenstates (Supplementary Fig. 42) reveal that the $p_y$ component is shifted to higher energies, increasing the relative contribution of $p_x$ to the lowest-energy optical transition. This confinement-induced redistribution yields a calculated DOLP of ~35% (Fig. 5g), whereas weakly confined structures retain nearly degenerate $p_x$ and $p_y$ components and exhibit negligible polarization (~1%, Fig. 5h).

Second, beyond the dipole anisotropy, electromagnetic effects further reinforce the polarization anisotropy. The pronounced shape anisotropy of GC@QB/2S induces asymmetric internal electric field redistribution due to the high dielectric contrast between CQWs and its surrounding low index media, as revealed by finite-difference time-domain (FDTD) simulations (Fig. 5i). Although this electric field anisotropy is significantly weakened by the ultrathin slab geometry in conventional CQWs, our strain-relieved shell-growth strategy increases the CQW thickness to ~5.2 nm, which markedly enhances the dielectric-induced internal field anisotropy. Thickness-dependent simulations (Fig. 5j and Supplementary Fig. 46) reveal a progressive enhancement of field anisotropy with increasing thickness (green dashed line), reaching a level that can independently account for ~44% polarization at 5.2 nm (without considering intrinsic dipole transition anisotropy, red dashed line). When combined with the confinement-induced dipole anisotropy discussed above, the calculated overall DOLP quantitatively well matches the experimentally observed values.

Collectively, as depicted in Fig. 5k, strong lateral confinement establishes the initial dipole transition anisotropy, which is subsequently magnified by the thickness-enhanced electric field anisotropy induced by the dielectric contrast. The synergistic action of these two effects results in the exceptionally high DOLP observed in GC@QB/2S.

## Conclusion

In summary, we establish quasi-2D CQWs as a reliable room-temperature single-photon platform via a wavefunction-engineered VQD framework. Individual CQWs were uniquely tailored to host a single, highly spatial localized bandedge state confined within the exciton coherent area. A universal strain-relieved quantum-barrier shell architecture was developed under the VQD strategy, resolving the longstanding thickness-confinement trade-off in quasi-2D CQWs. Collectively, the resulting VQD-CQWs exhibit superior performance as a new class of quantum emitters, including a narrow emission linewidth, near-blinking-free emission (on-time >99.5%), fluence-insensitive single-photon purity with $g^{(2)}(0)$ ~ 0.041 and pronounced linear polarization with a DOLP up to 73% at room temperature, arising from the synergistic effects of lateral-confinement-induced transition-dipole anisotropy and thickness-enhanced electric-field anisotropy. We systematically characterized and analyzed VQD-CQWs and their single-photon properties via aberration-corrected STEM, statistical single-particle spectroscopy, and ultrafast TA spectroscopy, along with comprehensive theoretical calculations and optical simulations. Beyond showcasing their excellent single-photon performance, these combined structural, spectroscopic, and theoretical analyses unravel the underlying blinking mechanism, bandedge exciton saturation dynamics, and the physical origin of linear polarization. Together, these advances provide a solid foundation for quasi-2D colloidal system as scalable, homogenous, high-performance single-photon sources for quantum technologies.

## Online content

Any methods, additional references, Nature Portfolio reporting summaries, source data, extended data, supplementary information, acknowledgements, peer review information; details of author contributions and competing interests; and statements of data and code availability are available at xxxx.

## Methods

### Chemicals

Cadmium nitrate tetrahydrate ($Cd(NO_3)_2 \cdot 4H_2O$, 99.997%, trace metals basis), sodium myristate (≥99%), cadmium acetate ($Cd(OAc)_2$, anhydrous, 99.995%), selenium (Se, 99.99%, trace metals basis), sulfur (S, 99,998%, trace metals basis), zinc acetate ($Zn(OAc)_2$, 99.999%, trace metal basis), 1-octanethiol (≥98.5%), 1-octadecene (ODE, technical-grade), oleic acid (OA, 90%), and oleylamine (OAm, 70%) were purchased from Sigma-Aldrich and used without any further purification.

### Precursor preparation

Before the VQD-CQWs synthesis, cadmium myristate ($Cd(Myr)_2$) powder, 0.1 M S-in-ODE solution, and 0.1 M 1-octanethiol-in-ODE solution were prepared. Specifically, $Cd(Myr)_2$ was prepared following a modified literature protocol. Cadmium nitrate tetrahydrate (3.69 g) and sodium myristate (12.4 g) were dissolved separately in methanol (120 mL and 750 mL, respectively) and then combined under vigorous stirring. After 1 h, the precipitated product was collected by vacuum filtration using a Buchner funnel, washed repeatedly with methanol to remove residual impurities, and dried under vacuum at 50 °C overnight. 0.1 M S-in-ODE solution was prepared by dissolving 32 mg of S in a mixture of 10 mL ODE and 300 μL OA, assisted by ultrasonication.

### Synthesis of VQD-CQWs

Below we describe the synthesis protocols for the VQD-CQWs series studied in this work, including the 4.5-monolayer GC, GC@QB, GC@QB/1S and GC@QB/2S. For the GC synthesis, $Cd(Myr)_2$ (230 mg), Se powder (17 mg) and ODE (20 mL) were loaded into a 50 mL three-neck flask, degassed at room temperature for 30 min, and then placed under $N_2$ before heating to 230 °C. At 200 °C, $Cd(OAc)_2$ (60 mg) was rapidly added, and after 45 s the in-plane CdSe-to-CdS compositional transition was initiated by injecting 2.0 mL of 0.1 M S-in-ODE solution at 230 °C over 8 min. After that, the reaction mixture was cooled to room temperature and the resulting GC were precipitated with ethanol (5 mL), collected by centrifugation (4,000 rpm, 10 min), washed, and redispersed in hexane (5 mL).

To grow GC@QB featuring a ZnS shell with 2-monolayer thickness. 0.5 mmol of $Zn(OAc)_2$, 1 mL of OA, and 10 mL of ODE were mixed into a 50 mL three-neck flask, degassed under vacuum for 20 min, and purged with $N_2$ before being heated to 200 °C for 30 min to obtain a clear solution. After cooling to 60 °C, 1 mL of GC dispersed hexane was introduced, and residual hexane was removed under vacuum. The system was then placed under $N_2$ and heated to 300 °C. 1 mL of OAm was injected at 90 °C during the heating. At 165 °C, 0.1 M 1-octanethiol-in-ODE solution was injected at 8 mL/h until fully consumed. The reaction was then quenched in a water bath, followed by the addition of 10 mL of hexane at 60 °C. The GC@QB were precipitated with 10 mL of ethanol, washed twice, and finally redispersed in hexane.

GC@QB/1S and GC@QB/2S with one or two growth cycles of $Cd_yZn_{1-y}S$ gradient-shell were synthesized under conditions analogous to those used for GC@QB, except that a mixed Cd/Zn precursor (0.06 mmol $Cd(OAc)_2$ and 0.34 mmol $Zn(OAc)_2$) was employed instead of pure $Zn(OAc)_2$. After heating to form a clear Cd/Zn precursor solution, the CQWs from the preceding reaction, dispersed in hexane, were injected at 60 °C. The injection temperature, injection rate of the S precursor, as well as the purification procedure of the final products, were kept identical to those described for GC@QB. The formation of the gradient CdS-to-ZnS shell is due to the higher reactivity of the Cd precursor relative to the Zn precursor, as demonstrated elsewhere.

**Structural characterizations**

Atomic-resolution imaging and EDS analysis were conducted on a JEM-ARM200F spherical aberration-corrected STEM (0.10 nm TEM resolution; 0.078 nm STEM resolution) equipped with an Oxford Instruments EDS detector. Prior to measurements, all samples were beam-showered for 2 h to remove residual organic contamination.

XRD characterizations were carried out on a Bruker AXS D8 Advance diffractometer using Cu Kα radiation. Diffraction peaks were collected over a 2θ range of 20~70° at room temperature.

**Ensemble optical characterizations**

Ensemble absorption and PL spectra of the CQWs were measured using a Shimadzu UV-1800 spectrophotometer and a Shimadzu RF-5310 PC spectrofluorophotometer, respectively. The PL quantum yields were determined under 405 nm laser excitation in an integrating sphere, with emission signals collected by an Ocean Optics S4000 spectrometer.

**Theoretical calculations**

The electronic structure was calculated within the effective mass framework. Conduction-band electrons were modeled using a single-band approximation, whereas valence-band holes were described by a six-band Luttinger-Kohn $k \cdot p$ Hamiltonian to account for heavy-hole (HH), light-hole (LH), and split-off (SO) band coupling. The corresponding eigenenergies and envelope wavefunctions were obtained by solving the associated Schrödinger equations.

To quantify the impact of hole confinement on room-temperature transition dipole anisotropy and the resulting DOLP, eigenenergies were calculated for the structures schematized in Supplementary Fig. 41 and the results are shown in Supplementary Fig. 42. Thermal occupation of the electronic states was evaluated using Boltzmann statistics to determine their contributions to bandedge optical transitions. Transition matrix elements were then computed from the corresponding electron and hole wavefunctions (Supplementary Fig. 43-44), allowing for the extraction of in-plane orthogonal transition strengths and, consequently, the DOLP (see Supporting Information for computational details).

**Single-photon measurements**

Room-temperature single-photon measurements were performed on a customized inverted confocal microscope platform (Nikon Eclipse Ti), comprising an excitation source, confocal optics, and photon-correlation detection modules[47]. A picosecond pulsed laser driver (PDL 800-D) with a 475 nm laser head (repetition rate: 10 MHz) was used for excitation and focused on individual CQWs through a 100× oil-immersion objective lens with a numerical aperture of 1.30. The emission from the CQWs was separated from the excitation beam by a dichroic mirror and a long-pass filter (Semrock, 488LP), then coupled into a multimode fiber and directed to a Hanbury Brown and Twiss (HBT) setup for single-photon statistics characterizations. For all the measurements, samples were prepared by diluting the original solution by a factor of ~ $5 \times 10^5$ in toluene, followed by spin-coating onto ~170 μm cover glass substrates without any encapsulation.

**Ultrafast spectroscopy measurements**

Femtosecond TA and power-dependent PL measurements were performed using the same ultrafast laser system. The excitation source was a regenerative amplifier (Coherent Libra, 1 kHz repetition rate, ~50 fs pulse duration) seeded by a Ti:sapphire oscillator (Coherent Vitesse, 80 MHz). The 800 nm fundamental output was converted to 400 nm via second-harmonic generation from an optical parametric amplifier (OPA, Coherent OPerA Solo), and used for excitation.

For TA measurements, a portion of the 800 nm beam was routed through a delay stage and focused onto a Sapphire crystal to generate a white-light continuum probe spanning from 460 to 830 nm. Residual 800 nm fundamental was removed using a 750 nm short-pass filter. The pump and probe beams were spatially overlapped within a cuvette containing the CQW dispersion, and the TA spectra were recorded by detecting the transmitted probe signal with a fiber-coupled spectrometer integrated into a commercial HELIOS system (Ultrafast Systems).

For power-dependent PL measurements, the 400 nm excitation beam was focused onto the solid CQWs film sample, and the emission was collected in a back-scattering geometry. The collected signal was directed via a flip mirror either to a monochromator (Princeton Instruments Acton SP2500) coupled to a CCD detector (Princeton Instruments PIXIS:100) for steady-state and power-dependent PL measurements or to a streak camera (Optronis GmbH) for time-resolved analysis.

**Defocused imaging**

Defocused imaging was acquired using a wide-field, hyperspectral imaging microscope (HyperCube™) platform through a 100× oil-immersion objective lens with a numerical aperture of 1.30.

**FDTD simulations**

Optical simulations were performed using the commercial software Ansys Lumerical FDTD Solutions. The simulation configurations are presented in Supplementary Fig. 45. In these calculations, the CQW was modeled as a homogeneous dielectric body to evaluate the influence of dielectric contrast on the in-plane electric-field intensity anisotropy within the structure.

To record the spatial distribution of the in-plane electric-field intensity, a plane wave was employed as the excitation source, and a field monitor was positioned at the geometric mid-plane of the CQW. For far-field radiation simulations, an electric dipole emitter was placed at the center of the CQW. The far-field intensity distribution was obtained via near-to-far-field transformation to generate the corresponding polar plots. Two orthogonal dipole polarizations with equal strengths were simulated independently to obtain polarization-resolved far-field patterns, which were subsequently combined to produce the total far-field polar distribution.

## Data availability

All data are available in the manuscript or the Supplementary Information

## Acknowledgements

We gratefully acknowledge the support from the National Research Foundation, Singapore, under its Investigatorship Programme (NRF-NRFI10-2024-0003). We also acknowledge additional support from the Competitive Research Programme (NRF-CRP25-2020-0004) and from the Ministry of Education under its AcRF Tier 1 grant (RG152/24) and AcRF Tier 2 grant (MOE-T2EP50123-0001). Additionally, the authors would like to acknowledge the Facility for Analysis, Characterization, Testing and Simulation (FACTS) at Nanyang Technological University, Singapore, and in particular Dr. Tay Yee Yan and Dr. Andrew Wong, for their valuable and professional technical supports. We also acknowledge the support from the Nanyang NanoFabrication Centre (N2FC) at Nanyang Technological University, Singapore. H. V. D. would also like to acknowledge the support from the TUBA.

## Author contributions

X.L. and H.V.D. conceived the idea, H.V.D supervised the research at all stages. T.C.S. supervised spectroscopy characterizations. X.L. designed experiments, synthesized the CQWs, performed theoretical calculations, aberration-corrected STEM, XRD and optical simulations. B.W. carried out statistical single-photon spectroscopy, defocused imaging, and DOLP measurements. Y.Y. supported single-photon measurements and performed time-resolved PL measurements. P.L.H.M. made polarization-related calculations. Z.X. conducted TA spectroscopy. Y.Y. and L.D. made power-dependent PL measurements. V.K.S. provided technical project coordination and sourcing. X.L. analyzed the data and wrote the manuscript with inputs from H.V.D. and W.B.; H.V.D and T.C.S. revised and finalized the manuscript with inputs from all authors.

## Competing interests

The authors declare no competing interests.

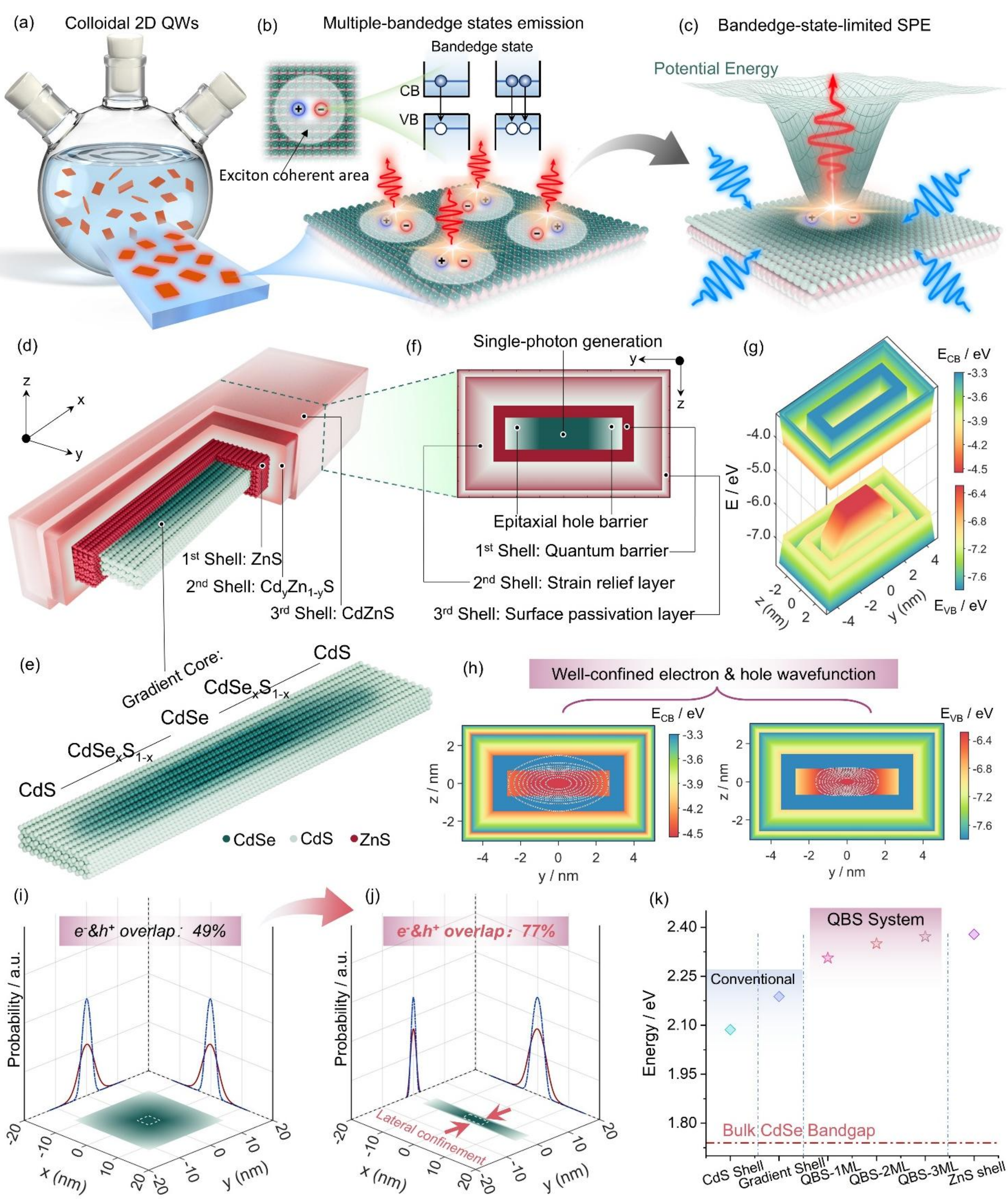

**Fig.1 | Volumetric quantum design for bandedge-state-limited single-photon emission in 2D CQWs. a**, Schematic illustration of CQWs combining 2D excitonic photophysics with colloidal chemistry. **b**, Illustration of multiple bandedge states within a single platelet arising from the limited room-temperature exciton coherent area. **c**, Conceptual illustration of the proposed single bandedge state enabled by an in-plane potential gradient. **d**, Schematic architecture of VQD-CQWs. **e**, In-plane structural design of the GC with a high aspect ratio.

**f**, yz cross-sectional view of the VQD-CQW architecture and the functional roles of each layer. **g**, Calculated conduction- and valence-band potential profiles along the yz cross-section. **h**, Calculated electron (left) and hole (right) probability density distributions projected onto the yz cross-section. **i-j**, Comparison of electron-hole probability density overlap for (i) a square-shaped GC and (j) a laterally confined high-aspect ratio GC. **k**, Calculated bandgap of CQWs with different shell architectures.

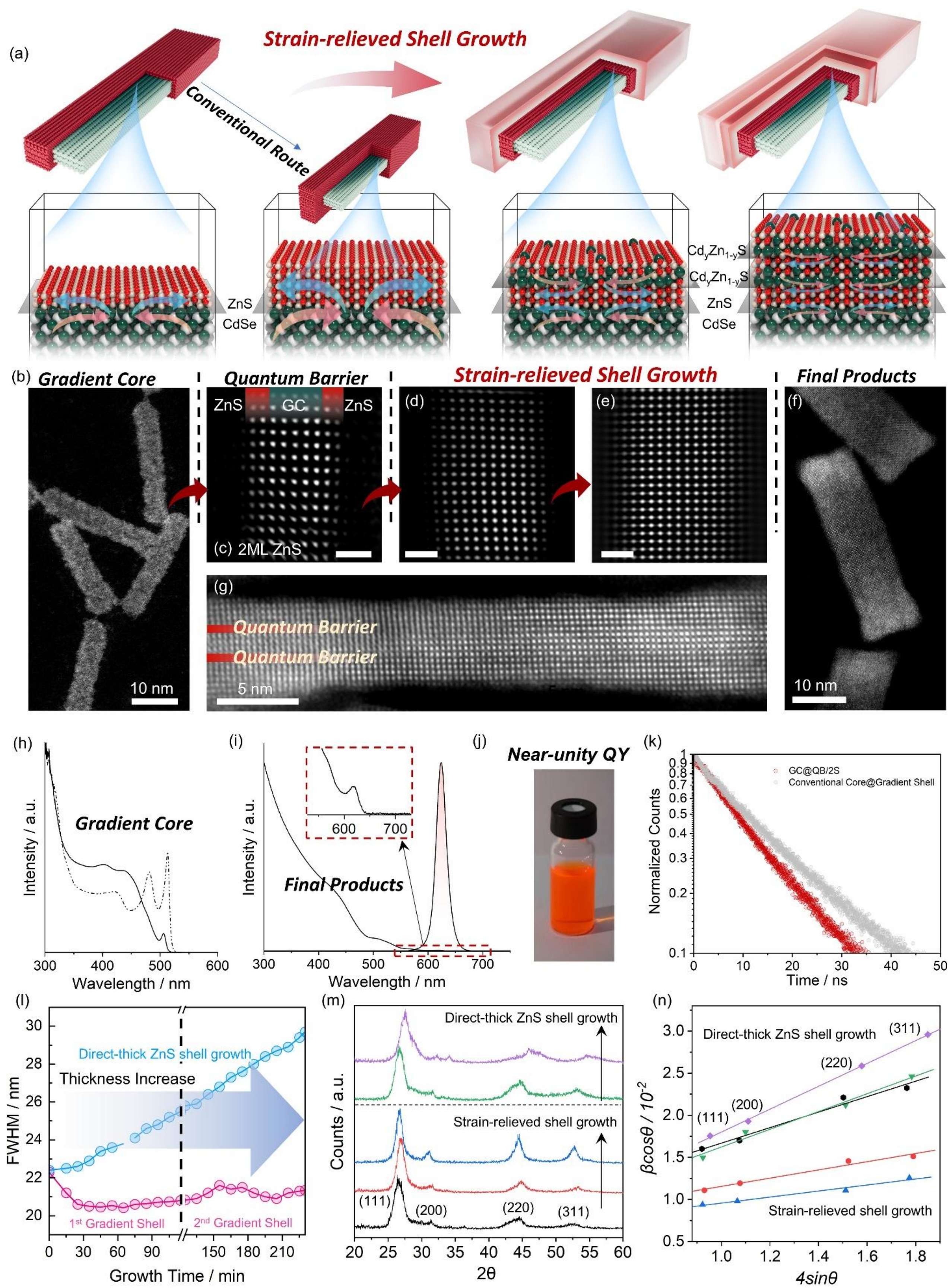

**Fig.2 | Atomic-scale validation of the strain-relieved QBS growth in VQD-CQWs. a**, Schematic illustration of the strain-relieved QBS growth process. **b-g**, Atomic-resolution

STEM characterizations of (b) GC, (c) GC@QB, (d) GC@QB/1S, and (e–f) GC@QB/2S. The ZnS quantum barrier is clearly resolved in (f) and (g), leveraging Z-contrast. Scale bar in (c-e): 1 nm. **h**, Absorption spectra of the GC (solid line) and pure CdSe nanoplatelets (dashed line). **i**, Absorption and PL spectra of GC@QB/2S, with the inset highlighting the pronounced excitonic absorption peak. **j**, Photograph of the as-made GC@QB/2S dispersion. **k**, Time-resolved PL decay of GC@QB/2S (red dots) compared with conventional core@gradient shell CQWs (grey dots). **l**, Evolution of the emission linewidth of CQWs grown via the conventional direct thick-ZnS route (blue) and the strain-relieved shell-growth strategy (pink). **m**, XRD patterns of CQWs prepared via the conventional direct thick-ZnS growth route (upper panel) and the strain-relieved shell-growth strategy (lower panel). **n**, Williamson–Hall analysis derived from the XRD measurements.

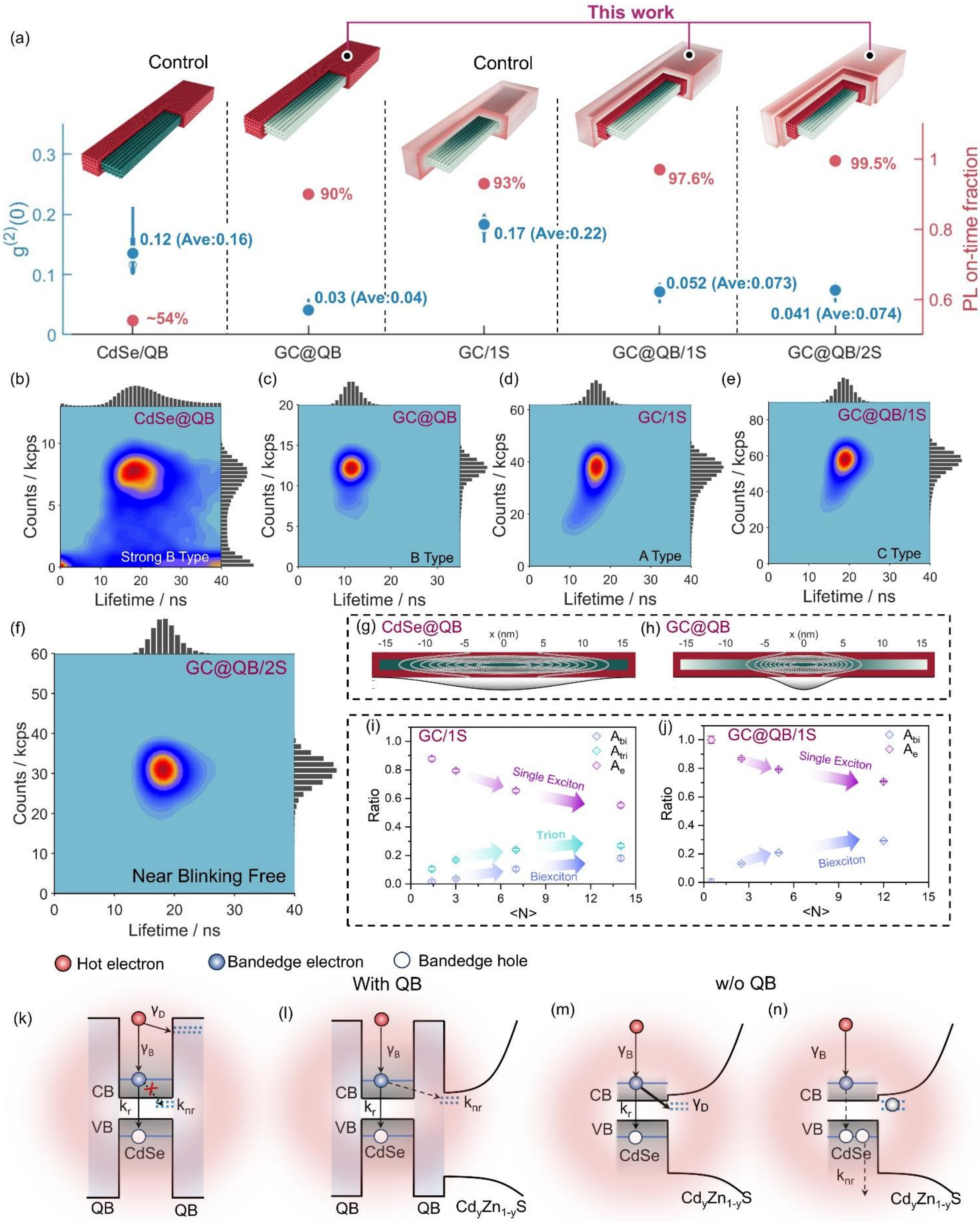

**Fig.3 | Near-blinking-free antibunching revealed by single-photon spectroscopy. a**, $g^{(2)}(0)$ (blue) and averaged PL on-time fractions (red) of the VQD-CQWs series and control samples. **b-f**, FLID mappings of (b) CdSe@QB, (c) GC@QB, (d) GC/1S, (e) GC@QB/1S,

and (f) GC@QB/2S, respectively. **g-h**, Calculated electron and hole probability density distributions for (g) CdSe@QB and (h) GC@QB, respectively. **i-j**, Extracted fractions of single-exciton, biexciton, and trion contributions in (i) GC/1S and (j) GC@QB/1S under varying excitation powers, deconvoluted from power-dependent TA measurements. **k-n**, Schematic illustrations of the blinking mechanisms in (k) GC@QB, (l) GC@QB/1S, and (m,n) GC/1S.

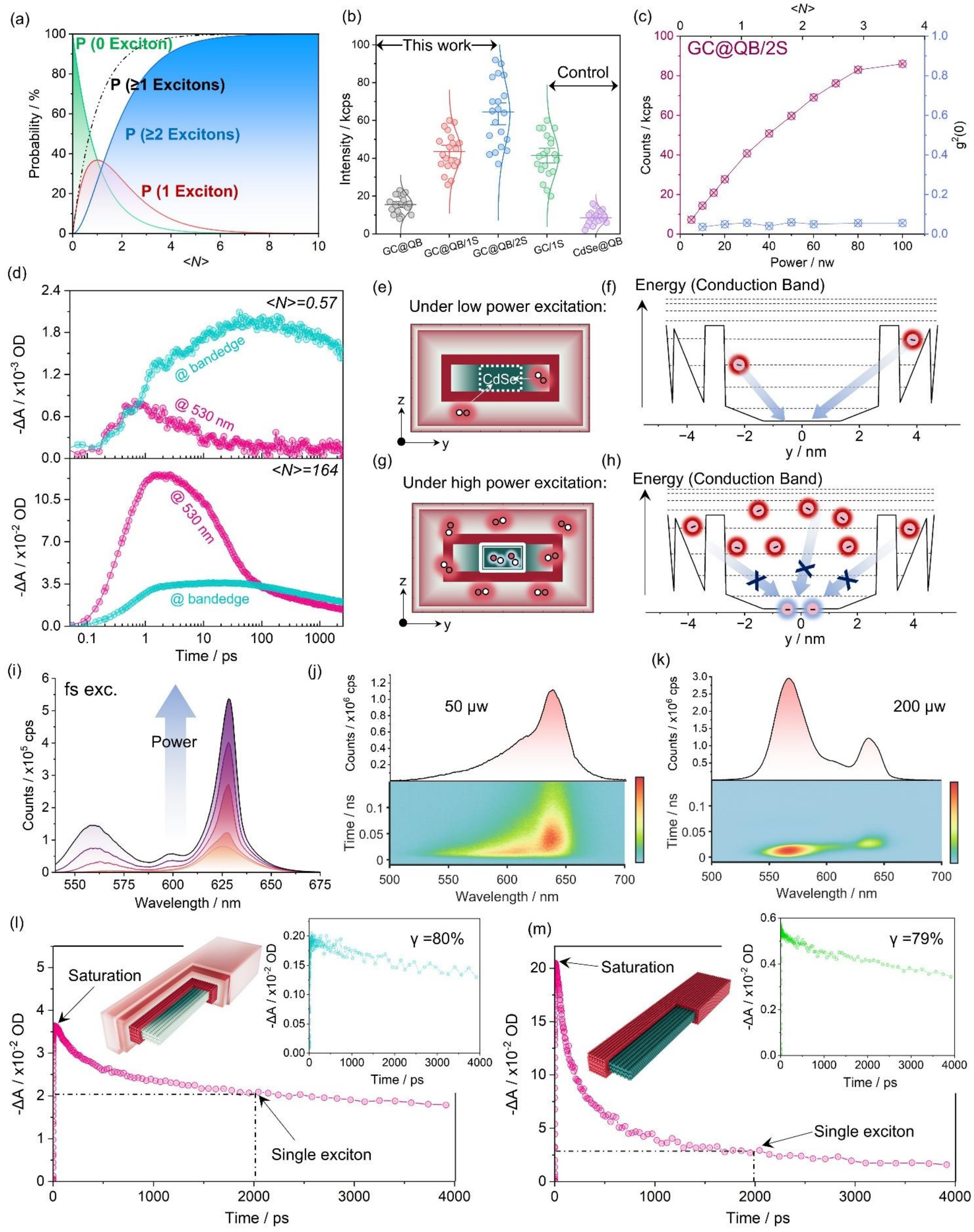

**Fig.4 | Unraveling the bandedge-state-filling bottleneck. a**, Calculated exciton population probabilities as a function of the average number of excitons ⟨N⟩ based on Poisson statistics.

**b**, Brightness of individual CQWs with different absorption cross-sections under identical excitation conditions. **c**, Power-dependent brightness (purple) and $g^{(2)}(0)$ (blue) of GC@QB/2S. **d**, TA dynamics of bandedge excitonic bleach signals (green) and higher-energy signals from shell regions (pink) in GC@QB/2S under low (⟨N⟩: 0.57, upper panel) and high (⟨N⟩: 164, lower panel) excitation fluences. **e-h**, Schematic illustrations of (e, f) efficient exciton transfer under low-power excitation and (g, h) the bandedge state-filling bottleneck mechanism under high-power excitation. **i**, Power-dependent PL spectra of a thick GC@QB/2S film under femtosecond laser excitation. **j-**k Time-resolved PL spectra recorded by a streak camera under (j) low and (k) high excitation power using the same sample and excitation source as in (i). **l-m**, TA dynamics of the bandedge excitonic bleach in (l) GC@QB/2S and (m) CdSe@QB measured at bandedge bleach saturation conditions. Insets show the corresponding TA dynamics under ultralow excitation fluences, ensuring single-exciton decay.

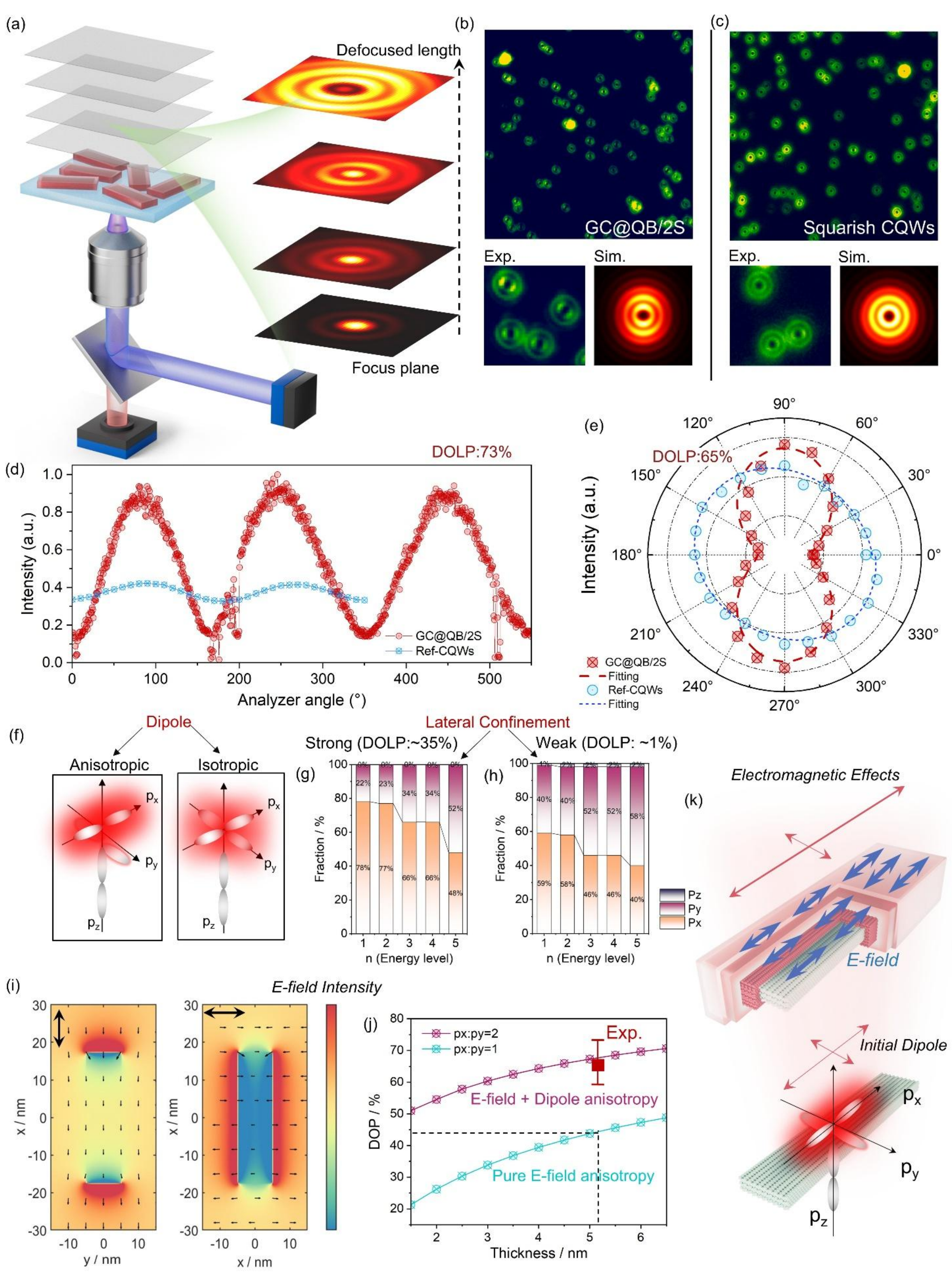

**Fig.5 | Origin of linear polarization emission in VQD-CQWs. a**, Schematic illustration of the defocused imaging technique, with simulated far-field radiation patterns of an individual

emitter possessing an isotropic dipole at different defocused lengths. **b-c**, Defocused microscopy images of (b) GC@QB/2S and (c) a square-shaped CQWs control sample. **d**, Analyzer-angle-dependent PL intensity of individual GC@QB/2S (red) and square CQWs (blue). **e**, Polar diagrams of the PL intensity for GC@QB/2S (red) and square CQWs (blue) individuals. **f**, Schematic illustration showing that strong lateral confinement along the y direction induces a quasi in-plane dipole anisotropy by lifting the energy of the $p_y$ component of Se 4p orbitals. **g-h**, Calculated contributions of orthogonal $p_x$, $p_y$, and $p_z$ components near the valence-band edge across different eigenstates. (g) Left panel: CQWs with strong lateral confinement. (h) Right panel: CQWs with the same aspect ratio but weak lateral confinement. **i**, FDTD-simulated electric-field distributions inside the CQW dielectric (GC@QB/2S), showing field-intensity anisotropy along the x direction (left) and y direction (right) arising from dielectric contrast. **j**, FDTD-simulated thickness dependence of DOLP with different intrinsic dipole anisotropies. **k**, Schematic illustration of the synergistic effects of dipole anisotropy and electric-field anisotropy leading to strong linear polarization in GC@QB/2S individuals.

# Supplementary Information

## Supplementary Note 1 | Theoretical calculation method.

### *Electron states calculations*

The conduction-band electron states are computed within the single-band effective-mass approximation. The composition-dependent Hamiltonian is given by:

$$H_e = -\nabla \cdot \left( \frac{\hbar^2}{2m^*(\mathbf{r})} \nabla \right) + E_c(\mathbf{r}),$$

where $m^*(\mathbf{r})$ denotes the composition-dependent electron effective mass and $E_c(\mathbf{r})$ represents the composition-dependent conduction-band potential.

The electron envelope functions are obtained by solving the corresponding effective-mass Schrödinger equation:

$$H_e \psi_e^{(n)} = E_e^{(n)} \psi_e^{(n)},$$

which yields the quantized conduction-band eigenenergies and wavefunctions.

### *Hole states calculations*

The valence-band hole states are described using the six-band Luttinger-Kohn $k \cdot p$ Hamiltonian for zinc-blende semiconductors, which captures the coupling among heavy-hole (HH), light-hole (LH), and split-off (SO) bands[1].

The Hamiltonian is constructed in the $(p_x, p_y, p_z) \otimes (\uparrow, \downarrow)$ basis and incorporates spatially varying material parameters.

The composition-dependent valence-band Hamiltonian reads:

$$H_h = H_{\mathrm{LK}} + H_{\mathrm{SO}} + E_v(\mathbf{r})$$

where $H_{\mathrm{LK}}$ denotes the Luttinger-Kohn kinetic-energy operator containing the parameters $L(\mathbf{r})$, $M(\mathbf{r})$ and $N(\mathbf{r})$. $E_v(\mathbf{r})$ represents the composition-dependent valence-band potential.

The orbital part of the Hamiltonian consists of two identical 3 x 3 blocks:

$$H_{\mathrm{LK}} = \begin{pmatrix} H_{3\times3} & 0 \\ 0 & H_{3\times3} \end{pmatrix}.$$

The 3 x 3 block in the $(p_x, p_y, p_z)$ basis reads:

$$H_{3\times3} = \begin{pmatrix} P_x & R_{xy} & R_{xz} \\ R_{yx} & P_y & R_{yz} \\ R_{zx} & R_{zy} & P_z \end{pmatrix}$$

with

$$\begin{aligned} P_x &= K\left[L(\mathbf{r})k_x^2 + M(\mathbf{r})(k_y^2 + k_z^2)\right], \\ P_y &= K\left[L(\mathbf{r})k_y^2 + M(\mathbf{r})(k_x^2 + k_z^2)\right], \\ P_z &= K\left[L(\mathbf{r})k_z^2 + M(\mathbf{r})(k_x^2 + k_y^2)\right]. \end{aligned}$$

and the off-diagonal coupling terms

$$R_{ij} = \frac{K}{2}\left(k_i N(\mathbf{r}) k_j + k_j N(\mathbf{r}) k_i\right), \quad i \neq j ,$$

where $K = \dfrac{\hbar^2}{2m_0}$ .

The spin-orbit coupling term $H_{\mathrm{SO}}$ mixes the spin blocks and is characterized by the local spin-orbit splitting $\Delta_{\mathrm{SO}}(\mathbf{r})$, which lifts the degeneracy among the heavy-hole, light-hole and the split-off band.

The hole envelope functions are obtained by solving the corresponding eigenvalue problem:

$$H_h \Psi_h^{(m)} = E_h^{(m)} \Psi_h^{(m)} ,$$

yielding quantized valence-band eigenenergies and six-component envelope wavefunctions,

$$\Psi_h^{(m)} = \left(p_x^{\uparrow}, p_y^{\uparrow}, p_z^{\uparrow}, p_x^{\downarrow}, p_y^{\downarrow}, p_z^{\downarrow}\right)^{\mathrm{T}} .$$

***Optical transition matrix elements***

The inter-band optical transition strength along the Cartesian direction $\alpha \in \{x, y, z\}$ is evaluated from the overlap between the electron envelope function and the corresponding hole orbital component.

For a given electron state $e$ and hole state $h$, the transition matrix element is defined as

$$M_\alpha^{(e,h)} = \left\langle \psi_e^{(e)} \middle| \ \psi_{h,\alpha}^{(h)} \right\rangle .$$

The polarization-resolved transition intensity for the electron-hole pair $(e,h)$ is therefore

$$P_\alpha^{(e,h)} = \left|M_\alpha^{(e,h,\uparrow)}\right|^2 + \left|M_\alpha^{(e,h,\downarrow)}\right|^2 .$$

Here $P_\alpha^{(e,h)}$ quantifies the polarization-resolved optical oscillator strength of each electron-hole pair.

***Boltzmann weighting of electronic states***

To incorporate thermal occupation effects at room temperature, a two-step Boltzmann weighting procedure is employed.

The electron occupation probability is given by

$$w_e^{(e)} = \frac{\exp\left[-(E_e^{(e)} - E_e^{\min})/(k_B T)\right]}{\sum_e \exp\left[-(E_e^{(e)} - E_e^{\min})/(k_B T)\right]}.$$

Similarly, the hole occupation probability reads

$$w_h^{(h)} = \frac{\exp\left[-(E_h^{(h)} - E_h^{\min})/k_B T\right]}{\sum_h \exp\left[-(E_h^{(h)} - E_h^{\min})/k_B T\right]}.$$

The electron-weighted transition intensity associated with each hole state is then defined as

$$\tilde{P}_\alpha^{(h)} = \sum_e w_e^{(e)} P_\alpha^{(e,h)}.$$

Finally, the thermally averaged global polarization components are obtained by summing over hole states:

$$P_\alpha^{\text{tot}} = \sum_h w_h^{(h)} \tilde{P}_\alpha^{(h)}.$$

This procedure ensures the Boltzmann-weighted contribution from both electron and hole states to the optical transition strengths.

***Calculation of degree of polarization***

Finally, the in-plane degree of polarization (DOP) is defined as

$$\text{DOP}_{xy} = \frac{P_x^{\text{tot}} - P_y^{\text{tot}}}{P_x^{\text{tot}} + P_y^{\text{tot}}}.$$

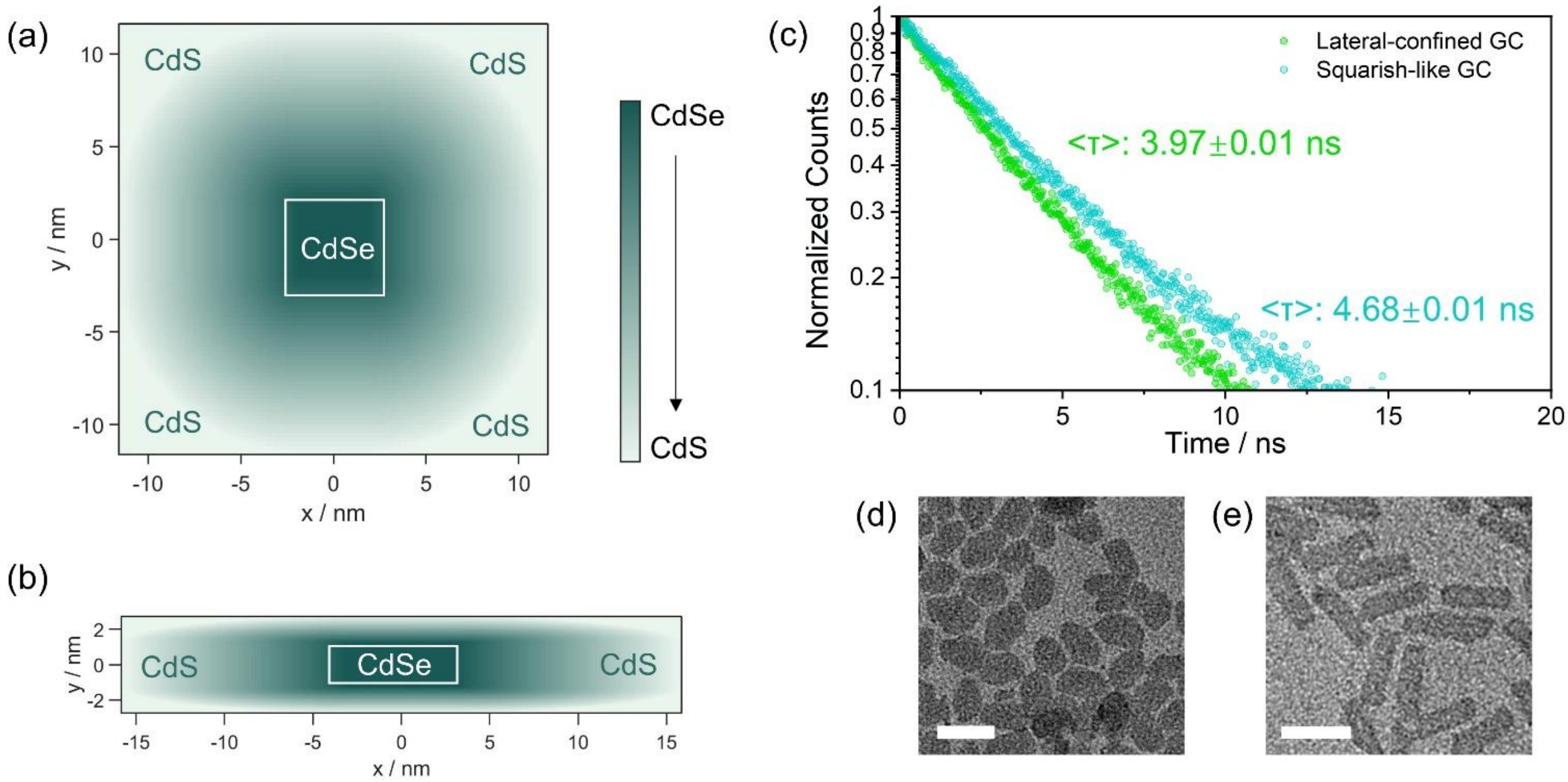

**Supplementary Fig. 1 | Experimental evidence of the wavefunction overlap in GCs with different in-plane geometries. a-b**, Schematic illustrations of the structural configurations of GCs with different in-plane geometries. Experimentally, the in-plane geometry is tuned following the reported protocol by adjusting the cation-to-anion precursor ratio and the cadmium acetate precursor (with or without hydration)[2, 3]. **c**, Fluorescence lifetime measurements of GCs with different in-plane geometries. The GC with a laterally strongly confined rectangular geometry exhibits a fluorescence lifetime of ~3.97 ns (green dots), whereas the GC with a more squarish geometry shows a lifetime of ~4.68 ns (cyan dots). **d-e**, TEM images of the synthesized (d) squarish GCs and (e) laterally strongly confined rectangular GCs, respectively (Scale bar: 20 nm).

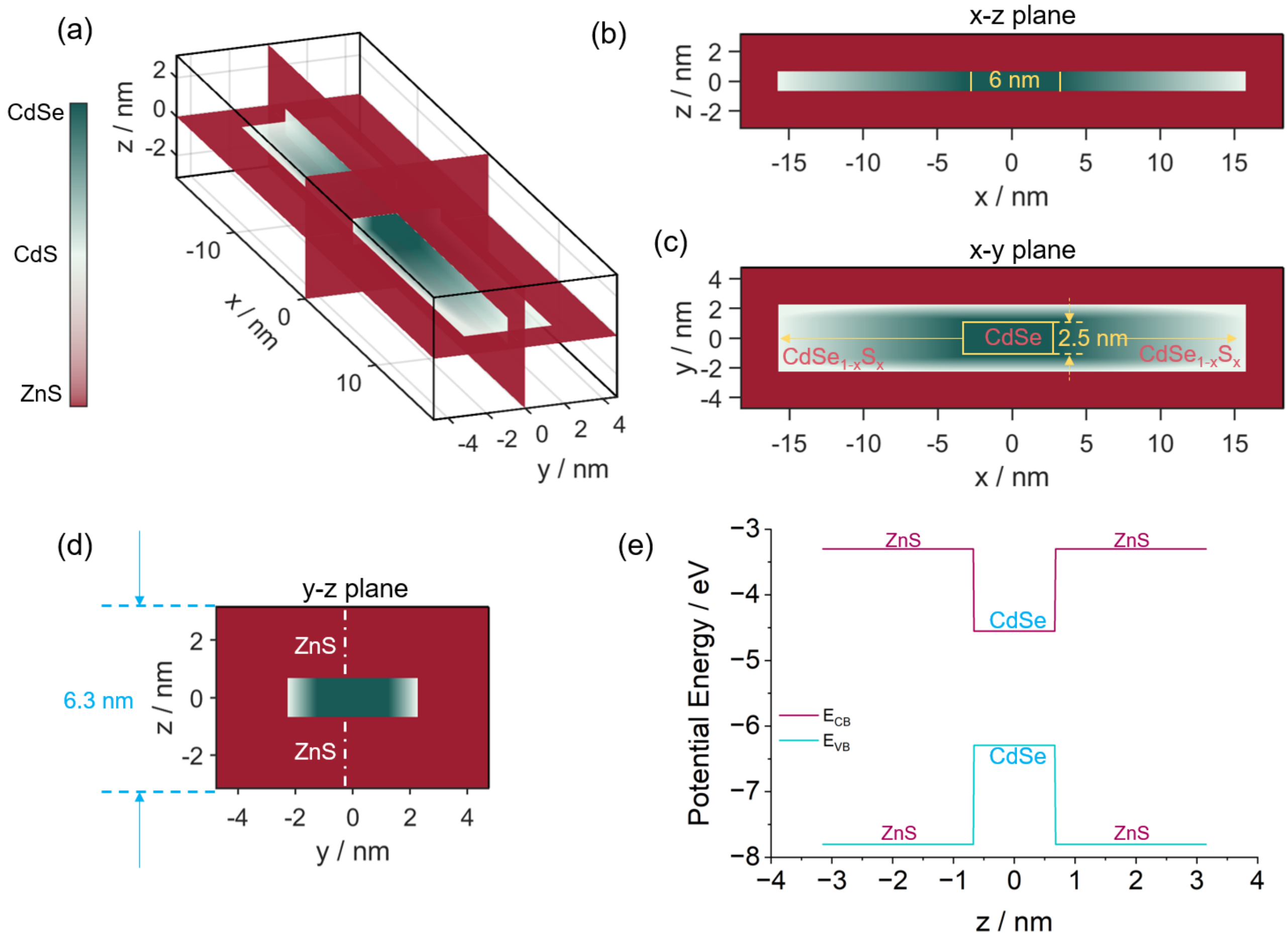

**Supplementary Fig. 2 | Structural parameters of ZnS-shell-based CQWs used for wavefunction calculations**. **a**, 3D compositional distribution of the heterostructure with orthogonal cross-sectional slices. **b-d**, Cross-sectional compositional distributions in the (b) xz, (c) xy, (d) yz planes, respectively. **e**, Conduction ($E_{CB}$) and valence ($E_{VB}$) band potential profiles in the y-z cross-section (y = 0).

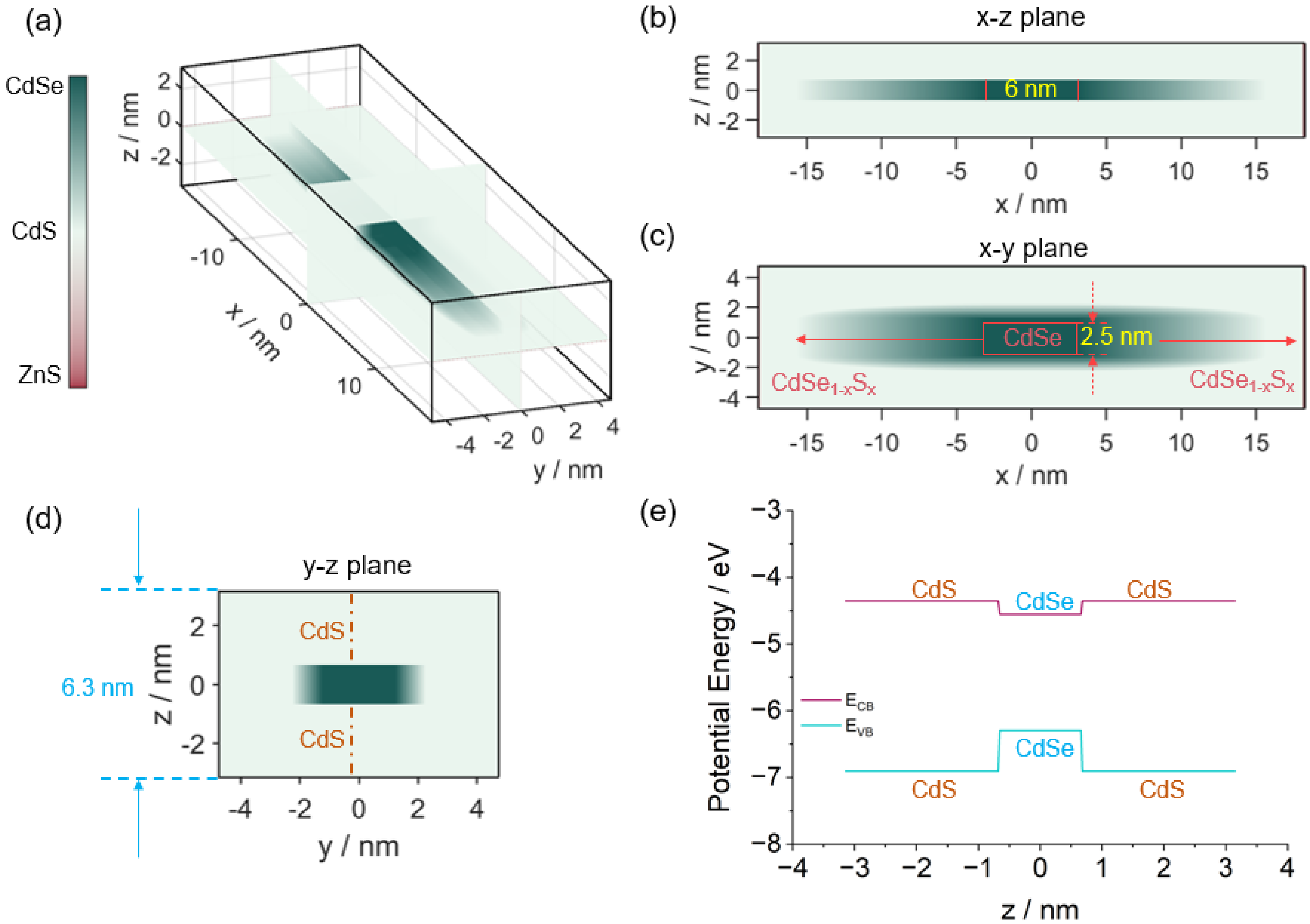

**Supplementary Fig. 3 | Structural parameters of CdS-shell-based CQWs used for wavefunction calculations**. **a**, 3D compositional distribution of the heterostructure with orthogonal cross-sectional slices. **b-d**, Cross-sectional compositional distributions in the (b) xz, (c) xy, (d) yz planes, respectively. **e**, Conduction ($E_{CB}$) and valence ($E_{VB}$) band potential profiles in the y-z cross-section (y = 0).

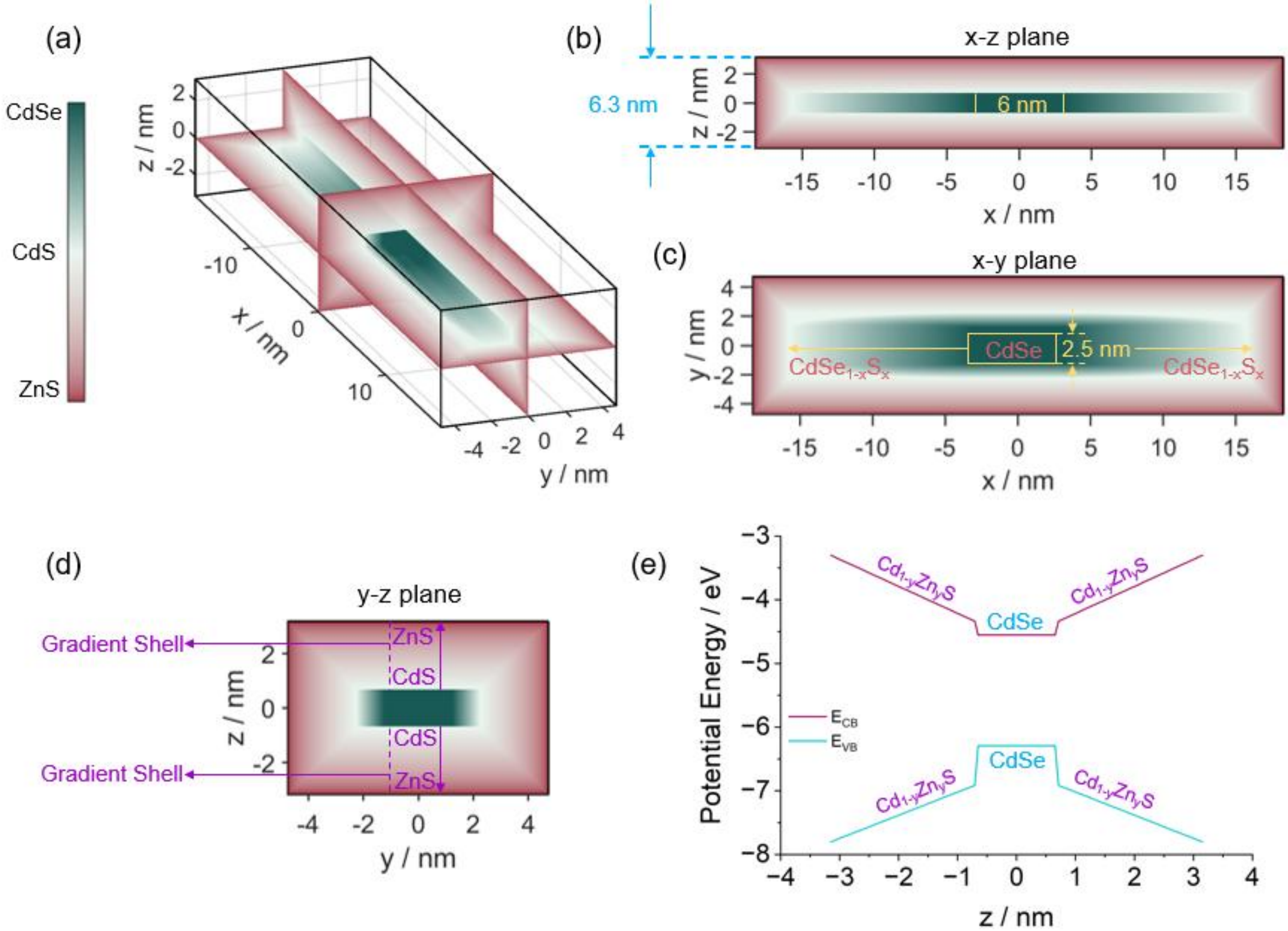

**Supplementary Fig. 4 | Structural parameters of $Cd_yZn_{1-y}S$ gradient-shell-based CQWs used for wavefunction calculations**. **a**, 3D compositional distribution of the heterostructure with orthogonal cross-sectional slices. **b-d**, Cross-sectional compositional distributions in the (b) xz, (c) xy, (d) yz planes, respectively. **e**, Conduction ($E_{CB}$) and valence ($E_{VB}$) band potential profiles in the y-z cross-section (y = 0).

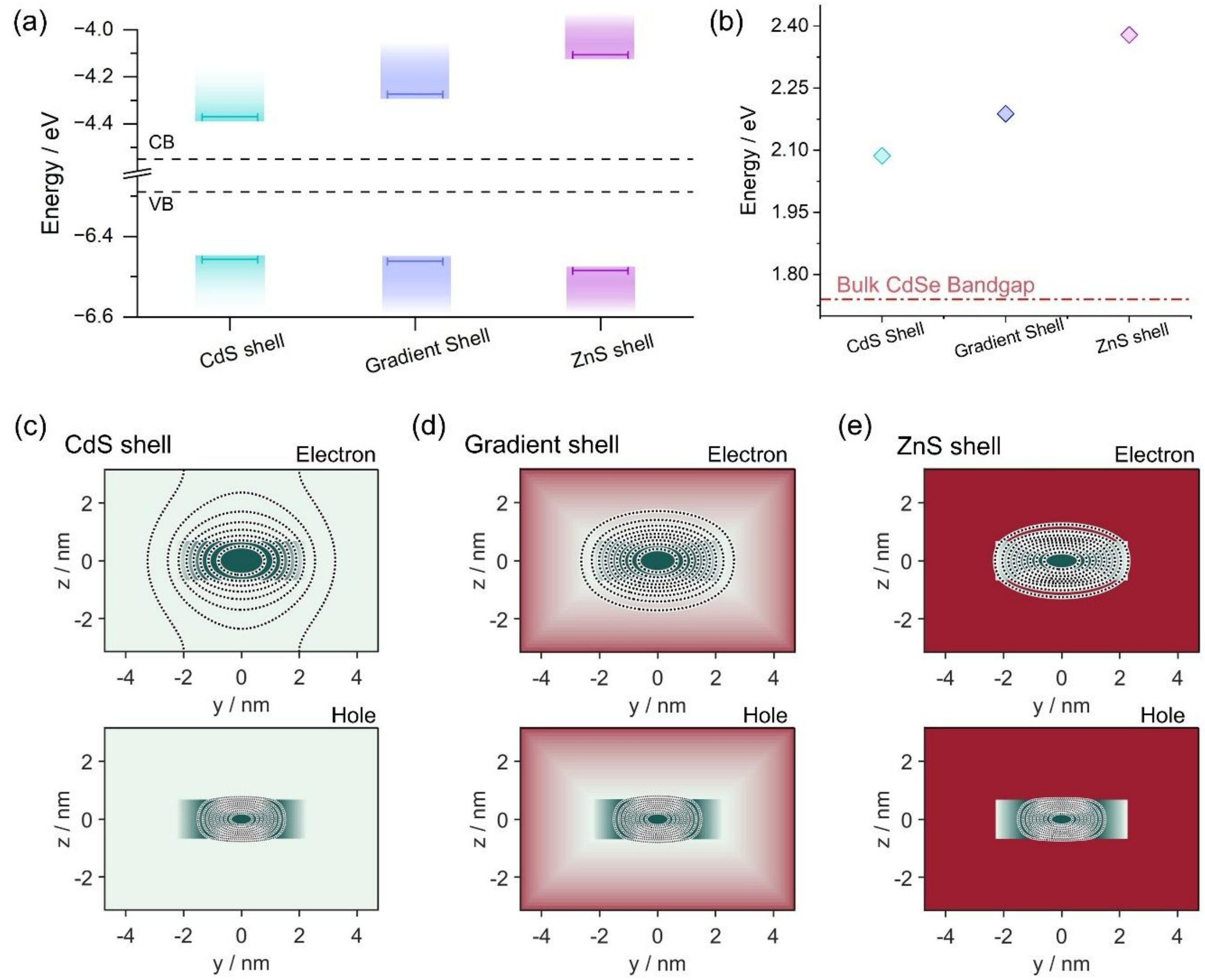

**Supplementary Fig. 5 | Theoretical calculation results of CQWs with three conventional shell configurations (CdS, $Cd_yZn_{1-y}S$ and ZnS). a,** Calculated conduction- and valence-bandedge energies relative to bulk CdSe. **b**, Bandgap comparison of CQWs with the three different shells. **c-e**, Calculated electron (top) and hole probability density (bottom) distributions of CQWs with CdS, $Cd_yZn_{1-y}S$ and ZnS shells, respectively.

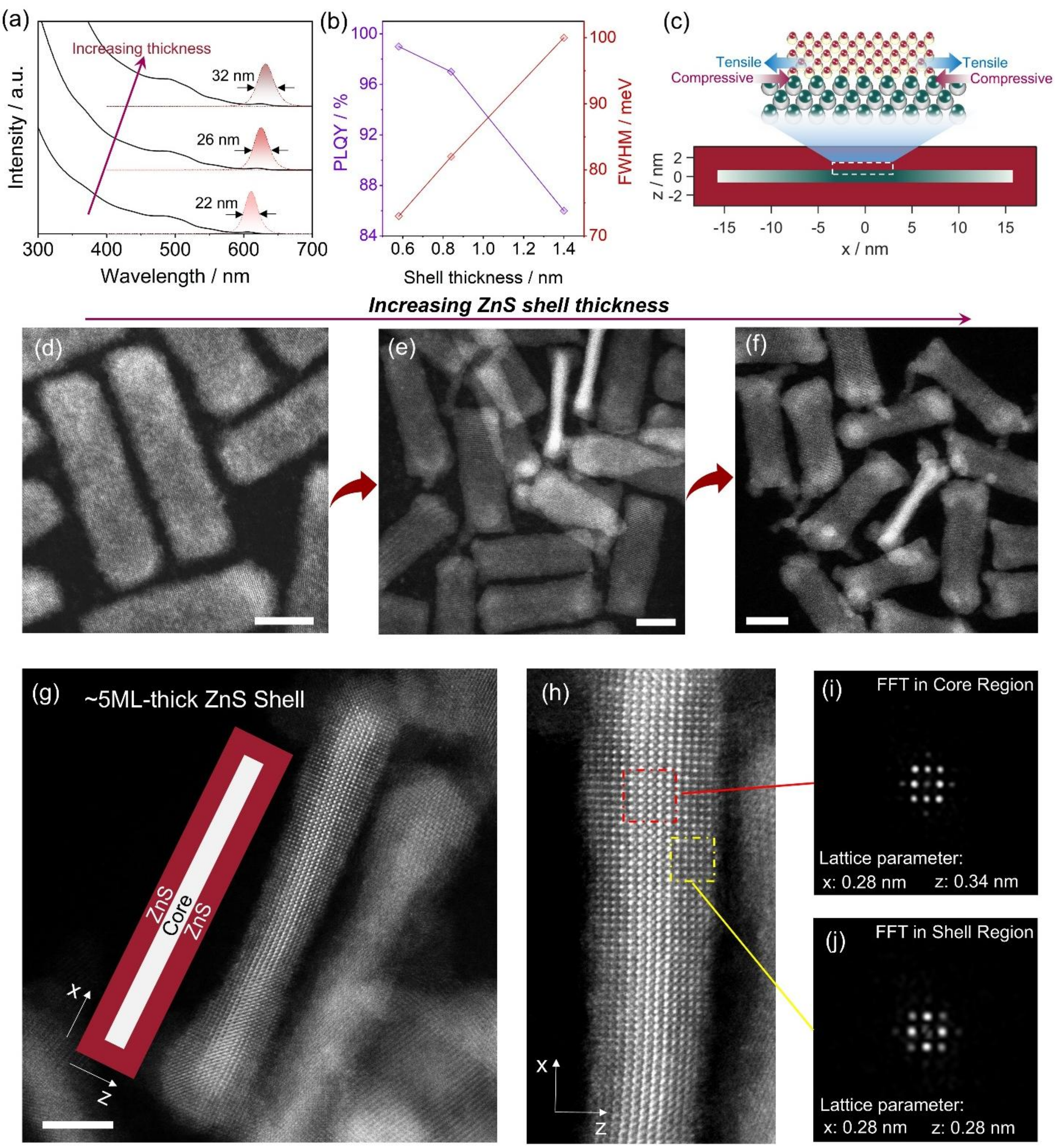

**Supplementary Fig. 6 | Experimental investigations of direct thick ZnS shell growth. a,** Evolution of absorption and PL spectra of the CQWs during ZnS shell growth. **b**, Shell-thickness-dependent emission linewidth and quantum yield (QY) of ZnS-shell-based CQWs. **c**, Schematic illustration of strain accumulation at the CdSe-ZnS interface due to the large lattice mismatch (~12%). **d-f**, Atomic-resolution TEM images of ZnS-shell-based CQWs with increasing shell thickness (scale bar: 10 nm). The ZnS shell growth follows a Stranski-Krastanov (S-K) mode, exhibiting coherent and uniform shells at early stages (d), followed by pronounced non-uniformity beyond a critical thickness, as shown in (e) and (f). **g**, Cross-sectional atomic-resolution TEM image of an individual platelet with a thick 5-monolayer ZnS

shell (Scale bar: 10 nm). **h-j**, Magnified cross-sectional TEM images and corresponding local Fourier transform patterns of the (i) core and (j) shell regions. Owing to the large lattice mismatch between CdSe and ZnS (~12%), the in-plane lattice constant of CdSe in the core region is compressed to ~0.282 nm along the in-plane direction, significantly deviating from its intrinsic value of ~0.32 nm. In contrast, the lattice spacing along the out-of-plane (z) direction is slightly expanded, likely arising from Poisson relaxation. The lattice distortion is substantially more pronounced in CdSe than in ZnS (j), reflecting the relatively softer nature of CdSe compared to the stiffer ZnS shell[4]. As a result, strain accumulation is primarily accommodated through in-plane lattice compression within the CdSe core. Combined with the increasing shell non-uniformity, this strain-induced lattice fluctuation leads to bandgap variations in the CdSe domain, thereby contributing to the pronounced PL linewidth broadening.

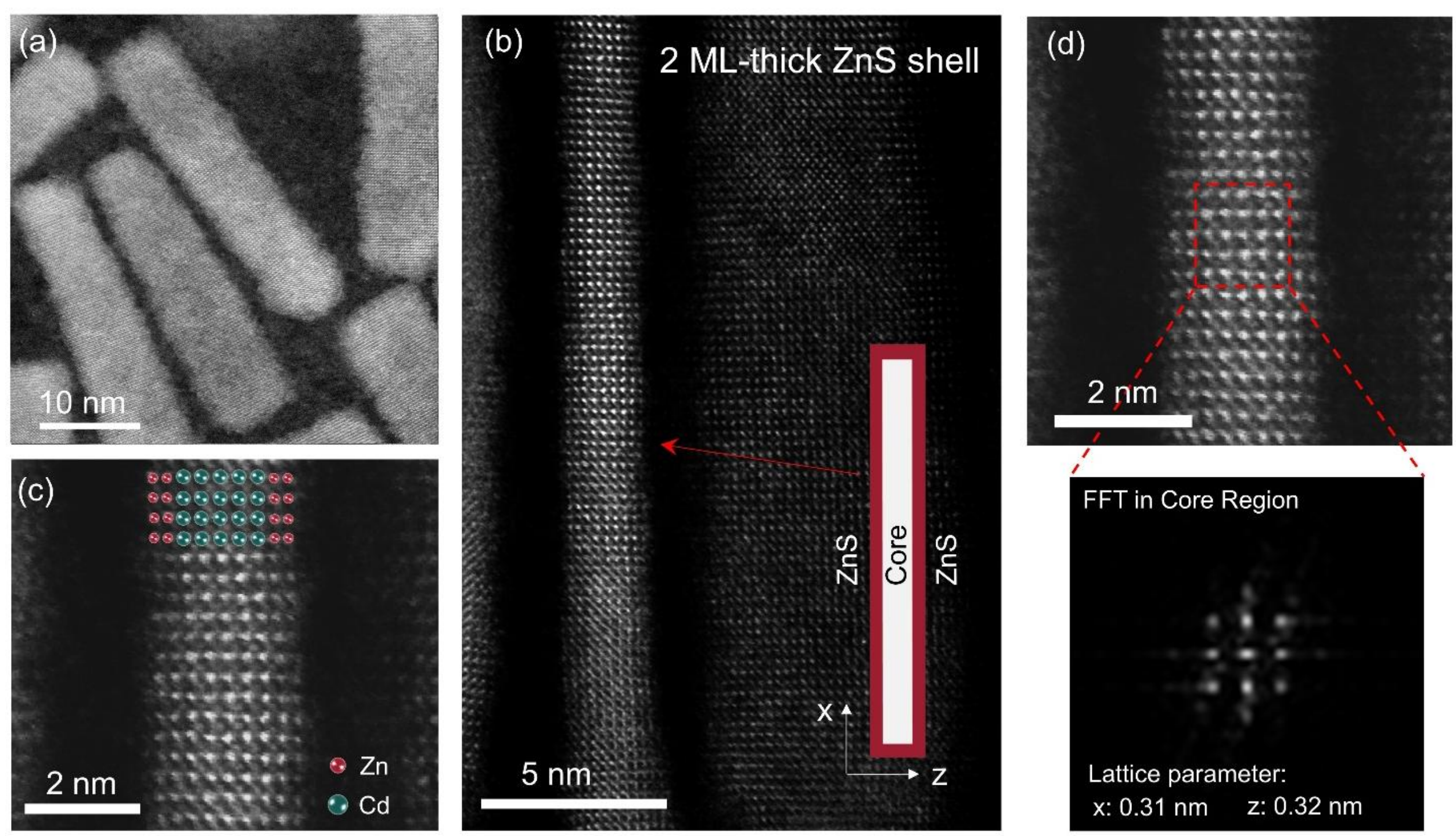

**Supplementary Fig. 7 | Preliminary structural characterization of thin ZnS shell (2-monolayer) growth. a,** Atomic-resolution TEM image of CQWs with a thin ZnS shell. **b-c**, Cross-sectional atomic-resolution TEM images of an individual platelet with an atomically smooth 2-monolayer ZnS shell. **d**, Magnified cross-sectional TEM image and corresponding local Fourier transform patterns of the CdSe core region. The CdSe lattice is slightly

compressed along the in-plane direction, with an in-plane lattice parameter of 0.31 nm, indicating that the accumulated strain remains within a controllable range.

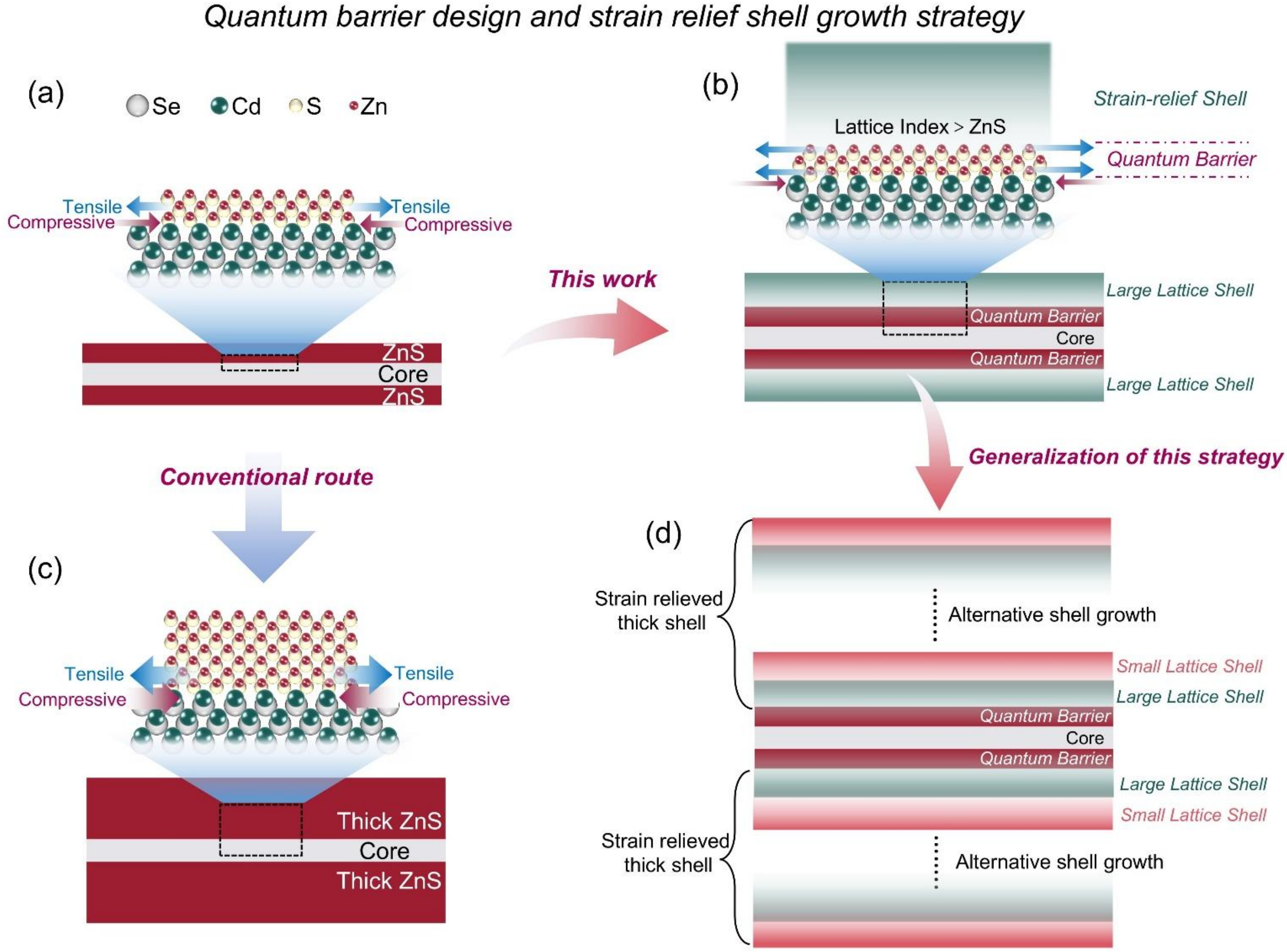

**Supplementary Fig. 8 | Schematic illustration of the quantum barrier design and the strain-relieved shell growth strategy. a-c,** Schematics of conventional direct thick ZnS shell growth and the associated accumulation of interfacial strain. **b**, Design of the quantum barrier shell (QBS) architecture. An atomically smooth, wide-bandgap ZnS wetting layer is embedded within the shell stack to serve as a quantum barrier. While preserving quantum confinement and structural uniformity of the core@shell configuration, subsequent epitaxial growth of a material with a larger lattice constant forms a core/barrier/outer-shell stacking structure with alternating lattice parameters. By symmetry, the non-uniform compressive strain imposed on the core can be redistributed and relaxed, thereby mitigating bandgap fluctuations and leading to PL spectral narrowing.

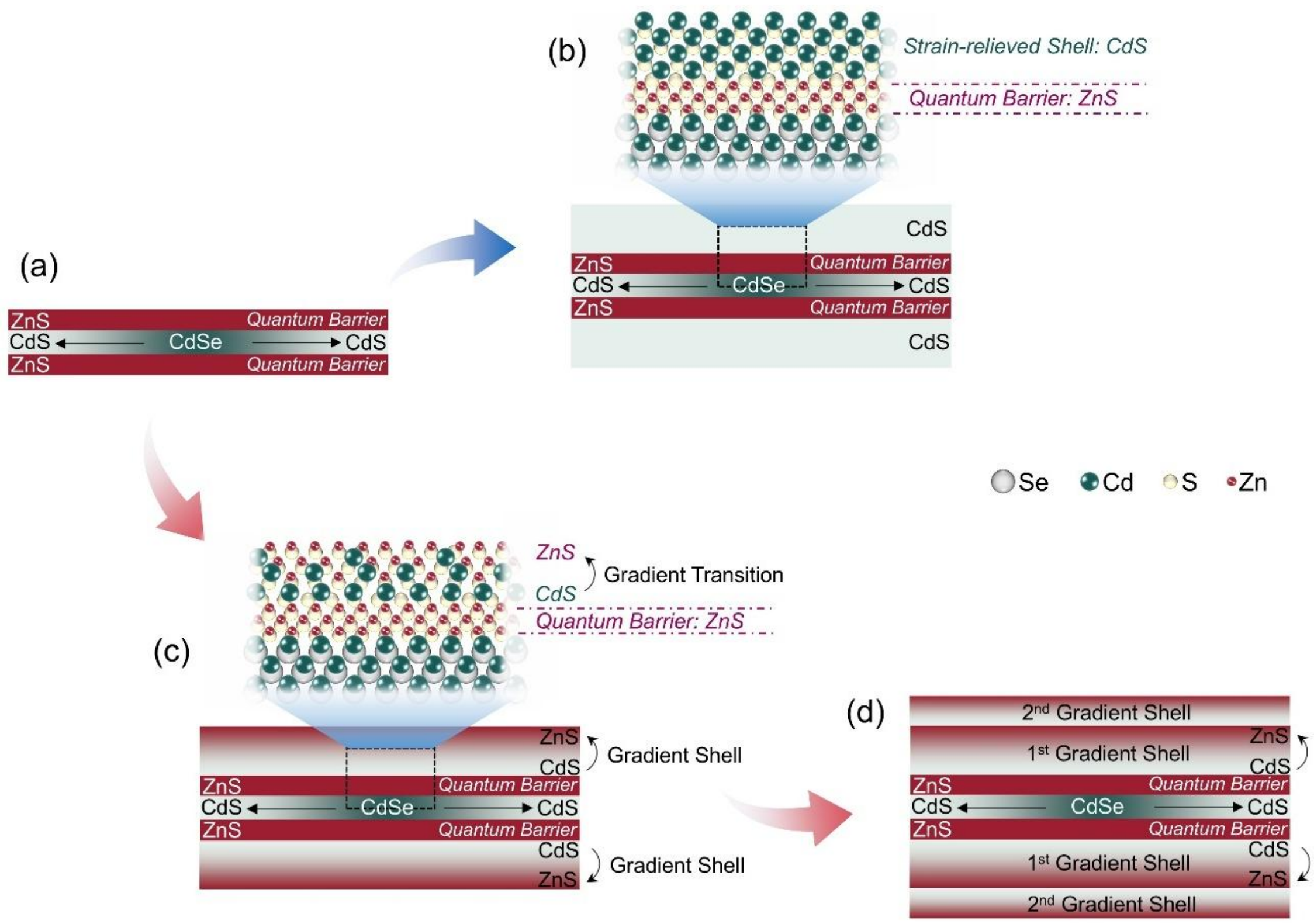

**Supplementary Fig. 9 | Implementation of the strain-relieved QBS strategy. a-c,** To realize the designed shell architecture, two outer-shell configurations are considered: (b) a CdS shell and (c) a gradient $Cd_yZn_{1-y}S$ shell. In both cases, a CdSe/ZnS/CdS-type stacking structure with alternating lattice parameters is formed. **d**, In the $Cd_yZn_{1-y}S$ shell case, the growth of $Cd_yZn_{1-y}S$ procedure can be iteratively repeated to further increase shell thickness while maintaining the strain-relieved shell architecture.

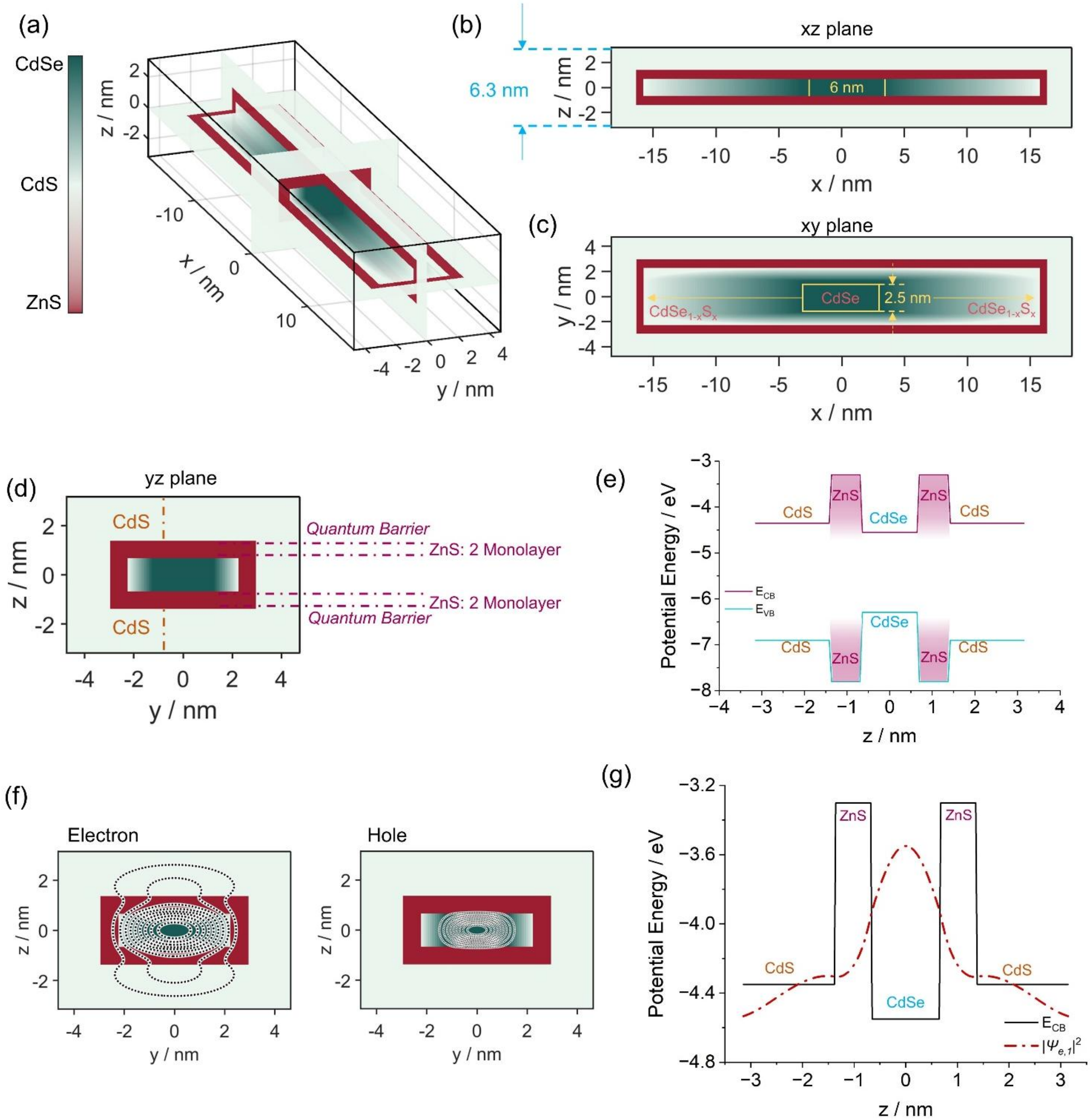

**Supplementary Fig. 10 | Structural parameters and wavefunction calculations of the strain-relieved QBS (CdS case). a**, 3D compositional distribution of the heterostructure with orthogonal cross-sectional slices. **b-d**, Cross-sectional compositional distributions in the (b) xz, (c) xy, (d) yz planes, respectively. **e**, Conduction ($E_{CB}$) and valence ($E_{VB}$) band potential profiles in the yz cross-section (y = 0). **f**, Contour plots of electron (left) and hole (right) probability density distributions projected onto the yz plane at x = 0. **g**, Electron probability density distribution overlaid with the potential profile in the yz cross-section (y = 0). As shown

in (g), substantial electron leakage across the ZnS quantum barrier arises from the shallow conduction-band potential of the CdS outer shell.

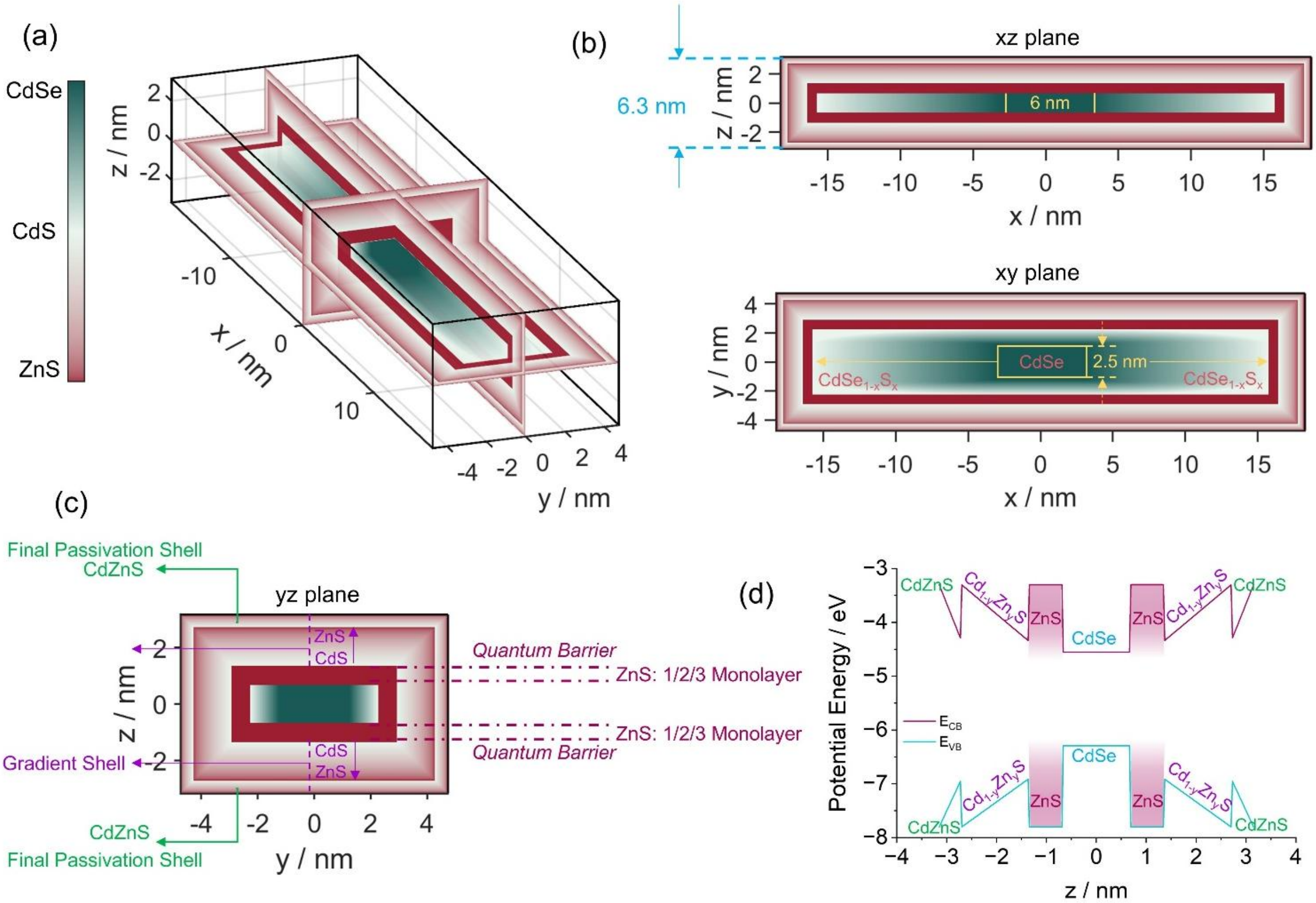

**Supplementary Fig. 11 | Structural parameters of the strain-relieved QBS (Gradient $Cd_yZn_{1-y}S$ case). a**, 3D compositional distribution of the heterostructure with orthogonal cross-sectional slices. **b-d**, Cross-sectional compositional distributions in the (b) xz, (c) xy, and (d) yz planes, respectively. **e**, Conduction ($E_{CB}$) and valence ($E_{VB}$) band potential distributions in the y-z cross-section (y = 0).

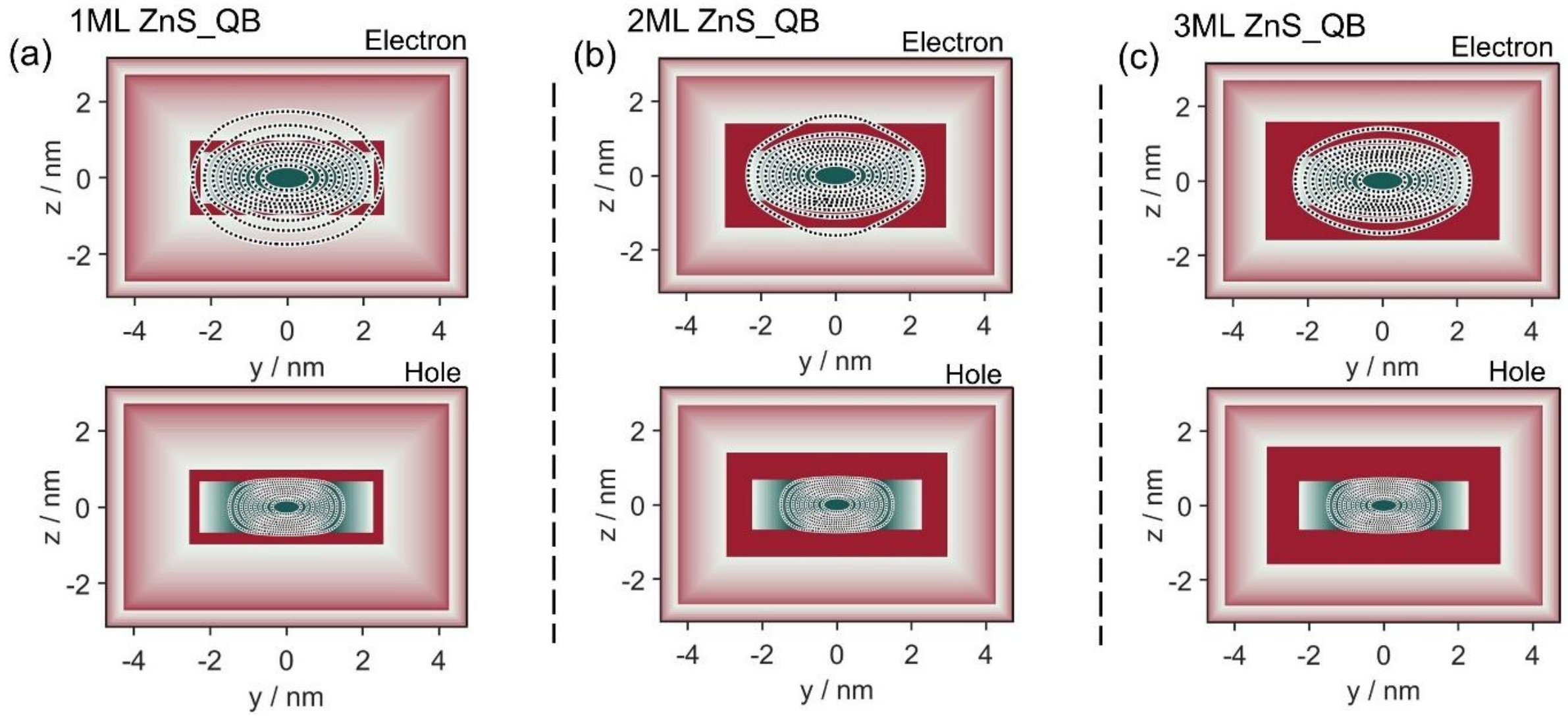

**Supplementary Fig. 12 | Effects of quantum-barrier thickness on wavefunction distribution.** Here the ZnS quantum-barrier thickness is varied between 1, 2, and 3 monolayers, corresponding to the smooth "wetting-layer" regime. **a-c**, Contour plots of electron (top) and hole (bottom) probability distributions projected onto the yz plane at x = 0 for (a) 1-monolayer, (b) 2-monolayer, and (c) 3-monolayer ZnS barriers, respectively.

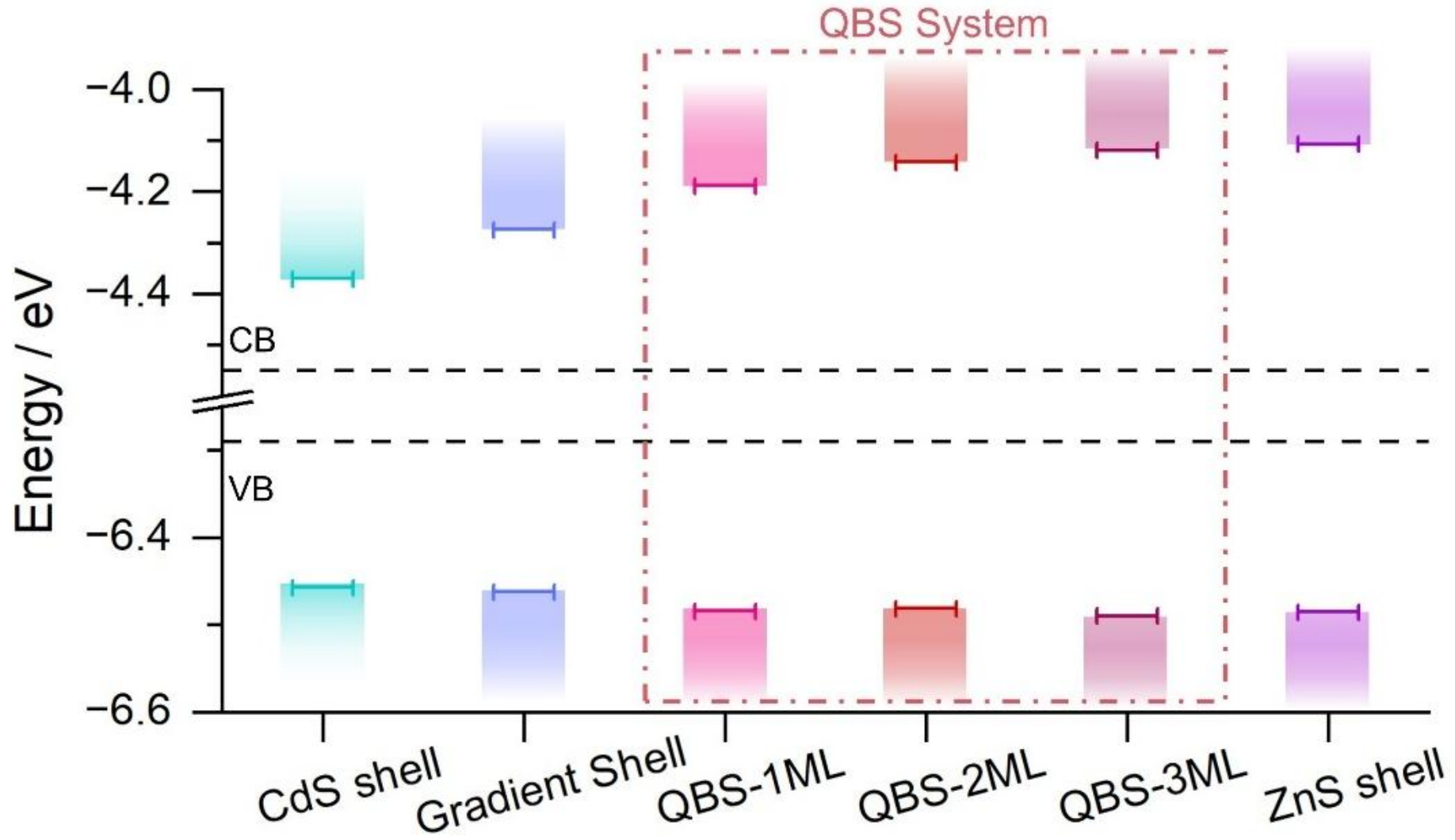

**Supplementary Fig. 13 | Calculated conduction- and valence-bandedge energies relative to bulk CdSe.** In the QBS system, a ZnS barrier thickness of ≥2 monolayers preserves quantum confinement at a level comparable to that of directly grown ZnS shell of the same thickness.

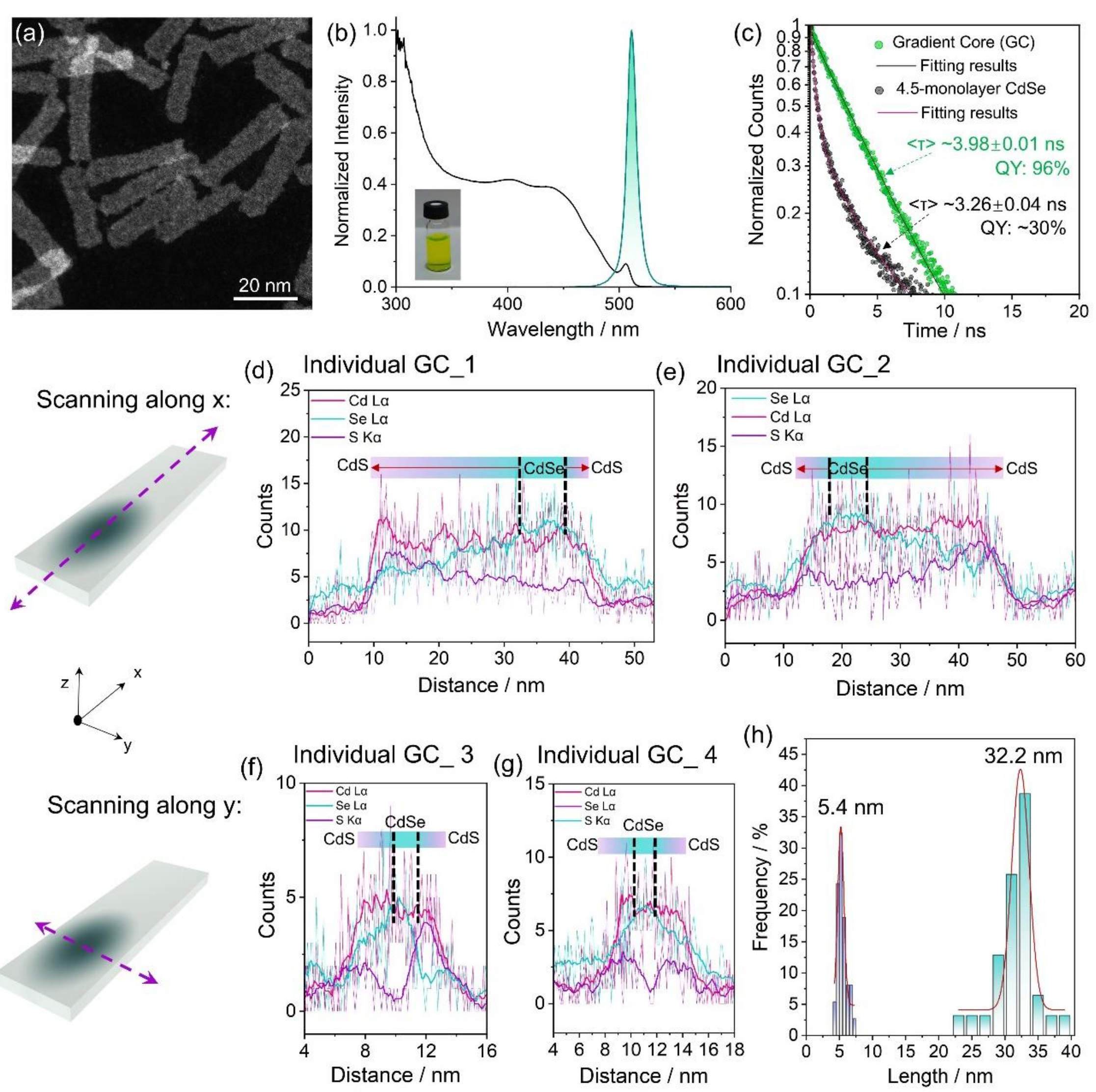

**Supplementary Fig. 14 | Structural and optical characterizations of gradient core (GC). a,** STEM image of the GC. **b**, Absorption and PL spectra of the GC. **c**, Comparison of fluorescence lifetimes between the GC and 4.5-monolayer pure CdSe without a gradient structure. The nearly mono-exponential PL decay of the GC, compared to pure CdSe, indicates improved passivation of peripheral defect states in the gradient structure. **d-e**, EDS line-scan profiles acquired under aberration-corrected STEM for two individual GCs along the long-axis (x, length) direction, showing that the CdSe region along x is approximately 6 nm. **f-g**, EDS line-scan profiles for two individual GCs along the short-axis (y, width) direction, indicating that the CdSe region along y is approximately 2.5 nm. **h**, Statistical distributions of the GC length and width. The average length and width are 32.2 nm and 5.4 nm, respectively.

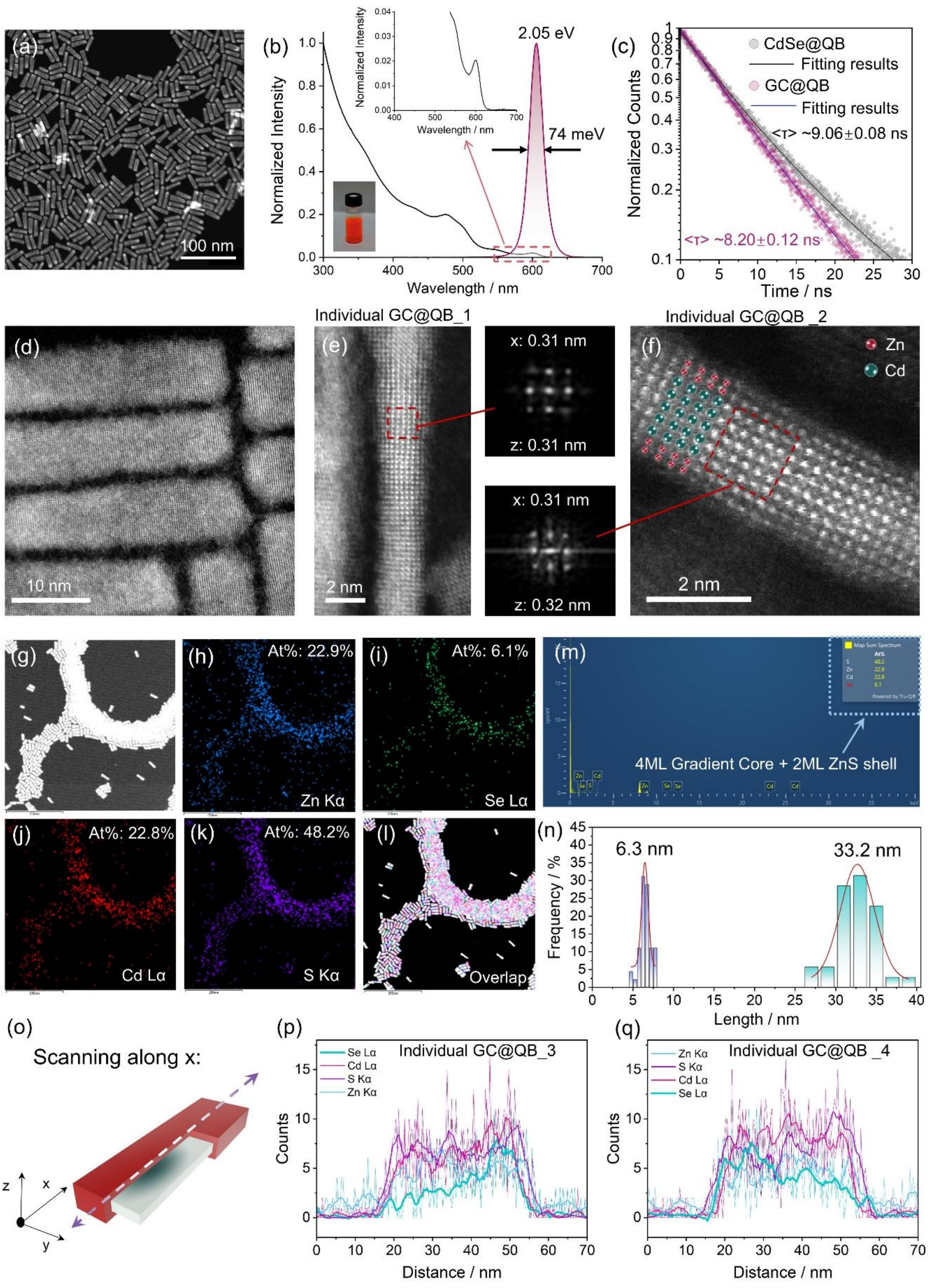

**Supplementary Fig. 15 | Structural and optical characterizations of GC@QB. a,** Low-magnification STEM image of GC@QB, showing uniform size and morphology. **b**, Absorption and PL spectra of the GC@QB. **c**, Comparison of fluorescence lifetimes between GC@QB and CdSe@QB with pure CdSe as the core. The shorter radiative lifetime observed in

GC@QB arises from enhanced electron-hole interactions within the laterally confined CdSe domain. **d**, Atomic-resolution STEM image of individual GC@QB. **e-f**, Magnified cross-sectional STEM images revealing the formation of a 2-monolayer ZnS shell. The corresponding local Fourier transform patterns of the CdSe core region exhibit preserved cubic symmetry characteristic of the zinc blende lattice, without evident lattice distortion. **g-m**, EDS elemental mapping of GC@QB clusters, confirming the presence and atomic percentages (At%) of Zn, Se, Cd, and S. The measured elemental ratios agree well with a GC@QB structure comprising a 5-monolayer Cd and a total of 4 monolayers of Zn. **n**, Statistical distributions of the GC@QB length and width. The average length and width are 33.2 nm and 6.3 nm, respectively. **o-q**, EDS line-scan profiles acquired under aberration-corrected STEM for two individual GC@QB platelets along the long-axis (x, length) direction, indicating a CdSe domain size of around 6 nm, consistent with that of the GC.

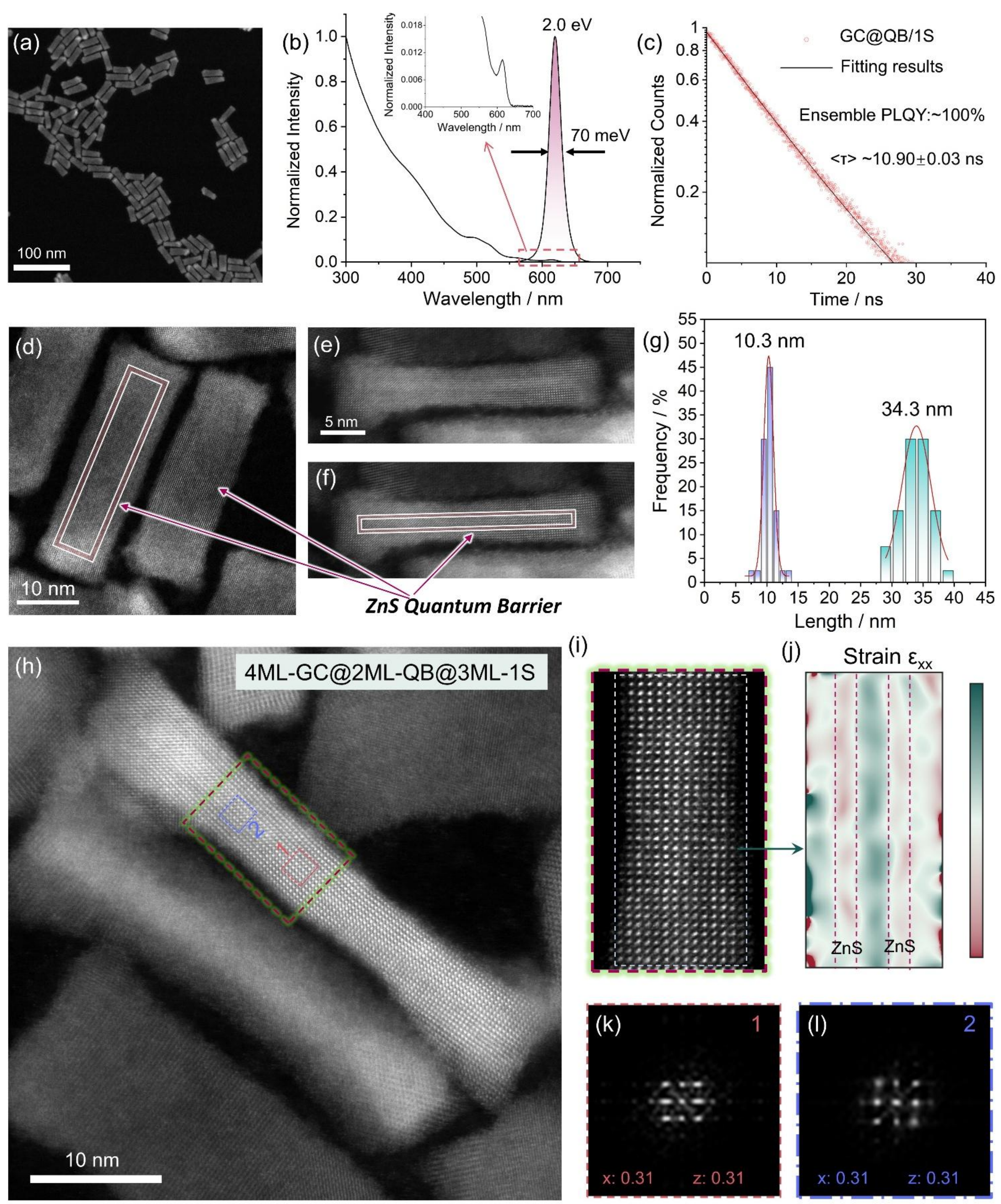

**Supplementary Fig. 16 | Structural and optical characterizations of GC@QB/1S. a,** Low-magnification STEM image of GC@QB/1S, displaying uniform size and morphology. **b**, Absorption and PL spectra of the GC@QB/1S. **c**, Fluorescence lifetime measurement of GC@QB/1S. **d-f**, Atomic-resolution STEM images of individual GC@QB/1S CQWs. Owing to the Z-contrast in high-angle annular dark-field (HAADF) STEM, the contour of the ZnS

quantum barrier can be distinguished, as highlighted in (d) and (f). **g**, Statistical distributions of the GC@QB/1S length and width. The average length and width are 34.3 nm and 10.3 nm, respectively. **h**, Atomic-resolution STEM image along the xz cross-section of an individual GC@QB/1S, revealing a structure consisting of a 4.5-monolayer GC, a 2-monolayer ZnS quantum barrier, and a 3-monolayer $Cd_yZn_{1-y}S$ gradient shell, yielding a total thickness of 4.8 nm. **i-j**, Local Fourier transform patterns extracted from two core regions in (h), confirming well-preserved cubic symmetry characteristic of the zinc blende lattice without discernible lattice distortion. **k–l**, Enlarged view of (k) the xz cross-section in (h) and corresponding (l) geometric phase analysis (GPA) map, further evidencing two regions with reduced lattice parameter within the shell, corresponding to the ZnS quantum barrier (dashed lines).

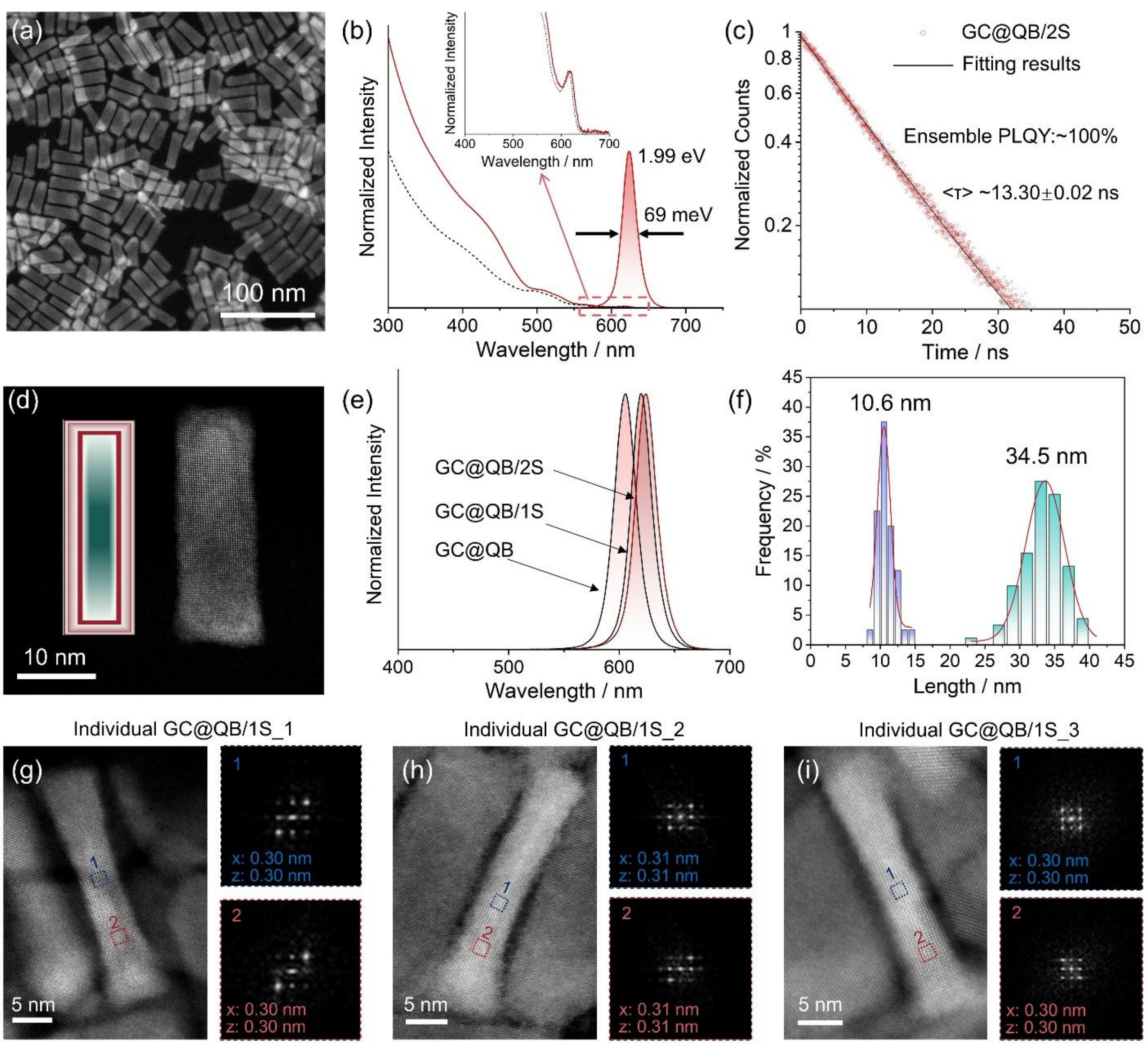

**Supplementary Fig. 17 | Structural and optical characterizations of GC@QB/2S. a,** Low-magnification STEM image of GC@QB/2S, indicating uniform size and morphology. **b**, Absorption and PL spectra of the GC@QB/2S. **c**, Fluorescence lifetime measurement of

GC@QB/2S, exhibiting near mono-exponential PL decay (PLQY: ~100%) and a PL lifetime of ~13.30 ns. **d**, Atomic-resolution STEM image of an individual GC@QB/2S. e, Comparison of PL spectra among GC@QB, GC@QB/1S, and GC@QB/2S. **f**, Statistical distributions of GC@QB/2S length and width, with average values of 34.5 nm and 10.6 nm, respectively. **g-h**, Atomic-resolution STEM images along the xz cross-section of three individual GC@QB/2S CQWs, with corresponding local Fourier transform patterns extracted from core regions, confirming well-preserved cubic symmetry characteristic of the zinc blende lattice without discernible lattice distortion.

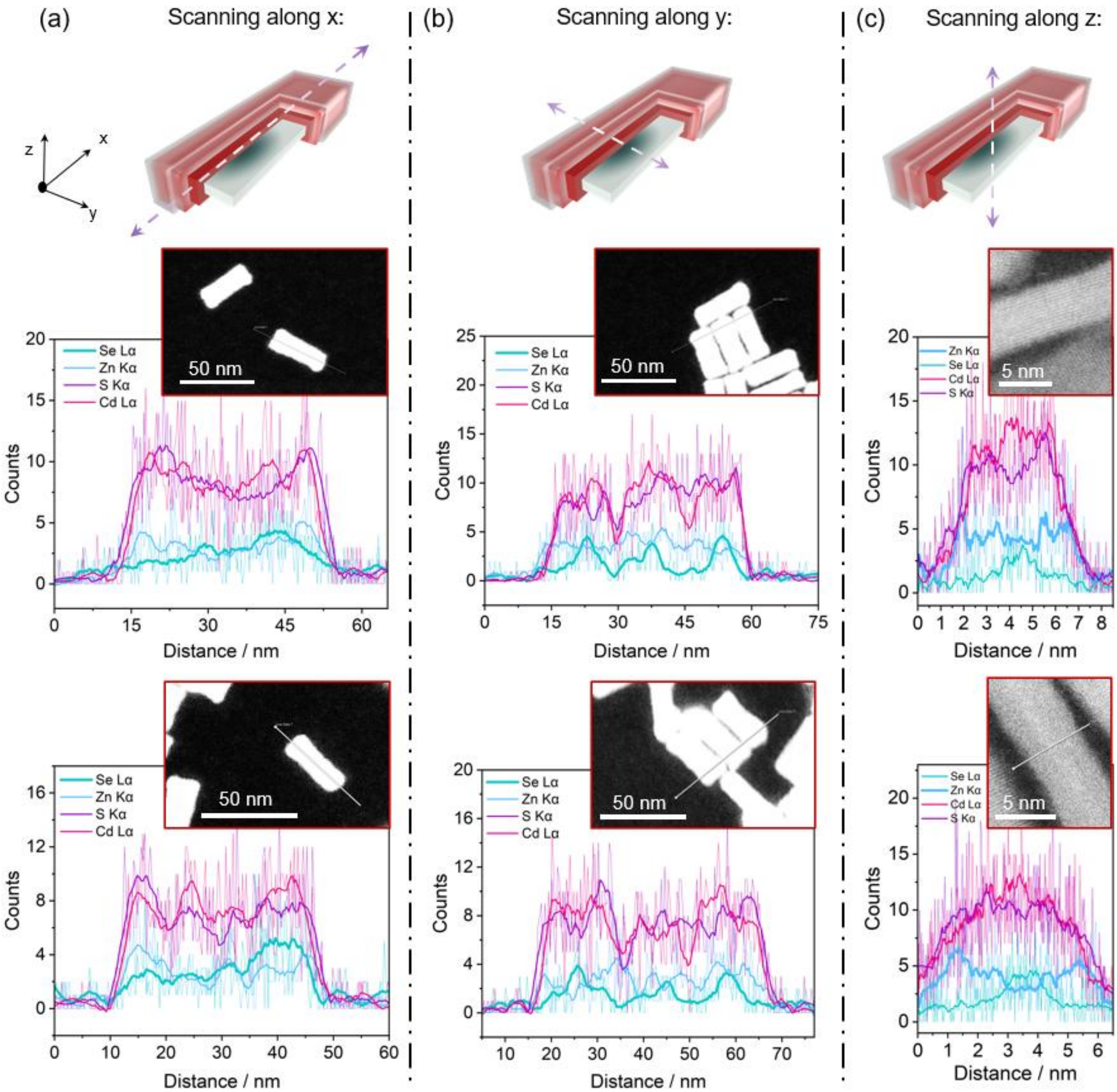

**Supplementary Fig. 18 | Additional structural characterizations of GC@QB/2S. a-c,** EDS line-scans acquired under aberration-corrected STEM along the (a) long-axis (x, length),

(b) short-axis (y, width), and (c) vertical (z, out-of-plane) directions. For each direction, two independent CQWs were analyzed to show reproducibility. The extracted CdSe core dimensions are consistent with the previous measurements in GC and GC@QB (length ~6 nm and width ~2.5 nm). Along the z direction, the shell composition agrees well with the designed QBS architecture, as evidenced by the characteristic variations in Zn, Cd, and S signal intensities across the vertical line-scan.

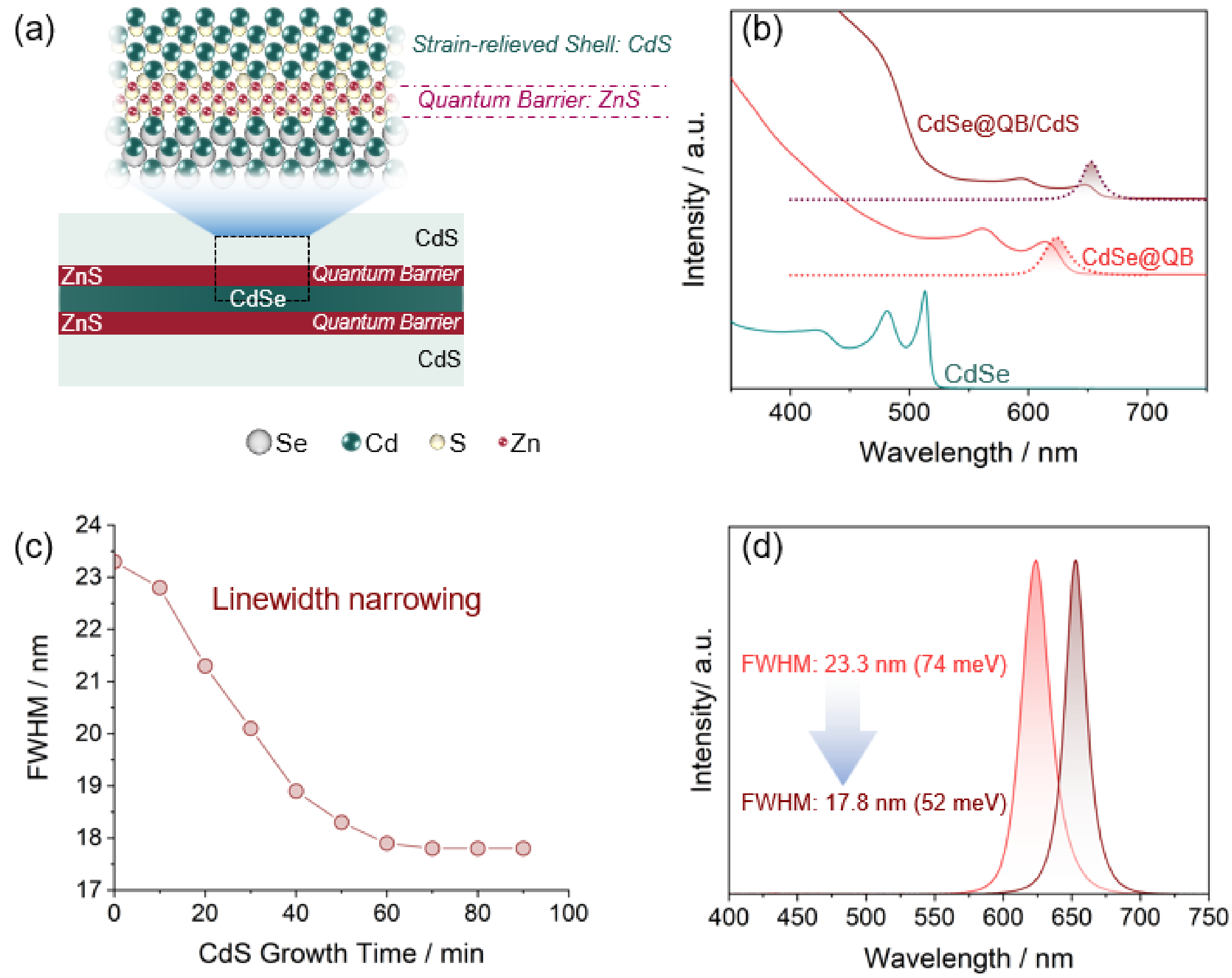

**Supplementary Fig. 19 | Emission linewidth narrowing in CdSe@QB/CdS. a,** Schematic illustration of the QBS architecture with CdS as the outer shell. **b**, Evolution of absorption spectra for CdSe core (green), CdSe@QB (pink), and CdSe@QB/CdS (red). **c**, Evolution of the emission linewidth during CdS outer-shell growth. **d**, Comparison of PL spectra before and after CdS outer-shell growth, exhibiting a linewidth reduction from 23.3 nm (CdSe@QB) to 17.8 nm (CdSe@QB/CdS).

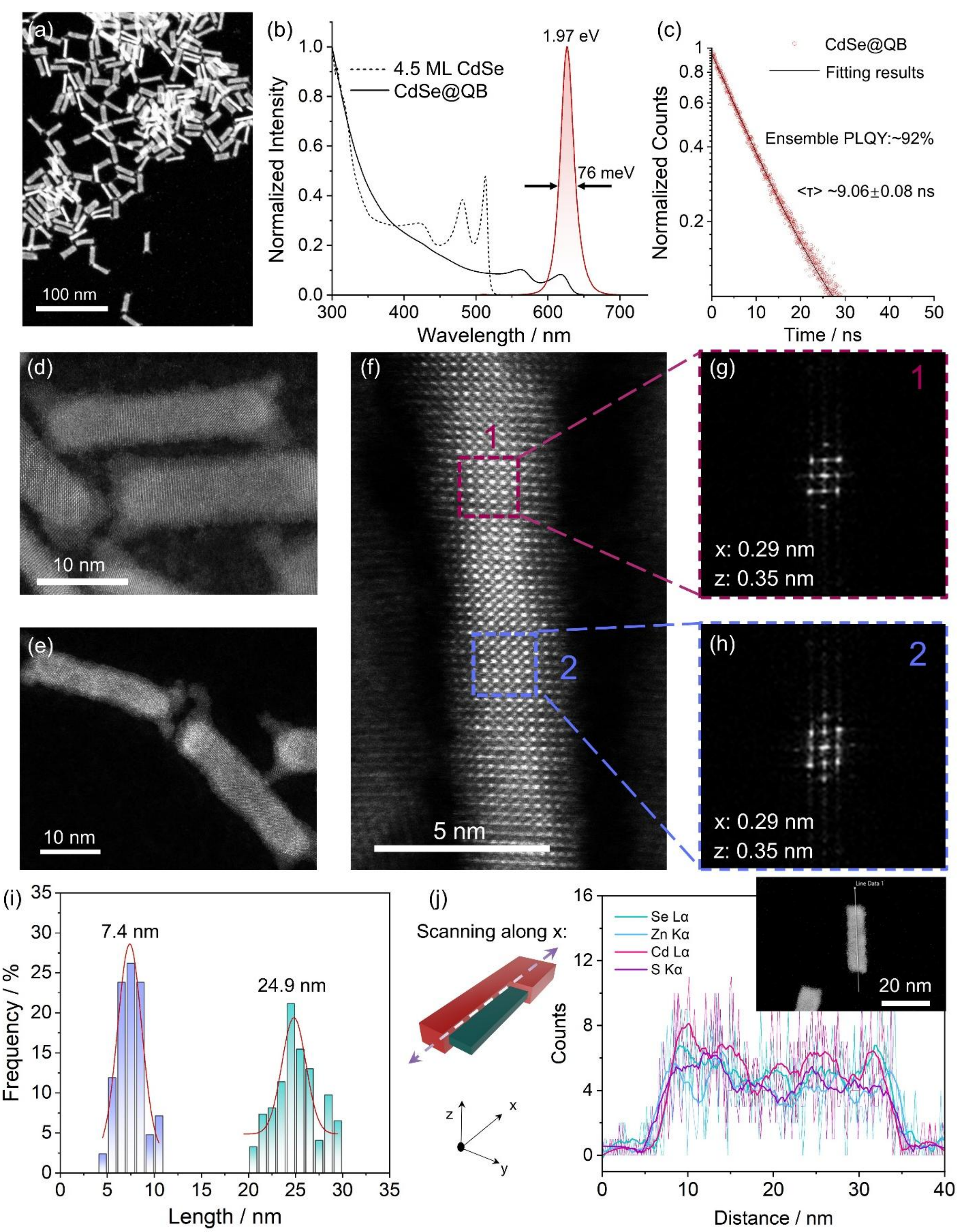

**Supplementary Fig. 20 | Structural and optical characterizations of the control sample CdSe@QB. a,** Low-magnification STEM image of CdSe@QB. **b**, Absorption and PL spectra of CdSe@QB (solid line). The black dashed line represents the absorption spectrum of the pure CdSe core. **c**, Fluorescence lifetime measurement of CdSe@QB. The decay curve is well fitted by a biexponential decay, yielding a radiative lifetime of ~9 ns. The QY of the

sample is ~92%. **d-e**, Atomic-resolution STEM images of individual CdSe@QB CQWs. **f**, Atomic-resolution STEM image along the xz cross-section of an individual CdSe@QB, revealing a structure composed of a 4.5-monolayer CdSe core and a 3-monolayer ZnS shell. **g-h**, Local Fourier transform patterns extracted from two core regions in (f), showing lattice distortion in the CdSe core induced by compressive stress from the ZnS shell. The in-plane lattice parameter of CdSe is reduced to 0.29 nm, while the out-of-plane lattice parameter expands to 0.35 nm. **i**, Statistical distributions of CdSe@QB length and width, with average values of 24.9 nm and 7.4 nm, respectively. **j**, EDS line-scan profiles acquired under aberration-corrected STEM along the long-axis (x, length) direction.

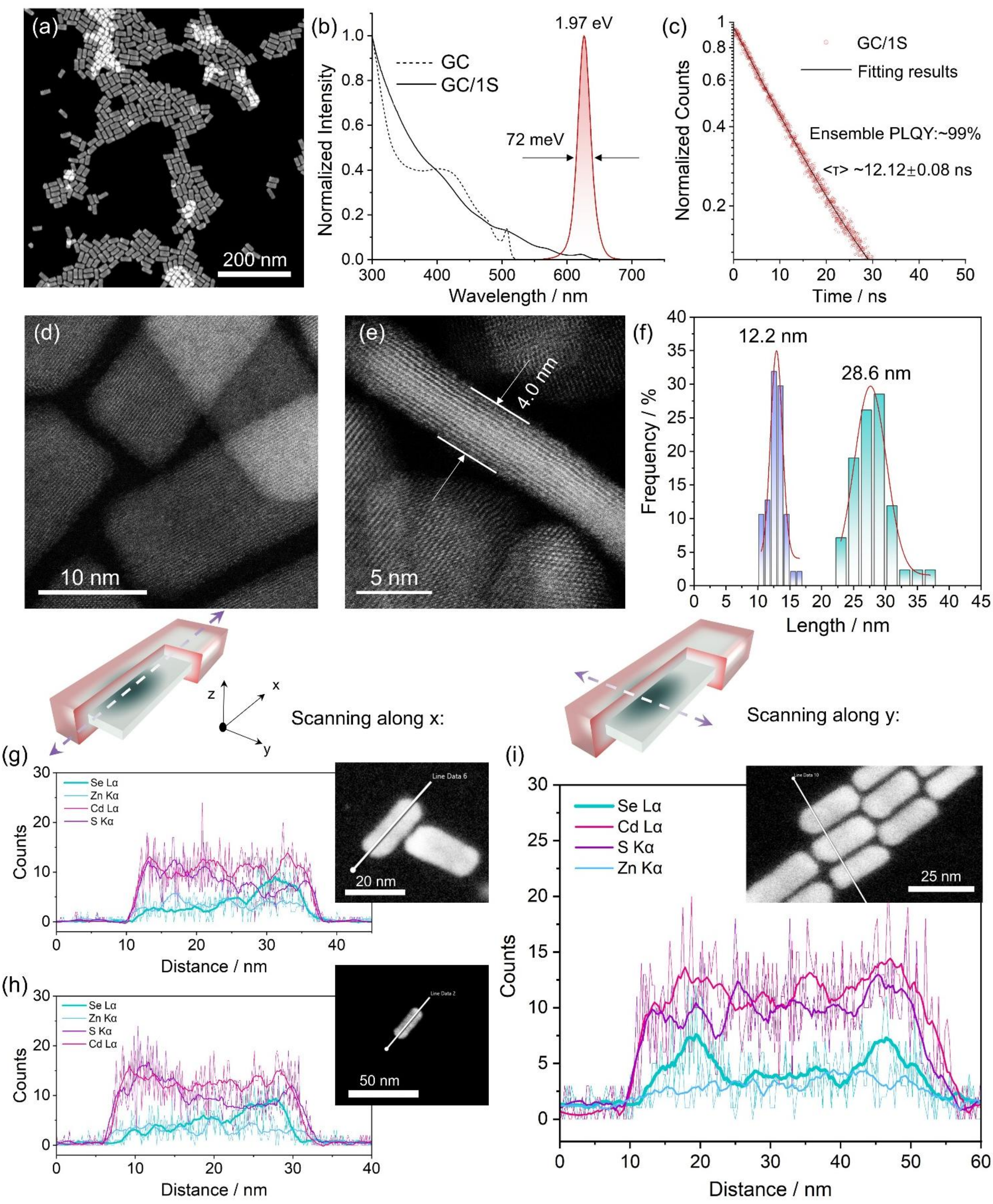

**Supplementary Fig. 21 | Structural and optical characterizations of the control sample GC/1S. a,** Low-magnification STEM image of GC/1S. **b**, Absorption and PL spectra of GC/1S (solid line). The black dashed line represents the absorption spectrum of the GC. **c**, Fluorescence lifetime measurement of GC/1S. The decay curve is well fitted by a mono-exponential decay, yielding a radiative lifetime of ~12.1 ns. The QY of the sample is ~99%.

**d-e**, Atomic-resolution STEM images of individual GC/1S CQWs with an overall thickness of ~4 nm. **f**, Statistical distributions of GC/1S length and width, with average values of 28.6 nm and 12.2 nm, respectively. g-i, EDS line-scan profiles acquired under aberration-corrected STEM along the (g-h) long-axis (x, length) direction and the short-axis (y, width) direction, indicating CdSe domain sizes of approximately 6 nm (x) and 3.2 nm (y), respectively.

**Supplementary Table 1 | Characterized parameters of the VQD-CQWs series and control samples.**

| | **Length (CdSe)** | **Width (CdSe)** | **Thickness** | **Length (overall)** | **Width (overall)** | **FWHM** | **QY** | **PL lifetime** |
|---|---|---|---|---|---|---|---|---|
| **GC@QB** | 6.0 nm | 2.5 nm | 2.2 nm | 33.2 nm | 6.3 nm | 74 meV | ~98% | 8.2 ns |
| **GC@QB/1S** | | | 4.8 nm | 34.3 nm | 10.3 nm | 70 meV | ~100% | 10.9 ns |
| **GC@QB/2S** | | | 5.2 nm | 34.5 nm | 10.6 nm | 69 meV | ~100% | 13.3 ns |
| **CdSe@QB** | 24.9 nm | 7.4 nm | 2.4 nm | 24.9 nm | 7.4 nm | 76 meV | 92% | 9.1 ns |
| **GC@/1S** | 6.3 nm | 3.0 nm | 4.0 nm | 28.6 nm | 12.2 nm | 72 meV | 99% | 12.1 ns |

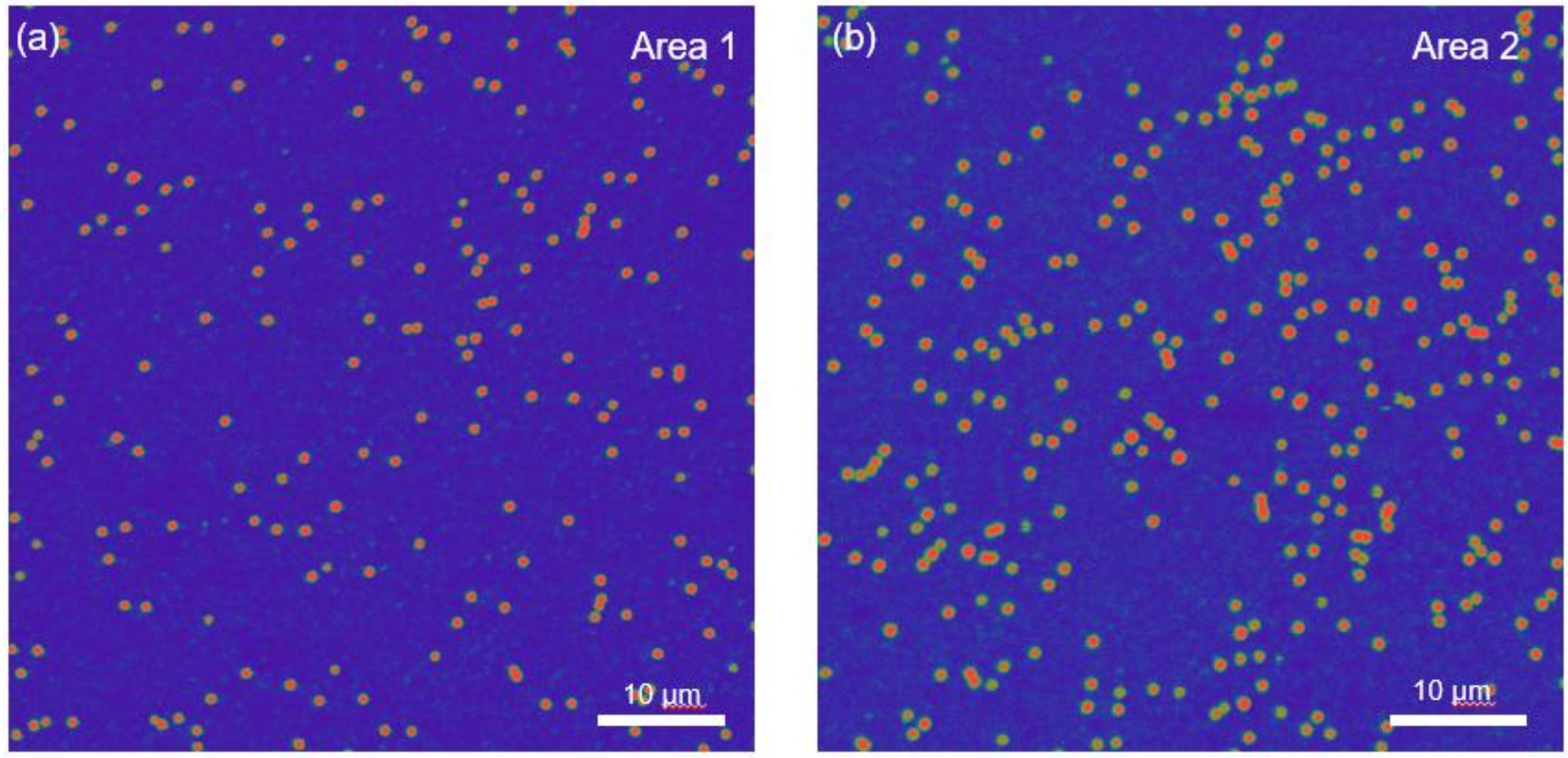

**Supplementary Fig. 22 | PL mapping of individual GC@QB/2S CQWs from two distinct areas.** a-b, Representative PL mapping of a dilute dispersion of GC@QB/2S CQWs spin-coated onto an ultrathin cover glass substrate without encapsulation.

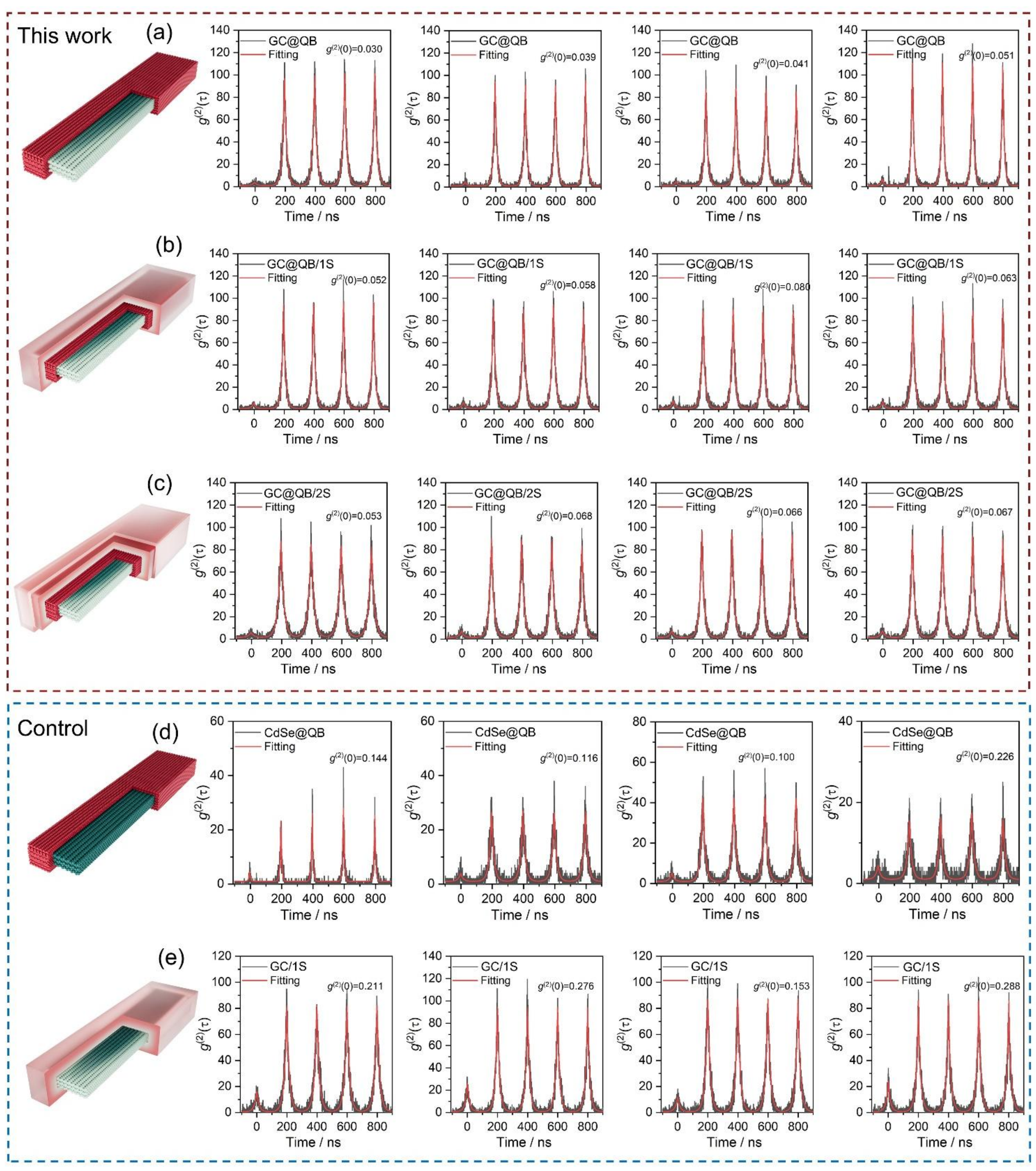

**Supplementary Fig. 23 | Additional second-order intensity correlation function $g^2(\tau)$ characterizations. a-e**, $g^2(\tau)$ measurements for (a) GC@QB, (b) GC@QB/1S, (c) GC@QB/2S, (d) CdSe@QB, and (e) GC@1S, respectively.

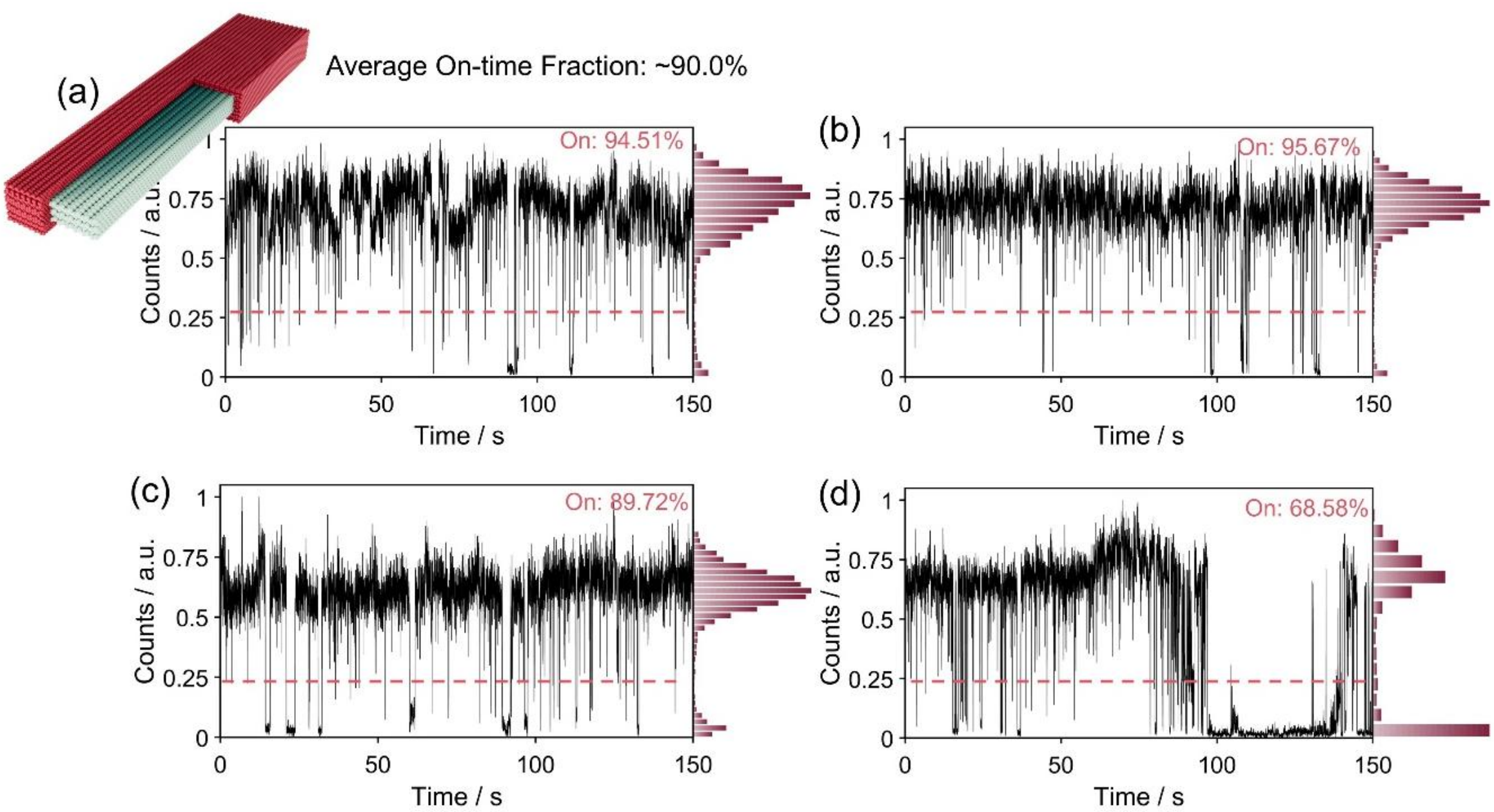

**Supplementary Fig. 24 | PL time traces (time bin: 20 ms) of GC@QB. a-d,** Representative PL time traces recorded from 4 individual CQWs from the same batch, with the on-time fraction shown in the corner of each panel.

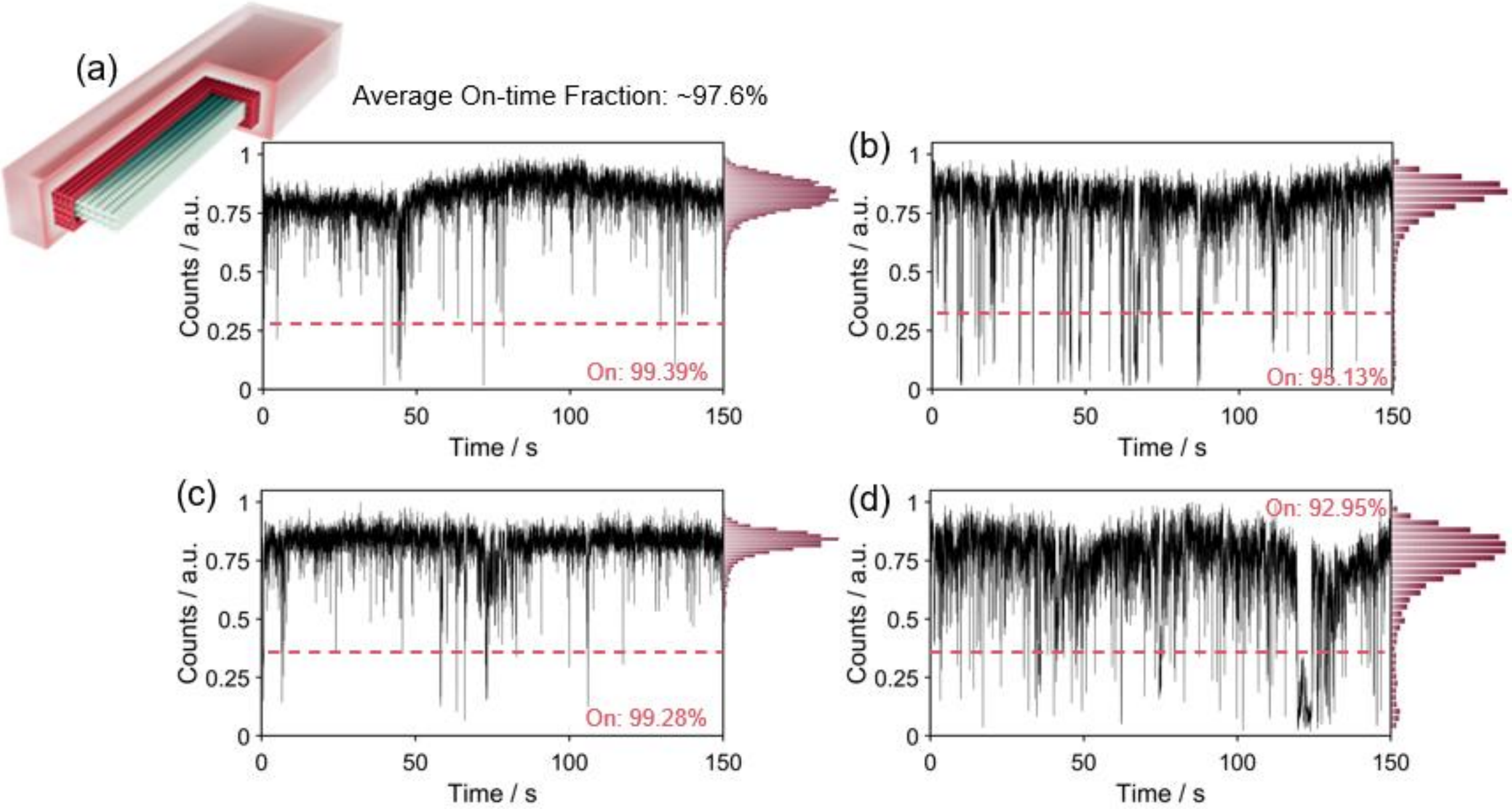

**Supplementary Fig. 25 | PL time traces (time bin: 20 ms) of GC@QB/1S. a-d,** Representative PL time traces recorded from 4 individual CQWs from the same batch, with the on-time fraction shown in the corner of each panel.

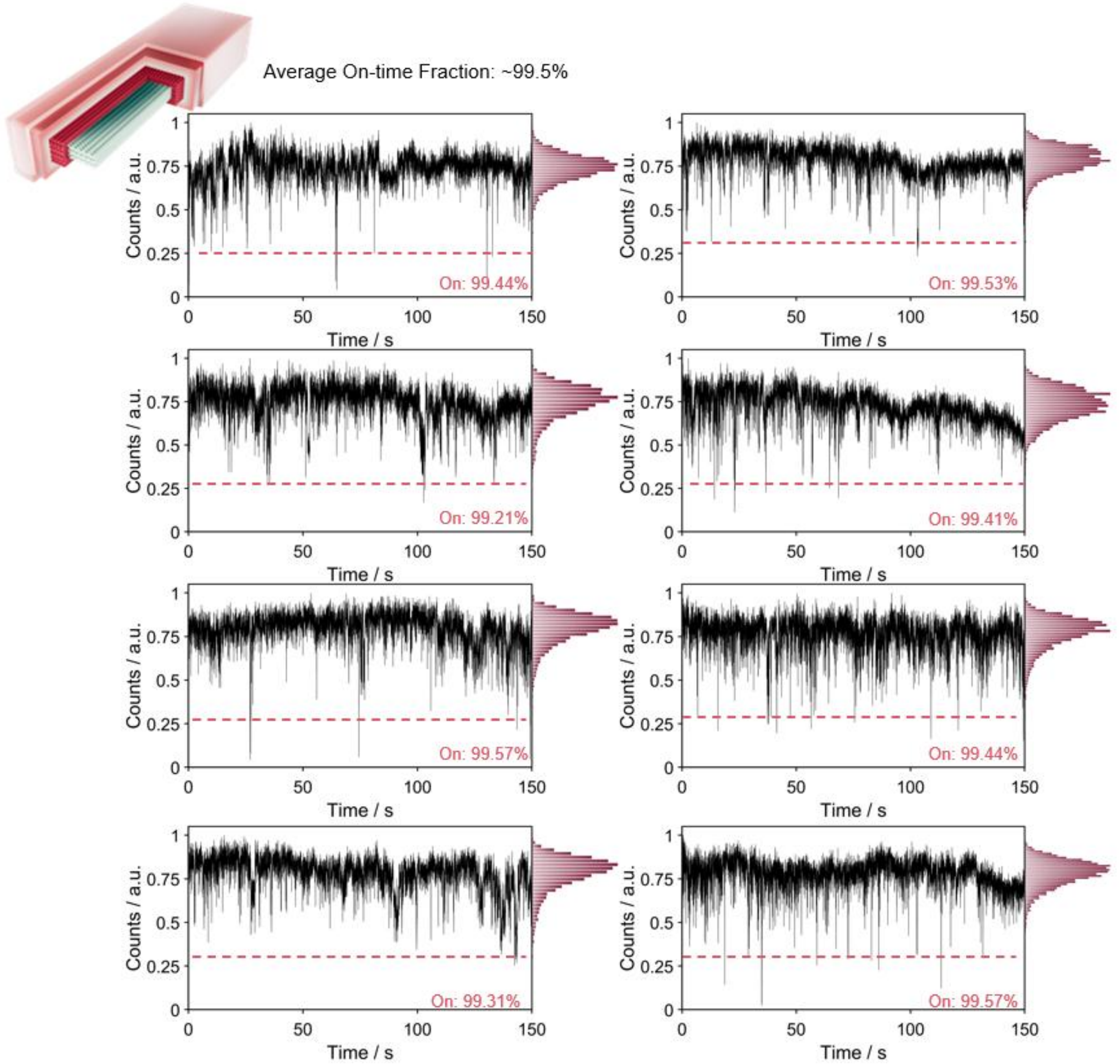

**Supplementary Fig. 26 | PL time traces (time bin: 20 ms) of GC@QB/2S. a-d,** Representative PL time traces recorded from 8 individual CQWs from the same batch, with the on-time fraction shown in the corner of each panel.

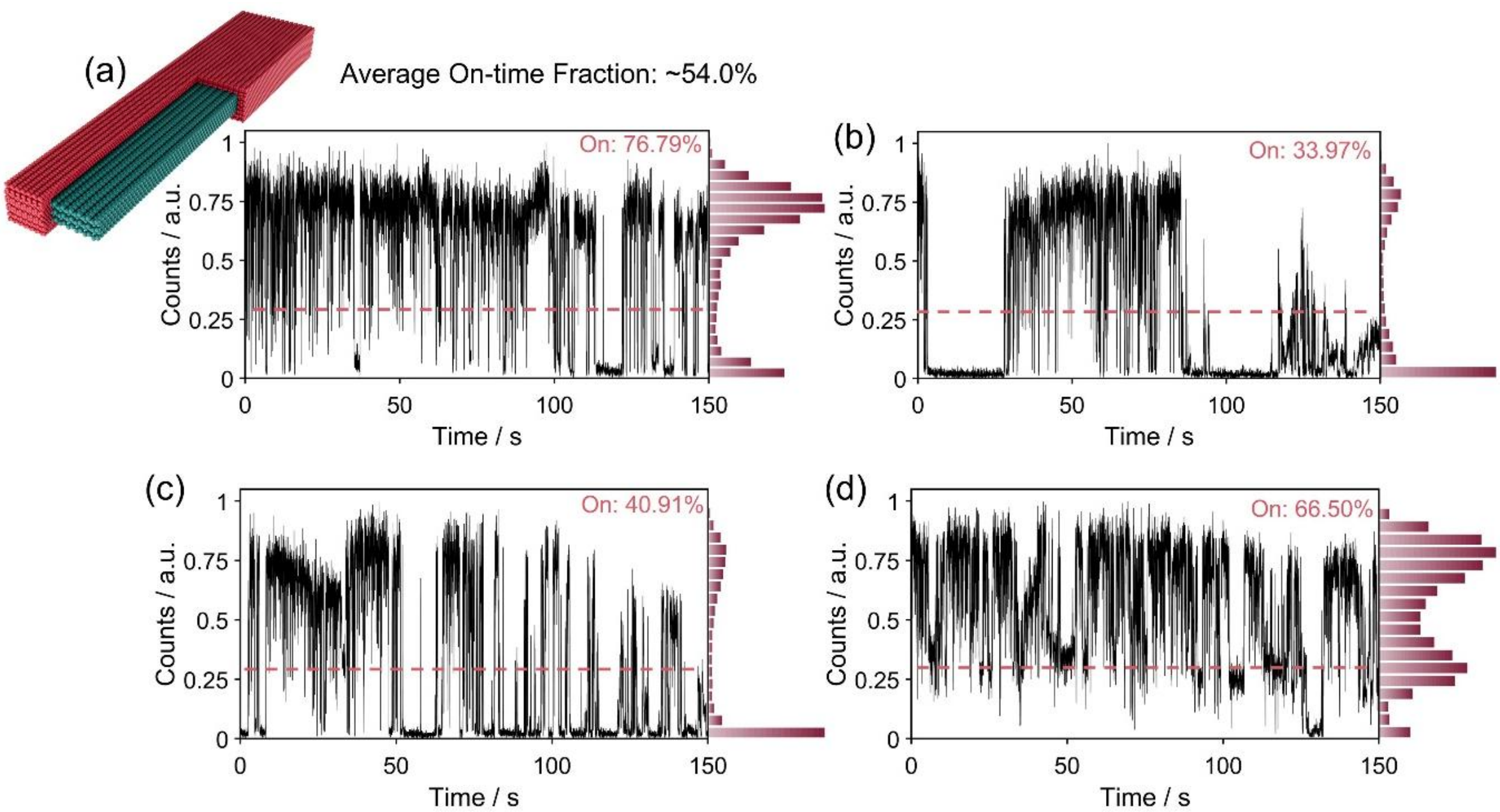

**Supplementary Fig. 27 | PL time traces (time bin: 20 ms) of CdSe@QB. a-d,** Representative PL time traces recorded from 4 individual CQWs from the same batch, with the on-time fraction shown in the corner of each panel.

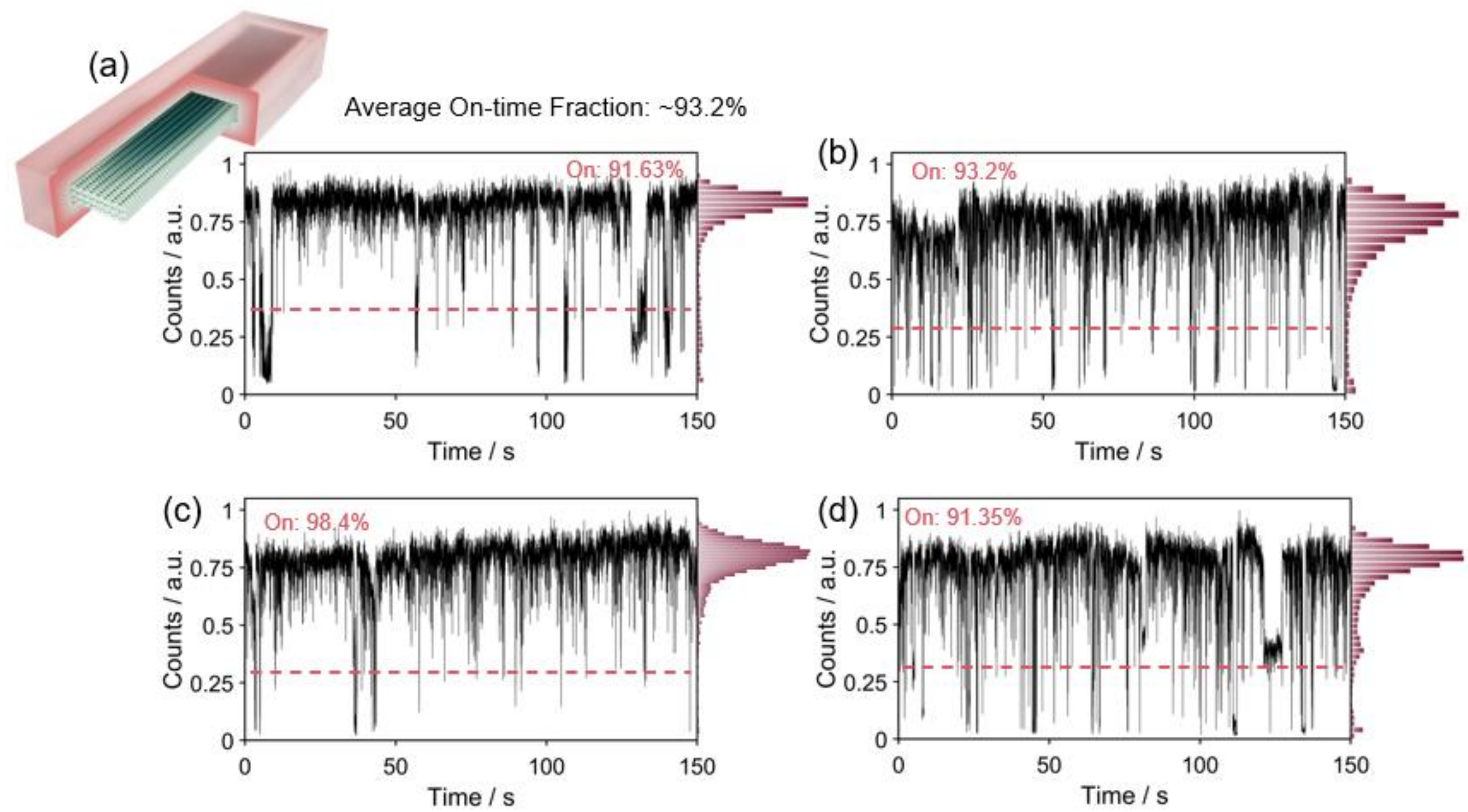

**Supplementary Fig. 28 | PL time traces (time bin: 20 ms) of GC/1S. a-d,** Representative PL time traces recorded from 4 individual CQWs from the same batch, with the on-time fraction shown in the corner of each panel.

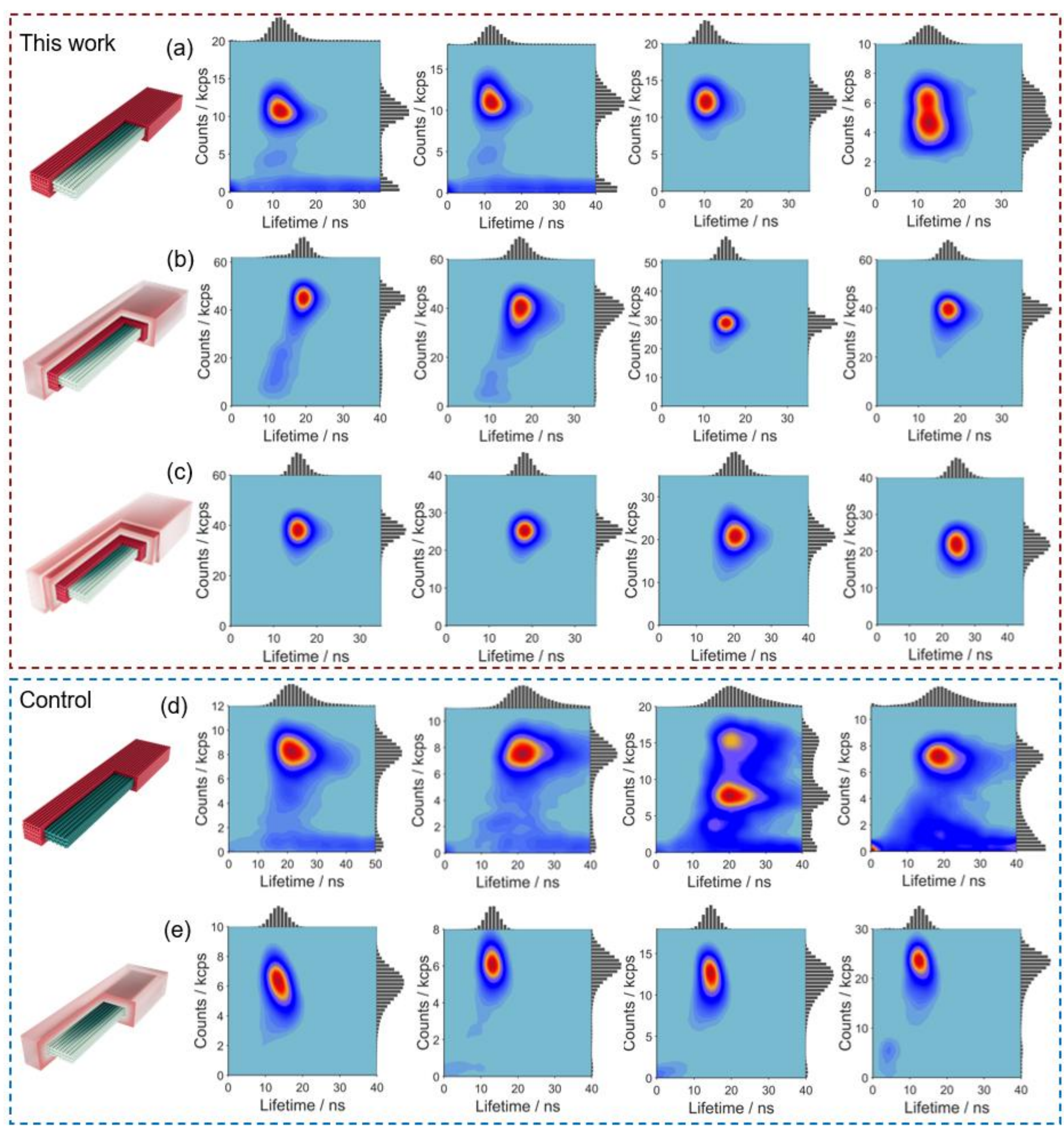

**Supplementary Fig. 29 | Additional FLID mapping characterizations. a-e**, FLID mapping results for (a) GC@QB, (b) GC@QB/1S, (c) GC@QB/2S, (d) CdSe@QB, and (e) GC@1S, respectively. The results indicate a high consistency of the blinking mechanism within the same batch of CQWs sharing the same architecture. In particular, GC@QB/2S consistently exhibits near-nonblinking behavior.

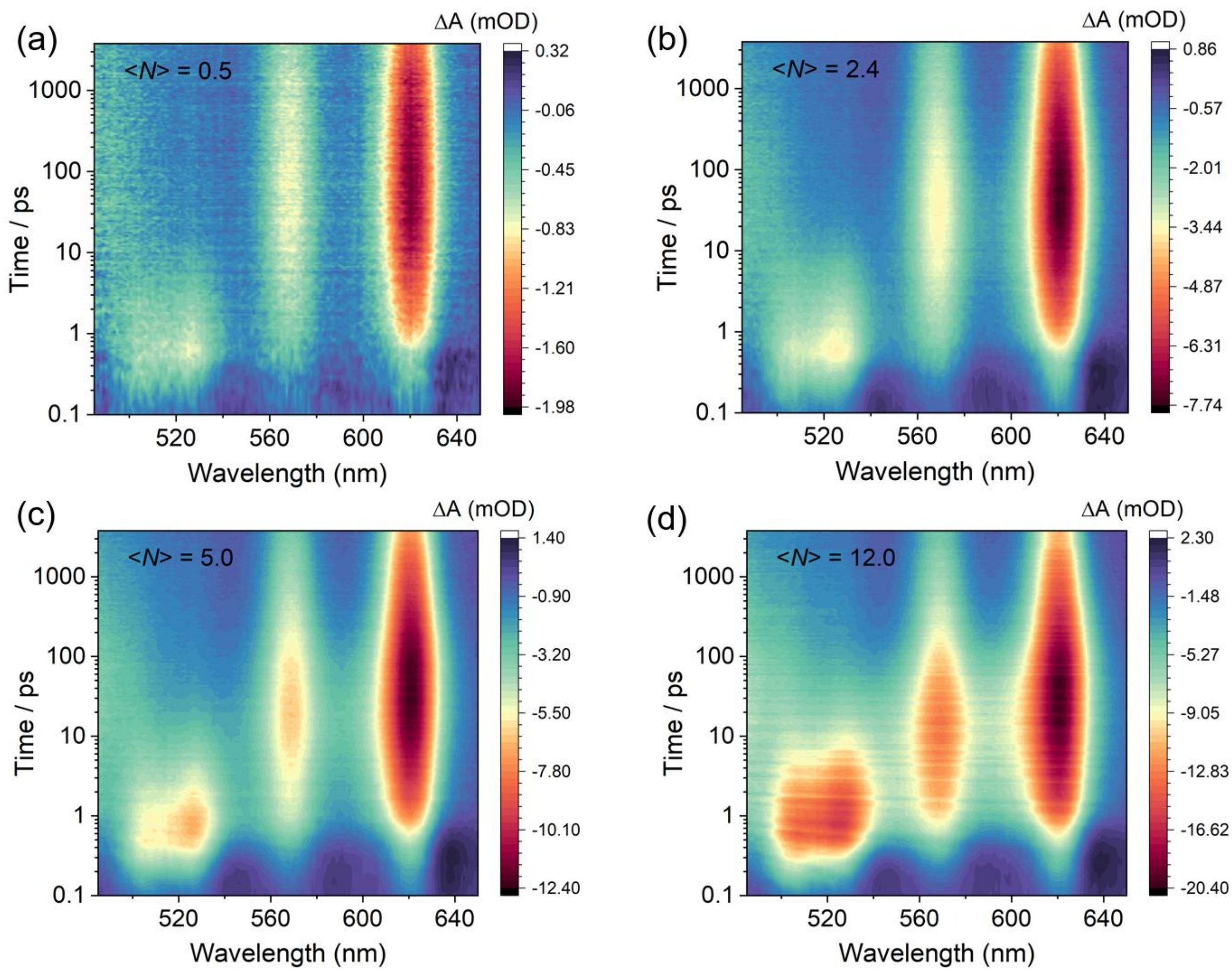

**Supplementary Fig. 30 | Representative power-dependent TA spectra of GC@QB/1S. a-d,** TA spectra of CdSe/QB recorded at pump fluences corresponding to (a) <N> = 0.5, (b) <N> = 2.4, (c) <N> = 5.0, and (d) <N> = 12.0. Here, <N> denotes the average number of photoexcited excitons per CQW.

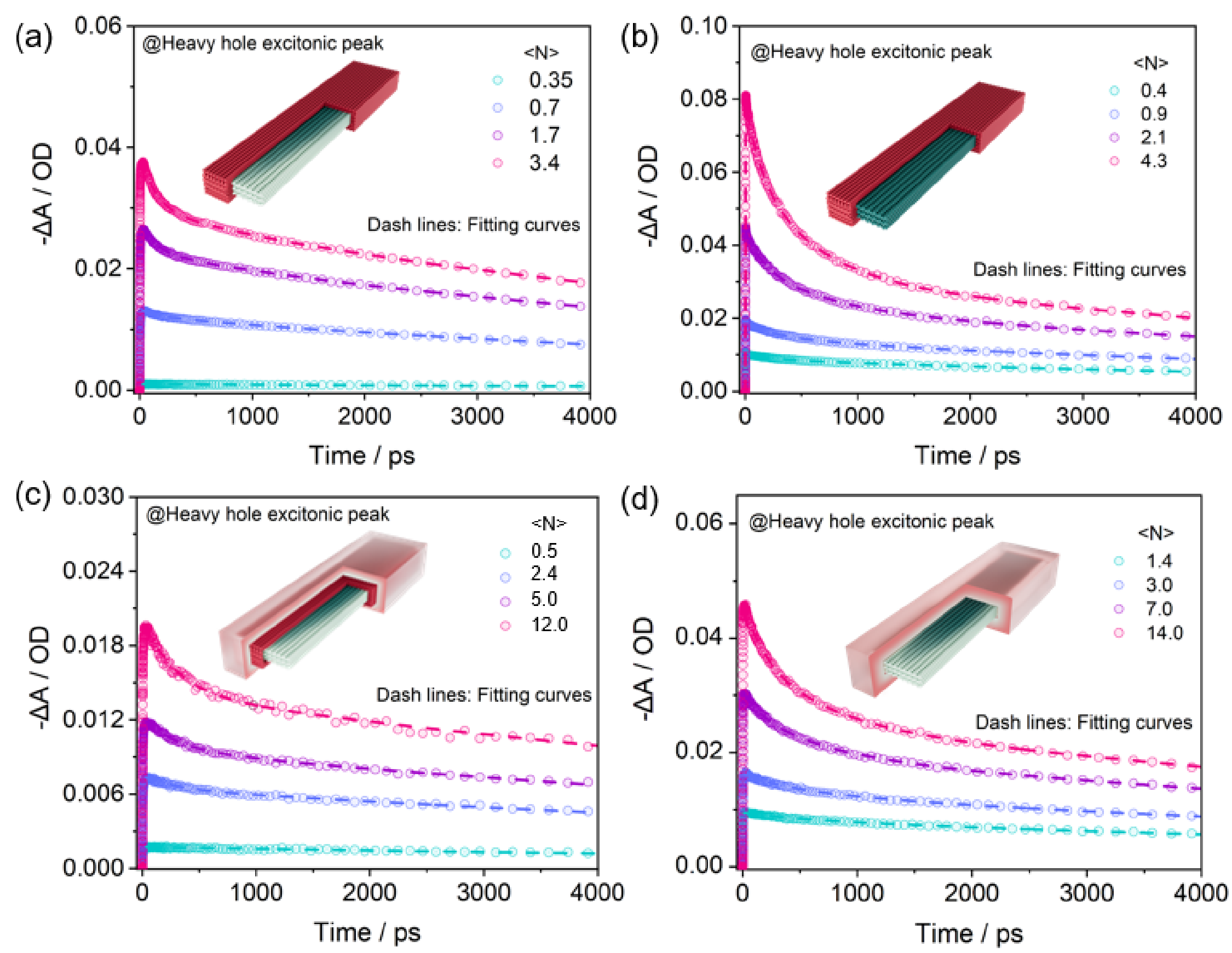

**Supplementary Fig. 31 | Power-dependent TA kinetics at the excitonic peak. a-d,** Power-dependent exciton dynamics of (a) GC@QB, (b) CdSe@QB, (c) GC@QB/1S, and (d) GC/1S, respectively.

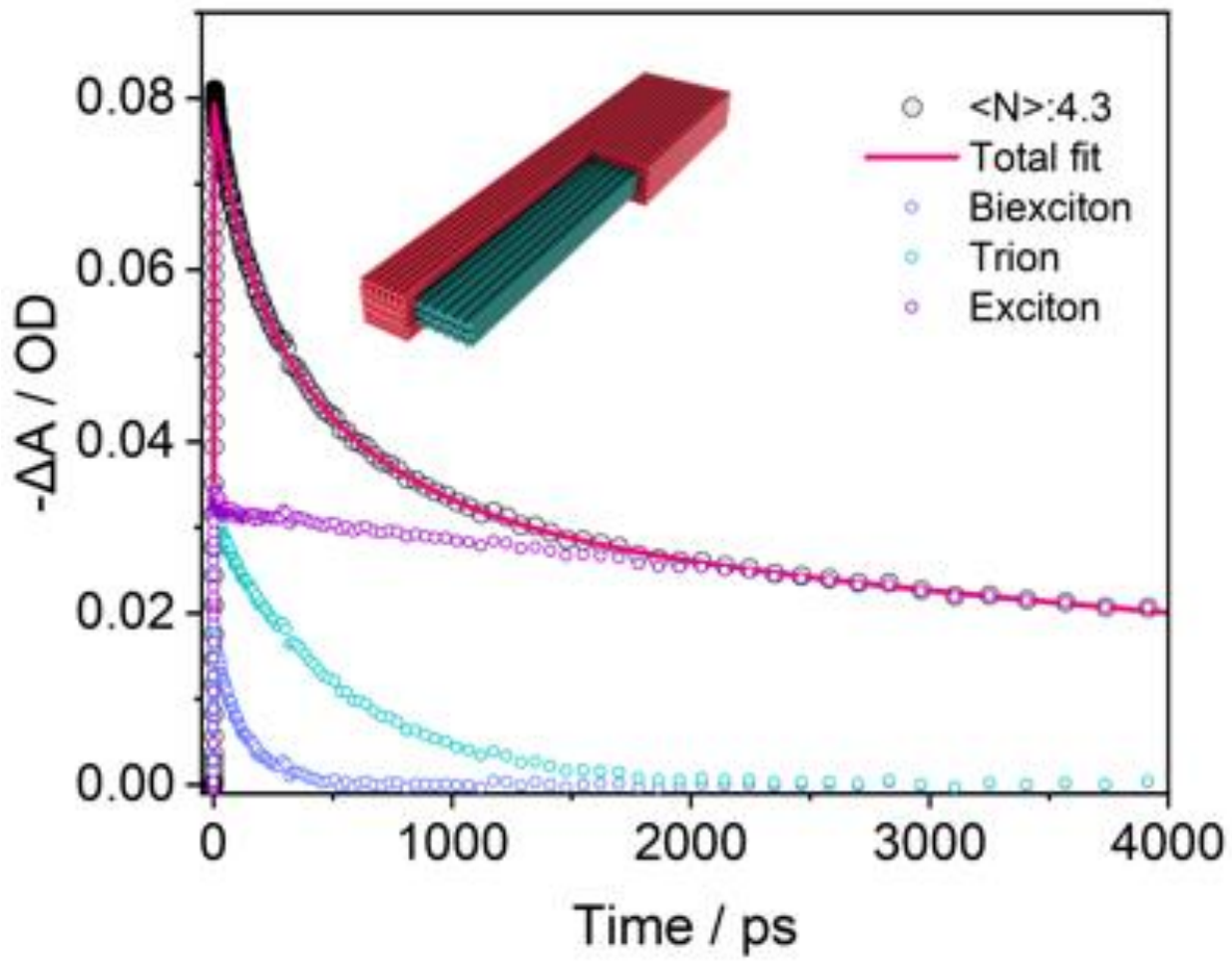

**Supplementary Fig. 32 | Representative deconvolution of the TA bleach kinetics into exciton, biexciton and trion contributions.** The kinetic trace is fitted with a tri-exponential function (red solid line) and subsequently deconvolved to extract the individual components.

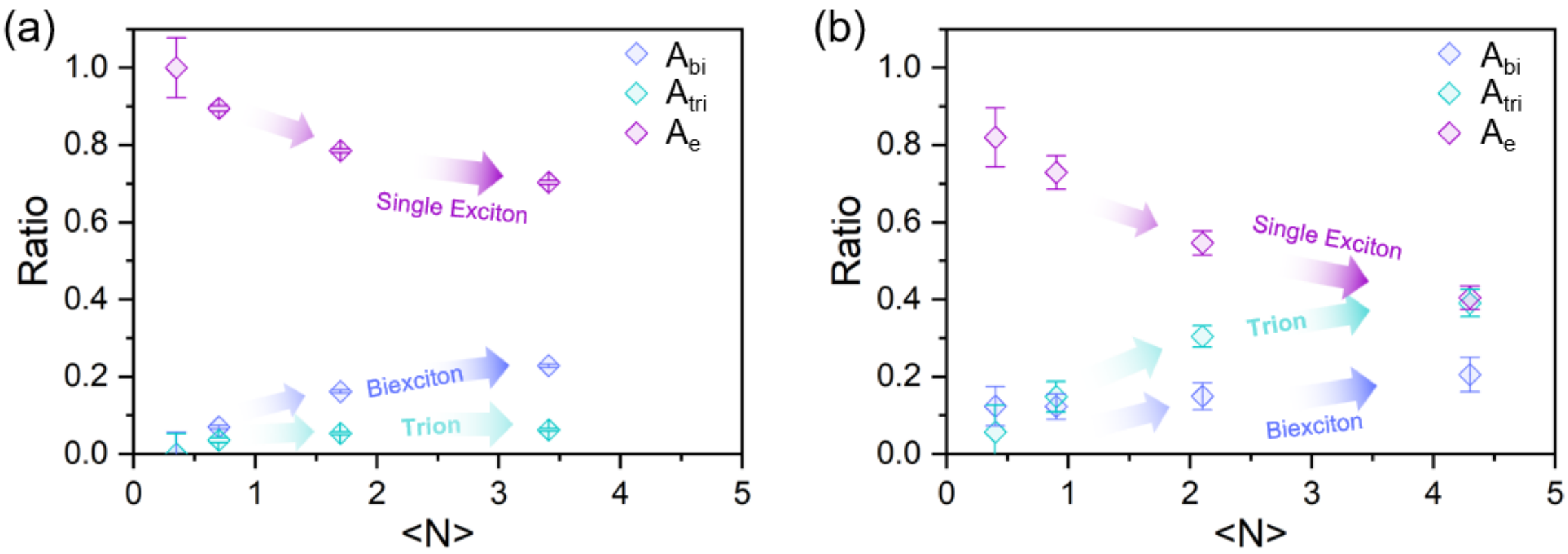

**Supplementary Fig. 33 | Additional extracted fractions of exciton, biexciton, and trion contributions**. **a-b**, Corresponding extracted fractions for (a) GC@QB and (b) CdSe@QB. respectively.

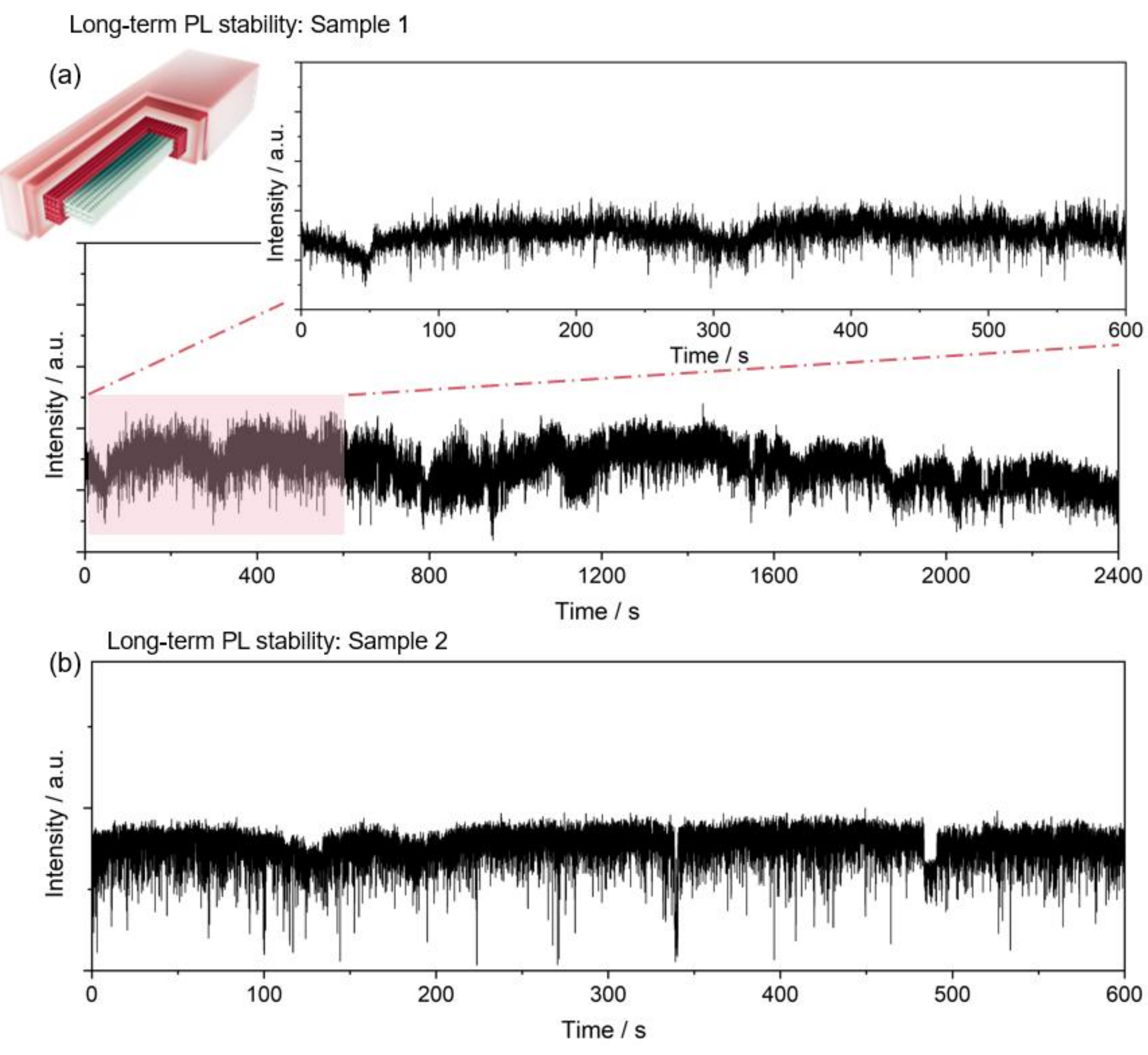

**Supplementary Fig. 34 | PL stability of individual GC@QB/2S. a-b,** Representative PL intensity time traces recorded under continuous excitation in air without any encapsulation. The GC@QB/2S individuals exhibit pronounced anti-bleaching characteristics and near-nonblinking behavior for durations exceeding 40 min. The slow intensity fluctuations observed

in the traces arise from periodic refocusing during long-term measurements, as environmental vibrations can gradually induce defocusing and thus affect the collection efficiency.

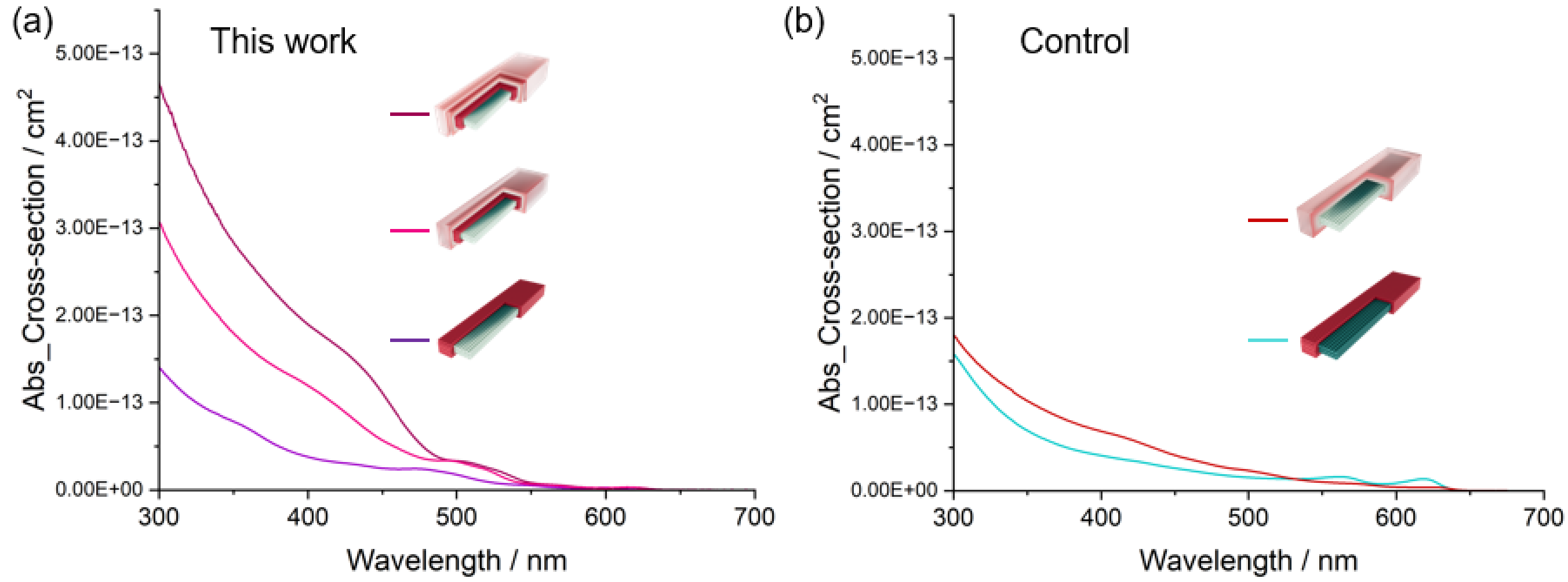

**Supplementary Fig. 35 | Determination of absorption cross-sections. a-b**, Wavelength-dependent absorption cross-sections of (a) the VQD-CQWs series in this work and (b) the control samples. The absorption cross-sections were determined following the method.[5] Briefly, dried CQW powders were completely dissolved in water using concentrated nitric acid, and the elemental concentrations were quantified by inductively coupled plasma optical emission spectroscopy (ICP-OES). The absorption cross-sections were then calculated based on the measured concentrations, absorption spectra, and structural information obtained from aberration-corrected TEM analysis.

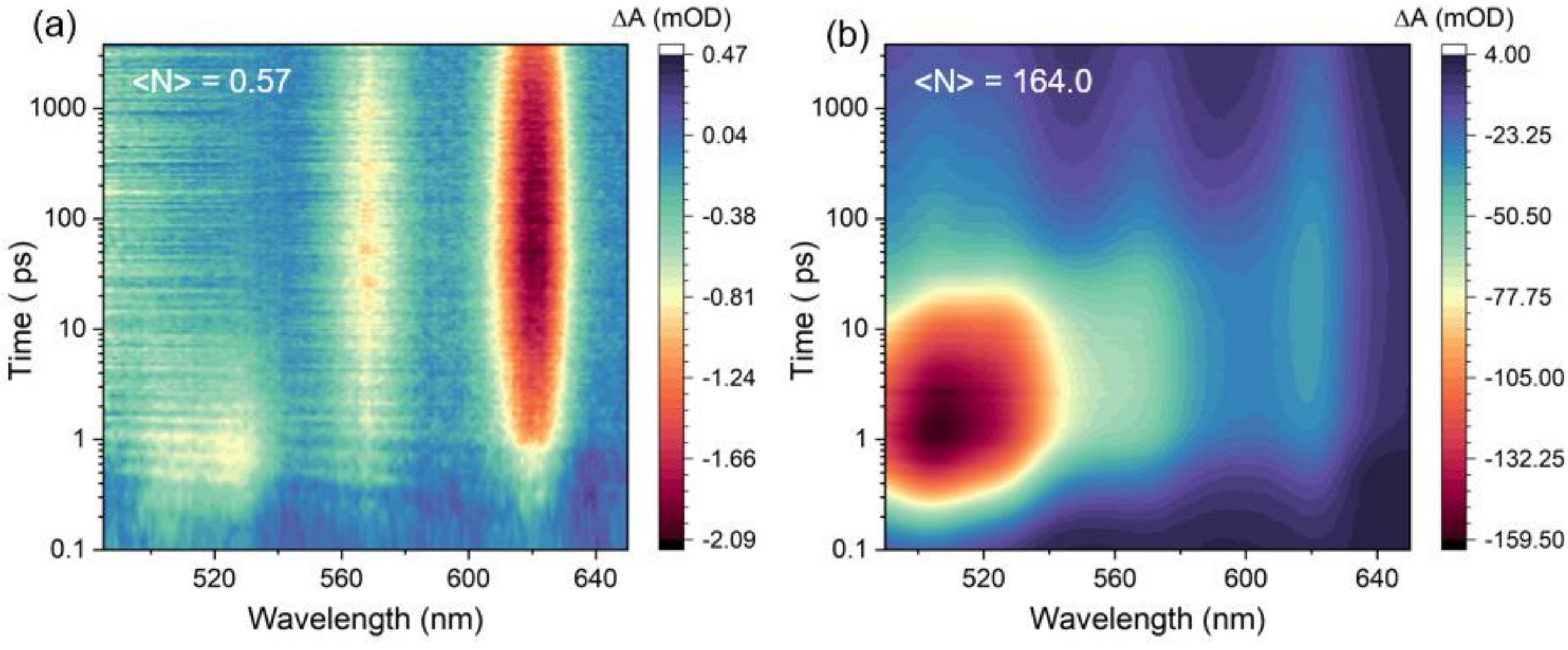

**Supplementary Fig. 36 | TA spectra of GC@QB/2S under low and high pump fluences. a-b**, TA spectra of GC@QB/2S dispersions excited at pump fluences corresponding to average exciton numbers (a) <N> = 0.57 and (b) <N> = 164.0, respectively.

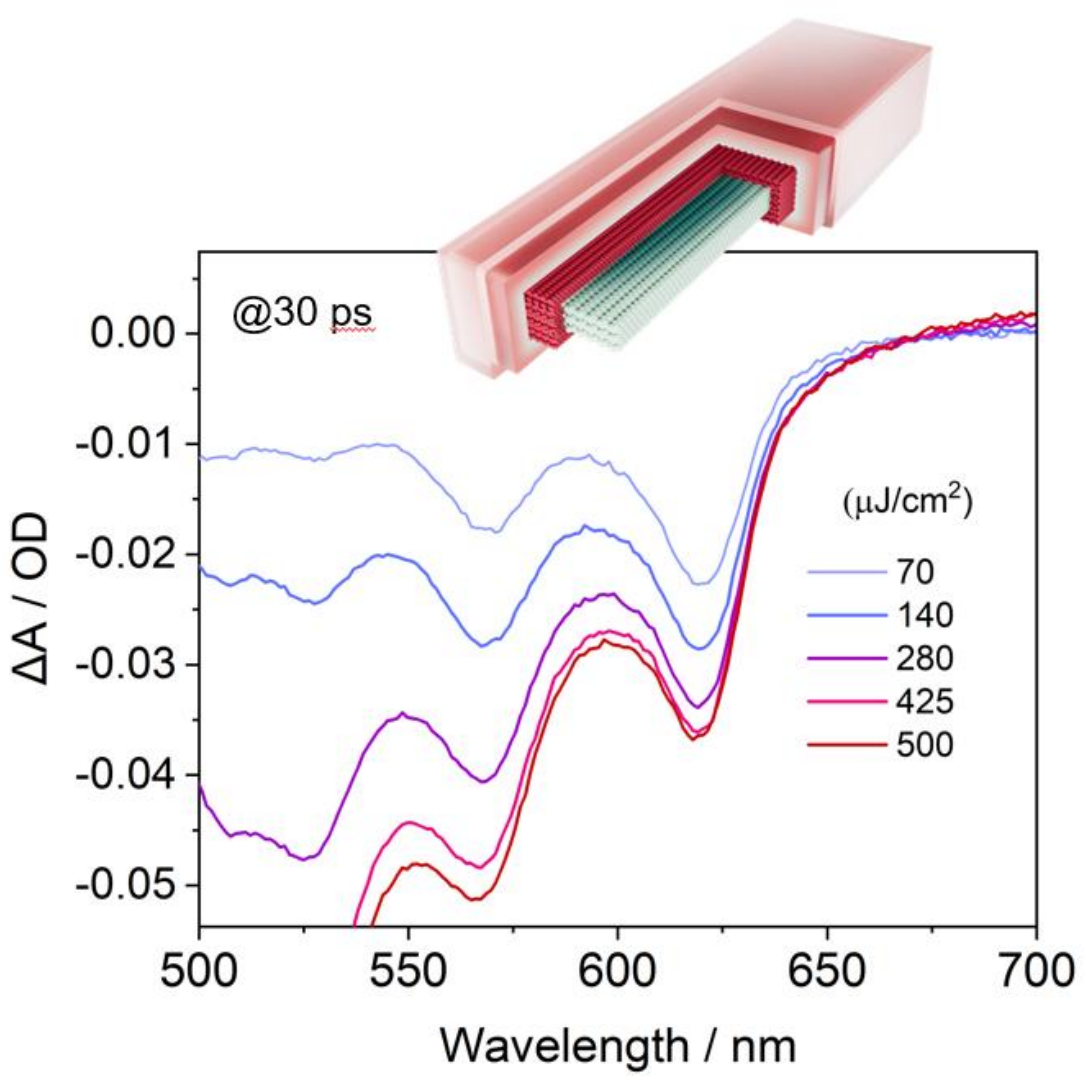

**Supplementary Fig. 37 | Saturation of the bandedge bleach under TA measurements.** TA spectra of GC@QB/2S recorded at the early delay time of 30 ps. The pump fluence was progressively increased to drive the bandedge bleach into saturation. The bleach amplitude reaches saturation at a pump fluence of ~430 μJ/cm$^2$.

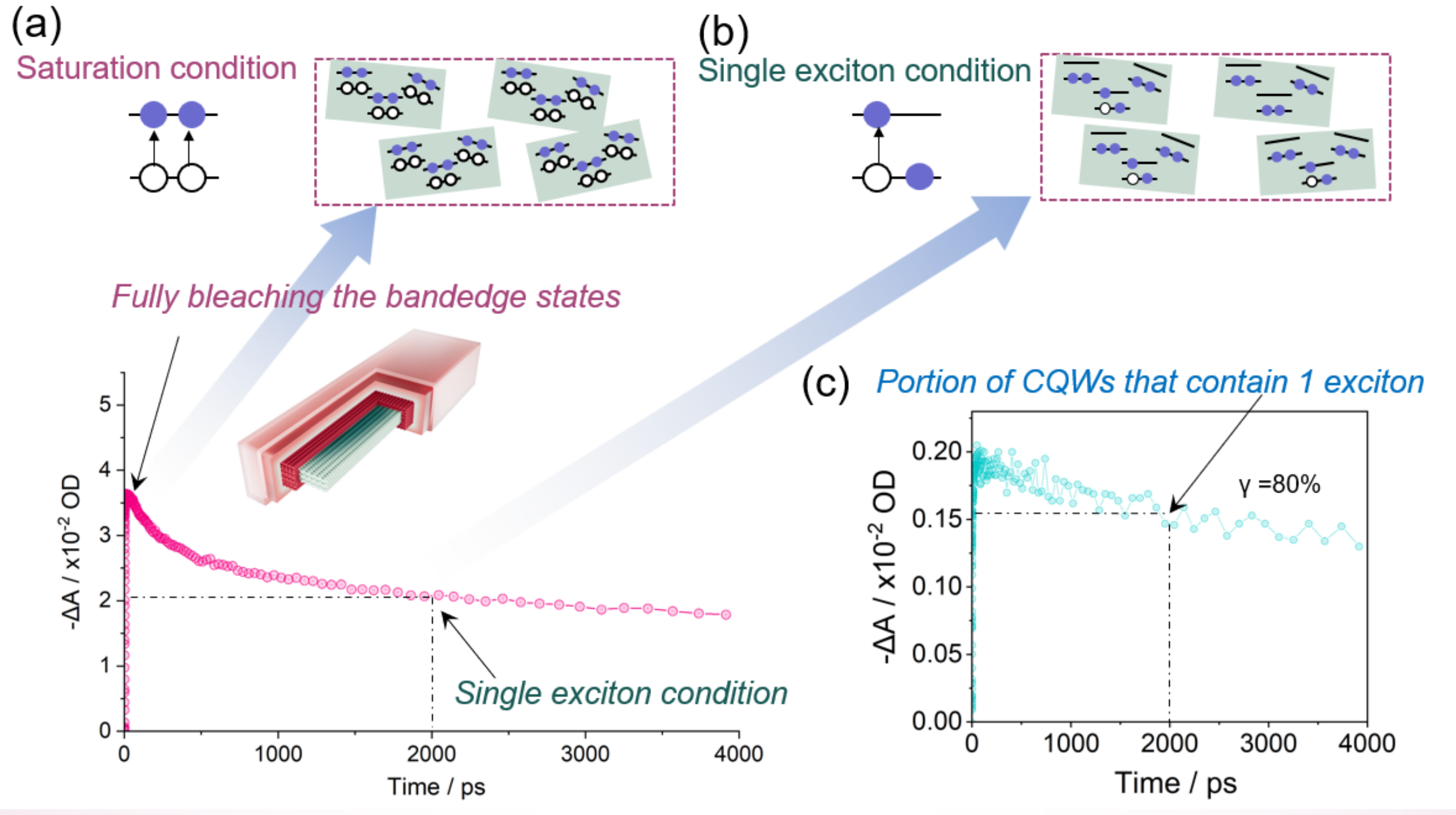

**Supplementary Fig. 38 | Determination of the number of bandedge states. a,** At early delay time immediately following exciton transfer, complete saturation (bleaching) of the bandedge excitonic peak indicates that the available bandedge states are fully occupied. **b**, At longer delay time (e.g., 2 ns in this study), multiexcitons have completely decayed, leaving on average fewer than one bandedge exciton per nanoplatelet in solution. **c**, At 2 ns, only some of the nanoplatelets host a single exciton. Therefore, by analyzing the bandedge exciton dynamics under ultralow excitation power, the fraction of nanoplatelets containing one exciton at 2 ns can be determined. With above information, using the relation shown at the bottom of the figure, the average number of bandedge states can thus be quantitatively calculated.

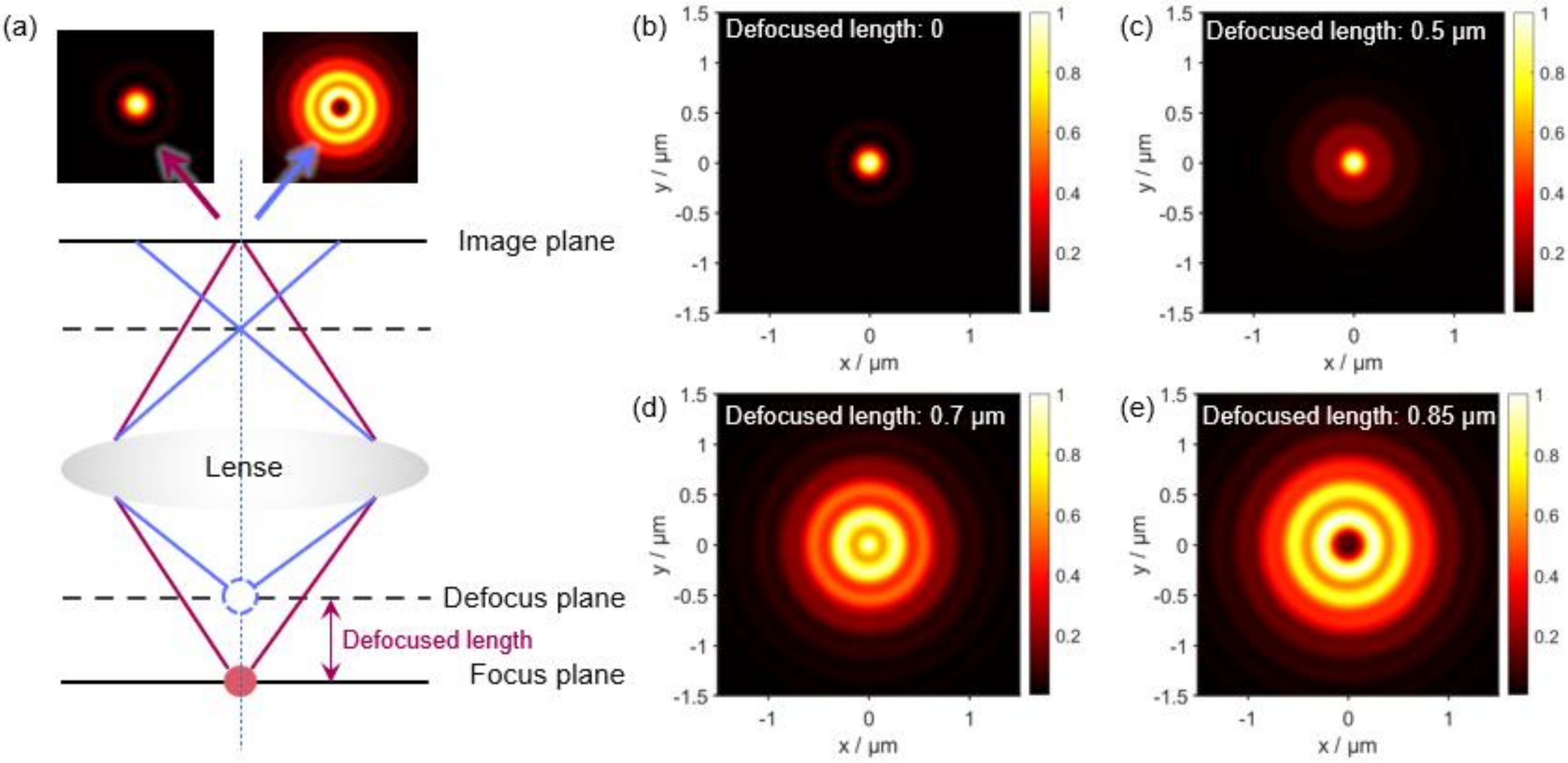

**Supplementary Fig. 39 | Optical simulations of defocused emission patterns at different defocus lengths. a**, Schematic illustration of the optical configuration underlying the formation of defocused patterns, for which the optical simulation of the defocused pattern was carried out following the method[6]. **b-e**, Simulated defocused emission patterns at defocus lengths of (b) 0 μm, (c) 0.5 μm, (d) 0.7 μm, and (e) 0.85 μm, respectively.

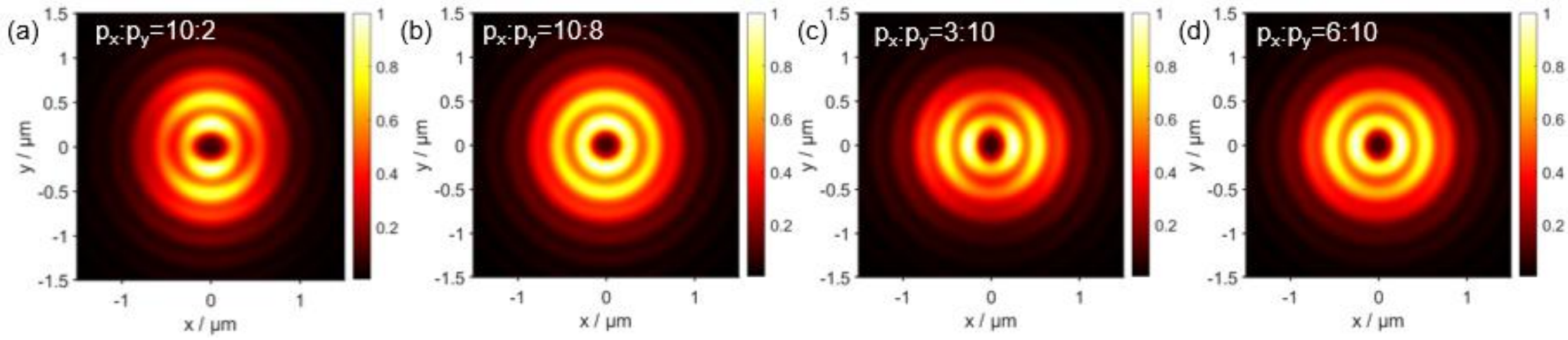

**Supplementary Fig. 40 | Simulated defocused emission patterns of an in-plane dipole with varying in-plane component ratios.** The simulated defocused emission patterns presented for four different in-plane component ratios. **a-d**, $p_x$ and $p_y$ denote the transition dipole moment components along the x- and y-directions in the emitter coordinate system.

.

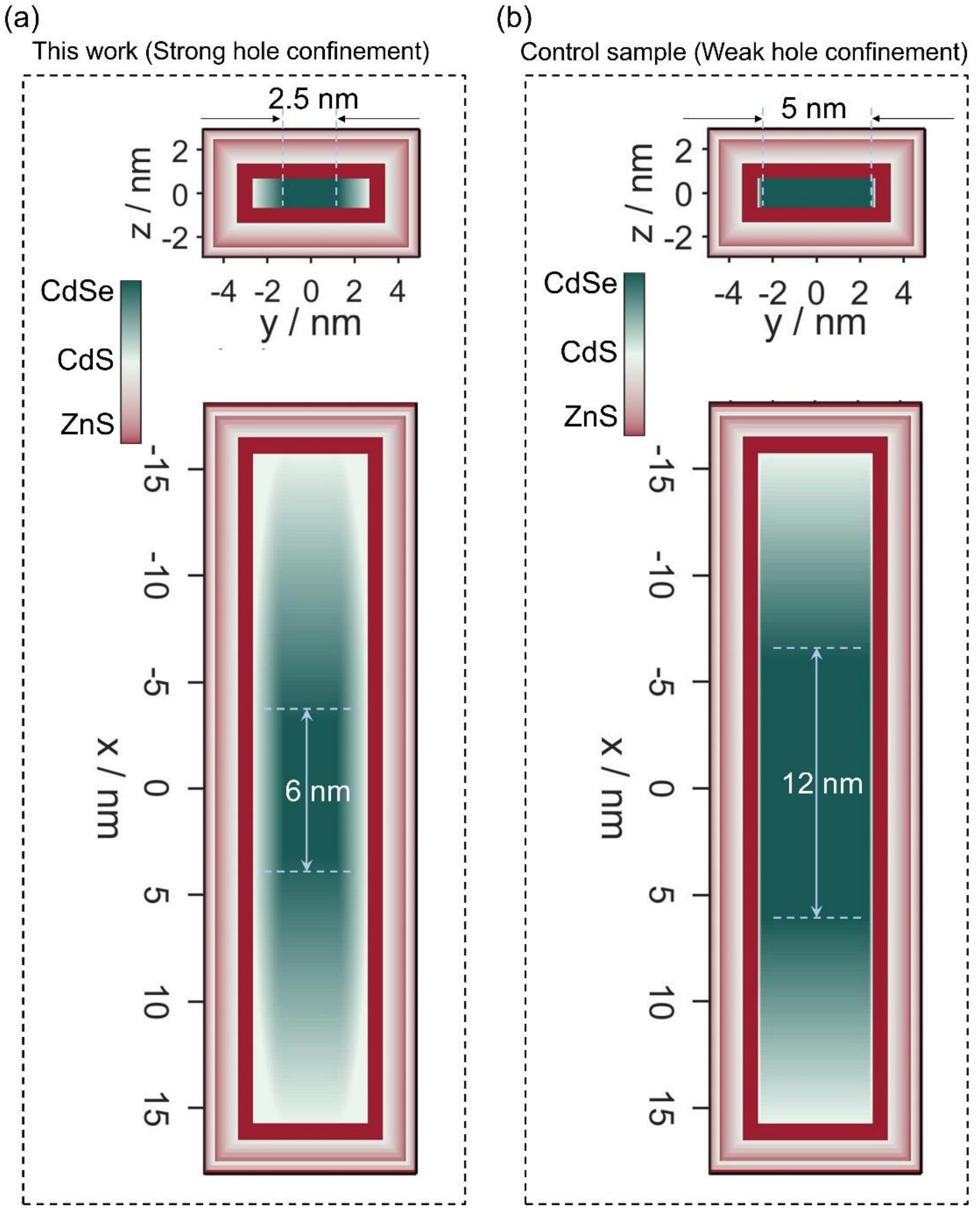

**Supplementary Fig. 41 | Structures used for theoretical evaluation of lateral hole confinement effects on dipole anisotropy. a,** Structure identical to GC@QB/2S, in which the CdSe core is laterally confined to 2.5 nm along the y direction. Owing to the relatively large valence-band offset between CdSe and CdS, this geometry imposes strong lateral confinement on the valence-band hole along the y axis. **b,** Control structure with overall CQW dimensions comparable to GC@QB/2S and identical CdSe aspect ratio. The CdSe lateral dimension along the y direction is increased to 5 nm, thereby reducing the lateral hole confinement strength.

Overall, the primary difference between the two structures lies in the degree of valence-band hole confinement along the y direction.

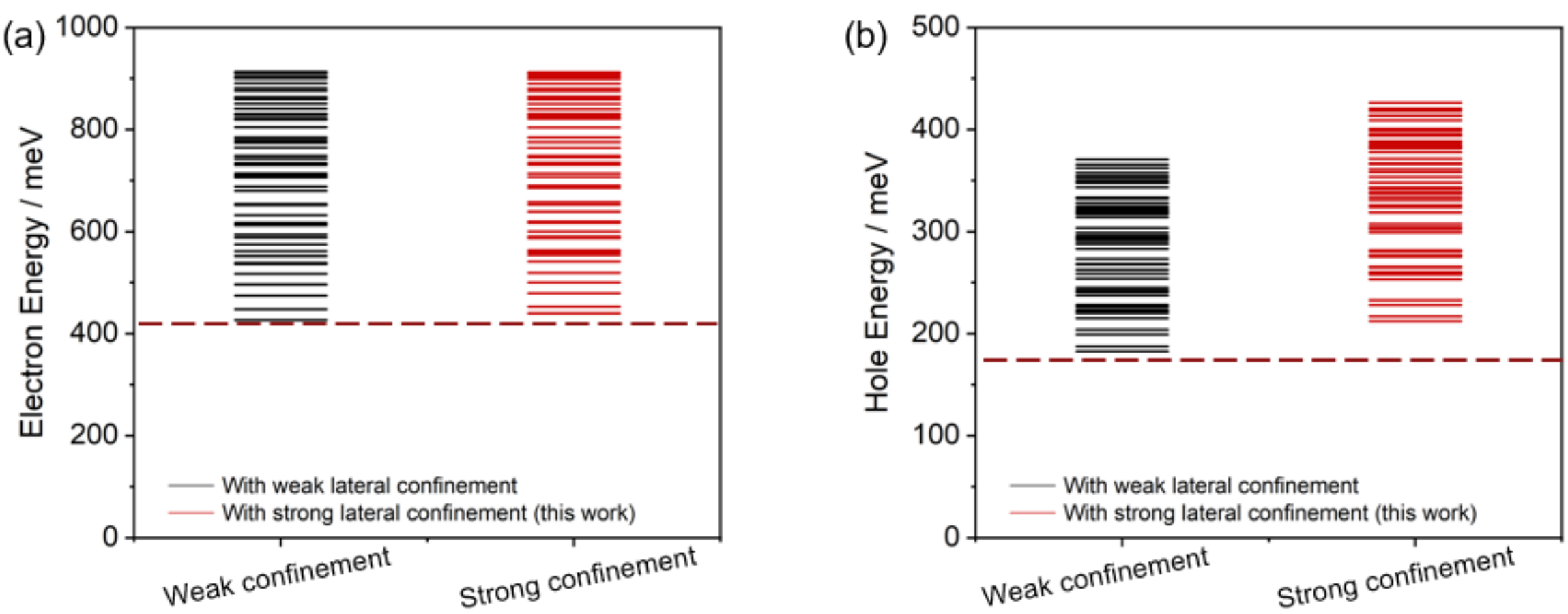

**Supplementary Fig. 42 | Calculated near-bandedge electron and hole eigenenergies, referenced to the CdSe bandedges. a-b,** Comparison of the eigenenergies of (a) electrons and (b) holes for the two structural models in Supplementary Fig. 41. The electron eigenenergies exhibit negligible variation between the two cases, whereas the GC@QB/2S structure with stronger lateral hole confinement shows a clear upward shift in the hole energy.

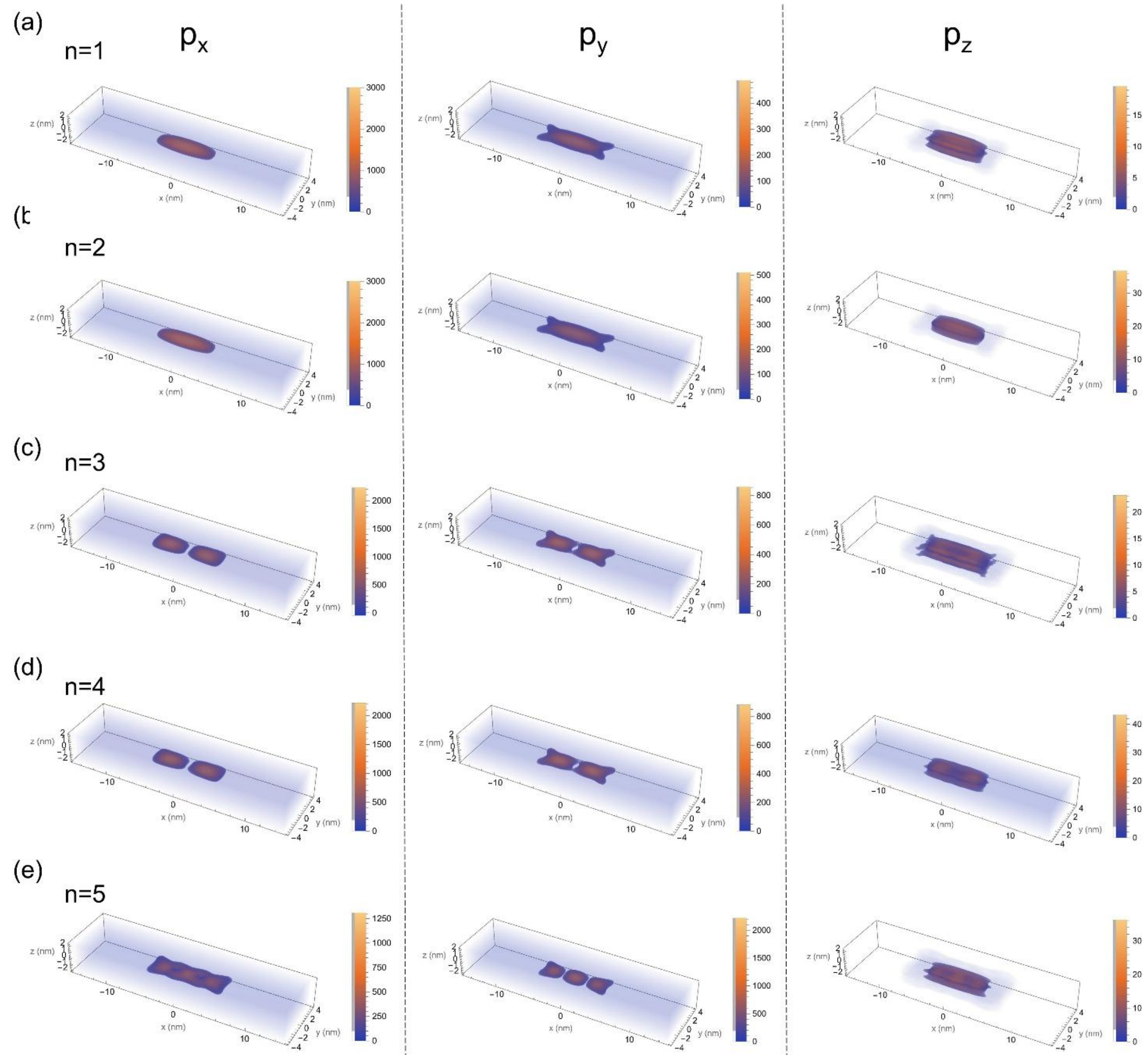

**Supplementary Fig. 43 | Calculated hole probability density distributions of GC@QB/2S. a-e,** Spatial distributions of the hole probability density for the 5 lowest eigenenergies, with projections onto the $p_x$, $p_y$, and $p_z$ orbital components. A total of the first 50 eigenstates were calculated, and only the lowest 5 eigenenergies are shown here for clarity.

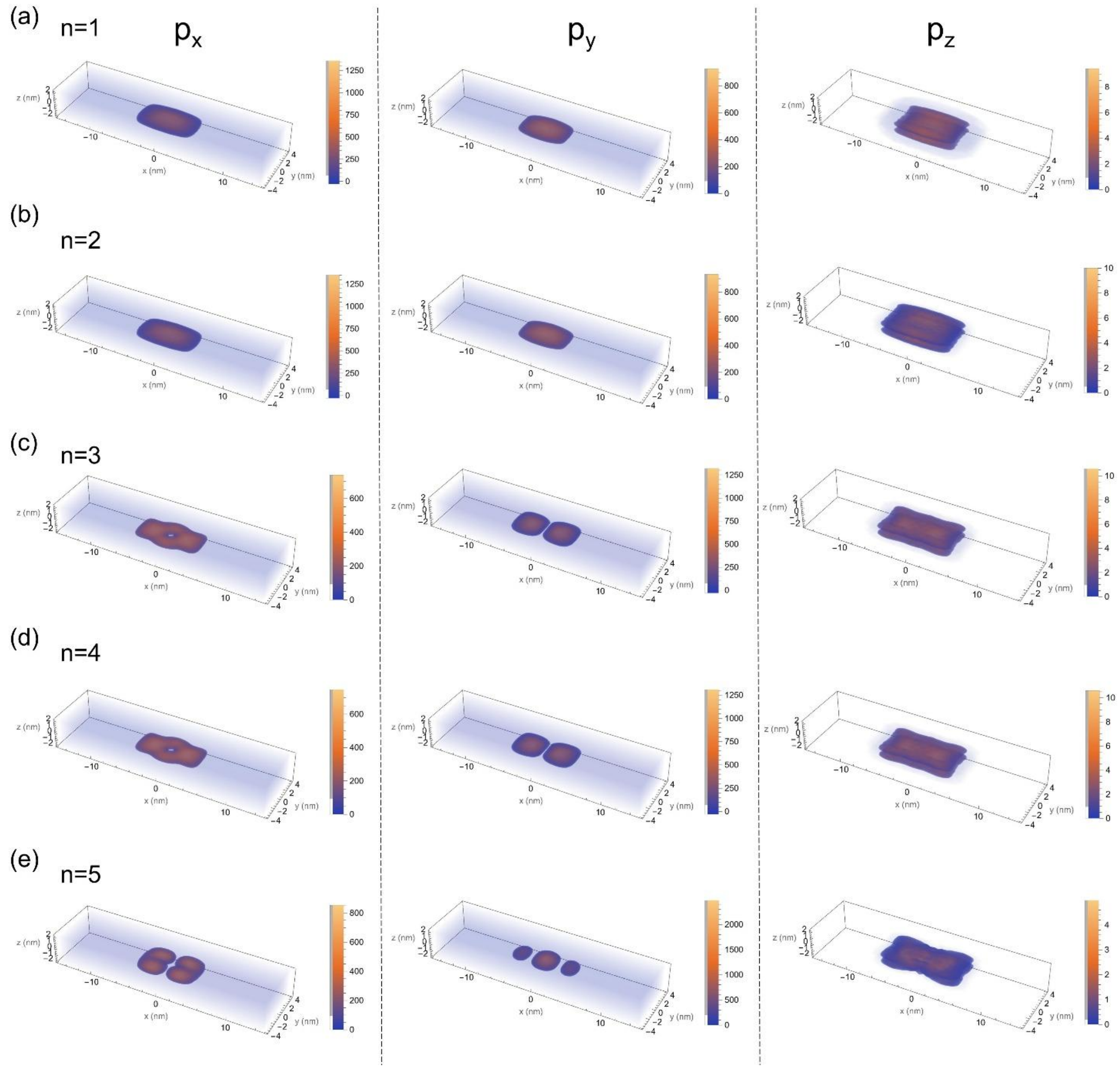

**Supplementary Fig. 44 | Calculated hole probability density distributions of the control model. a-e,** Spatial distributions of the hole probability density for the 5 lowest eigenenergies, with projections onto the $p_x$, $p_y$, and $p_z$ orbital components. A total of the first 50 eigenstates were calculated, and only the lowest 5 eigenenergies are shown here for clarity.

The hole eigenstates solved within the six-band k·p framework were further employed to compute the transition dipole matrix elements between conduction-band electrons and valence-band hole states. The resulting dipole intensities were weighted according to the Boltzmann occupation probabilities of each eigenstate, enabling quantitative evaluation of the degree of polarization (DOP).

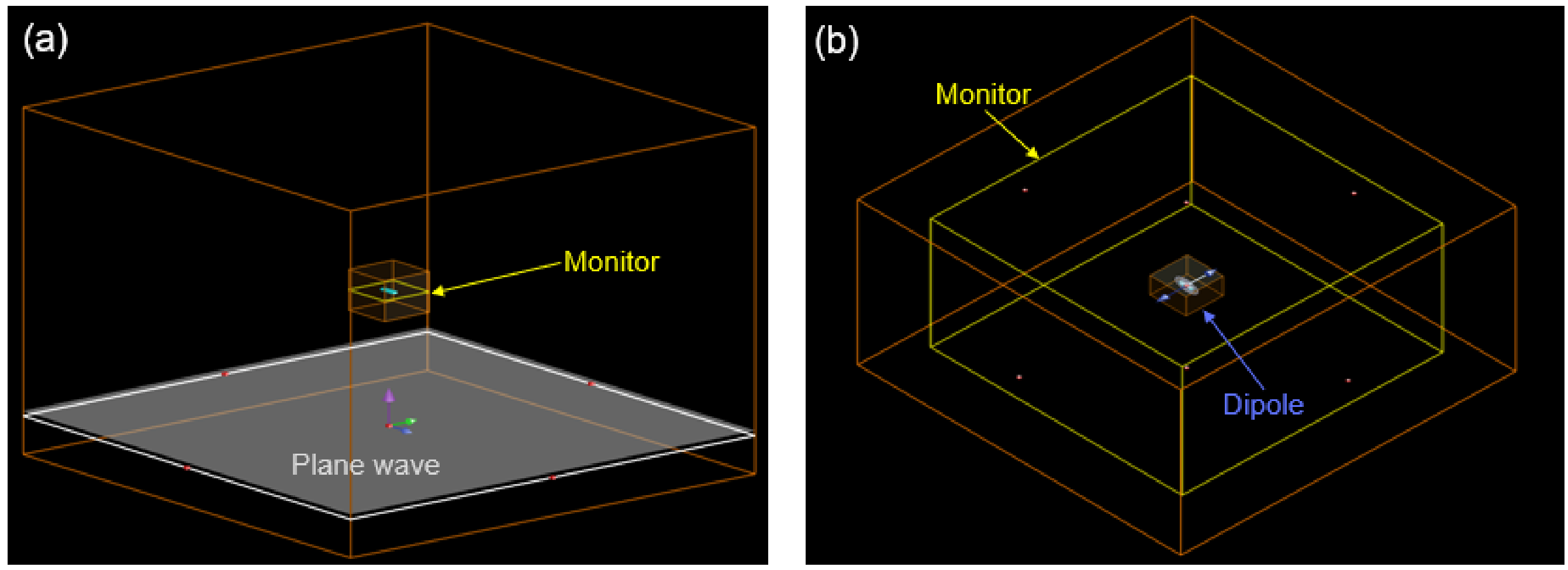

**Supplementary Fig. 45 | Configurations for FDTD optical simulations.** In simulations, the CQW was treated as a homogeneous dielectric object to evaluate how dielectric contrast modulates the anisotropy of the electric-field intensity inside the CQW. **a**, A plane wave was used as the excitation source, and a field monitor was placed at the geometric mid-plane of the CQW to record the spatial distribution of the in-plane electric-field intensity within the structure. **b**, For far-field radiation simulations, an electric dipole emitter was positioned at the CQW center, and the far-field intensity distribution was obtained using a near-to-far-field transformation to generate the corresponding polar plots.

Two orthogonal, equal-strength dipole polarizations were simulated separately to obtain the polarization-resolved far-field intensity patterns, which were then combined to yield the overall far-field polar distribution.

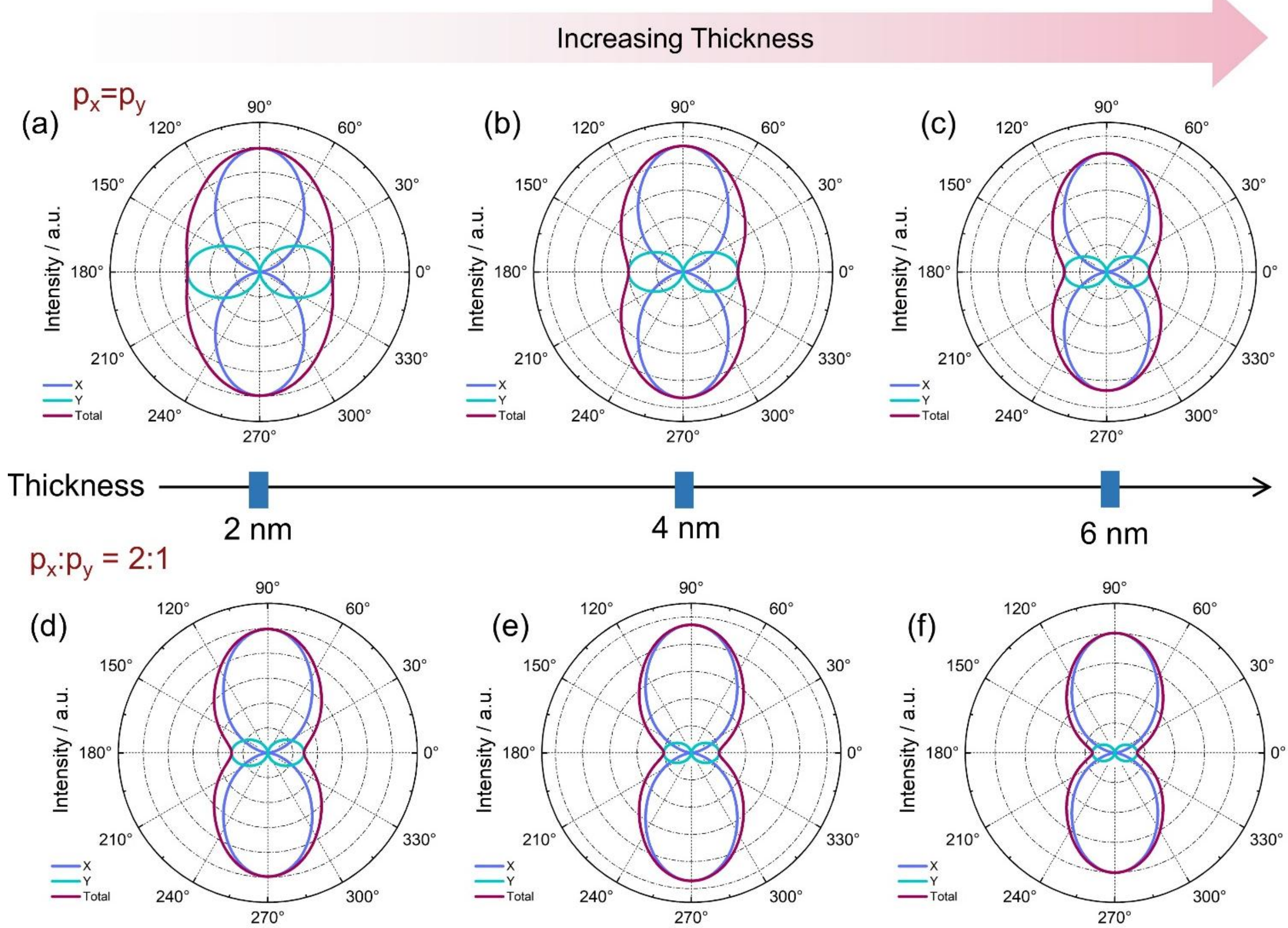

**Supplementary Fig. 46 | Thickness-dependent modulation of far-field polar intensity distributions. a-c**, Evolution of the far-field polar intensity distributions with increasing CQW thickness for an isotropic in-plane dipole ($p_x = p_y$). As the dipole anisotropy is absent in this case, the observed redistribution of the far-field intensity arises solely from dielectric-induced electric-field anisotropy within the CQW. **d-f**, Corresponding thickness-dependent evolution for an anisotropic in-plane dipole with $p_x:p_y = 2:1$. In this case, the far-field polar distribution is determined by the combined effects of intrinsic dipole anisotropy and thickness-dependent electric-field modulation.